\documentclass[prc,twocolumn,showpacs,showkeys,superscriptaddress,nofootinbib]{revtex4}
\usepackage{graphicx,amsmath,amssymb}
\def\be{\begin{eqnarray}}
\def\ee{\end{eqnarray}}
\def\bc{\begin{center}}
\def\ec{\end{center}}
\def\dsp{\displaystyle}

\def\rmF{{\rm F}}
\def\rmd{{\rm d}}

\def\om{\omega}
\def\Tc{{\mathcal T}_{\mathcal C}}

\def\uc{\underline{c}}
\def\ud{\underline{d}}

\newcommand{\lsim}{\stackrel{\scriptstyle <}{\phantom{}_{\sim}}}
\newcommand{\gsim}{\stackrel{\scriptstyle >}{\phantom{}_{\sim}}}

\newcommand{\op}[1]{%
    \fontdimen12\textfont3=2pt\fontdimen12\scriptfont3=1.4pt%
    \!\null\mathop{\vphantom{#1}\smash{#1}}\limits_{\sim}\null\!}

\def\piproj{\hat{\pi}}

\begin{document}
\title{
Superfluid nucleon matter in and out of equilibrium and weak interactions\footnote{The paper
is dedicated to the memory of A.B. Migdal on occasion of his the 100-th anniversary}}
\author{\firstname{E.E.}~\surname{Kolomeitsev}}\email{kolomeitsev@fpv.umb.sk}
\affiliation{Matej Bel  University, SK-97401 Banska Bystrica, Slovakia}%
\author{\firstname{D.N.}~\surname{Voskresensky}}\email{D.Voskresensky@gsi.de}%
\affiliation{ National Research Nuclear University "MEPhI", Kashirskoe Avenue 31, 115409 Moscow, Russia}%
\affiliation{GSI Helmholtzzentrum f\"ur Schwerionenforschung (GSI), Planckstr. 1, D-64291, Darmstadt, Germany}%
\date{\today}

\begin{abstract}
The Larkin-Migdal approach to a cold superfluid Fermi liquid is generalized for a non-equilibrium
system. The Schwinger-Keldysh diagram technique is applied. The developed formalism
is applicable to the pairing in the states with arbitrary angular momenta. We consider the white
body radiation problem by calculating probabilities of different direct reactions from a piece of
a fermion superfluid. The closed diagram technique is formulated in terms of the full Green's
functions for systems with the pairing correlation. The cutting rules are used to classify the
diagrams representing one-nucleon, two-nucleon, etc. processes in the matter. The important role
of multi-piece diagrams for the vector-current conservation is demonstrated. In the case of
equilibrated systems, dealing with dressed Green's functions, we demonstrate correspondence
between calculations in the Schwinger-Kadanoff-Baym-Keldysh formalism and the ordinary Matsubara
technique. As an example we consider neutrino radiation from the neutron pair breaking and
formation processes in case of a singlet pairing. Necessary correlation effects are included. The
in-medium renormalization of normal and anomalous vertices is performed.
\end{abstract}

\maketitle

\tableofcontents

\section{Introduction}\label{sec:intor}

\subsection{Historical remarks} \label{subsec:history}

The phenomenological theory of normal Fermi liquids at zero temperature was proposed by L.D.
Landau in Refs.~\cite{Landau58,LP1981}. A.B. Migdal made the very important observation that a
jump in the particle momentum distribution at the Fermi momentum corresponds to a pole of the
fermion Green's function in the normal Fermi liquid and is preserved even in the strongly interacting
system \cite{Mjump}. The presence of the pole contribution to the fermion Green's function allowed
Galitsky and Migdal to develop the Green's function formalism for many-body fermionic systems, see Ref.~\cite{Mqp}.
These concepts were first elaborated on example of low-lying
particle-hole excitations in Fermi liquids. A.B. Migdal was first who applied these methods to
description of various nuclear phenomena and constructed a closed semi-microscopic approach that
is usually called "Theory of finite Fermi systems"~\cite{M67a,M67}.

A general understanding of the phenomenon of superconductivity in case of the weak attraction
between fermions was achieved by J.~Bardeen, L.N.~Cooper and J.R.~Schriffer in Ref.~\cite{BCS57},
see also Ref.~\cite{Schriffer} for detailed exposition. Due to the sharpness of the Fermi surface
provided by presence of the Migdal's jump one can consider the fermions on the Fermi surface as
moving in an effective two-dimensional momentum space. It follows immediately that even a weak
attraction between two particles is sufficient to form a Cooper pair. As soon
as the pairing phenomenon is established one can follow different routes in description of the
superconductivity and superfluidity: In Ref.~\cite{Bog58} N.N.~Bogolyubov suggested a very
convenient transformation from the particle $\psi$ operators to the new operators of effective
excitations on top of the background of Cooper pairs. This transformation is broadly used in the
theory of superconductivity. L.P.~Gorkov developed the Green's function  formalism for superconducting
fermion systems with an electron-phonon interaction~\cite{Gor58}. Y.~Nambu introduced a matrix
formalism to the theory of superconducting metals \cite{Nam60} (Green's functions formulated in
the Nambu-Gorkov space). In Ref.~\cite{El60} G.M.~Eliashberg extended the A.B.~Migdal's theory of the strong
electron-phonon interaction in normal metals~\cite{MigEl} to include the Cooper pairs. This
approach can be used to describe strong coupling superconductors. Also A.B.~Migdal was the first
who rose the idea about a possibility of the neutron-neutron pairing and superfluidity in neutron
stars, which where hypothetical objects that time~\cite{M59}.

The Fermi liquid theory was then generalized by A.I.~Larkin and A.B.~Migdal for the description of
fermion superfluids at zero temperature~\cite{LM63}. Their formulation is more general than that
done in papers by Y. Nambu and L.P.~Gorkov,  since it allows for different interactions in the
particle-particle and particle-hole channels. A.J.~Leggett applied this formalism for the
superfluid $^3$He at a finite temperature~\cite{Leg65a,Leg65}. J.~Schwinger in
Ref.~\cite{Schw61}, L.P.~Kadanoff and G.~Baym in Ref.~\cite{KB62}, and L.P.~Keldysh in
Ref.~\cite{Kel64} developed the non-equilibrium diagram technique for the description of
non-equilibrium Fermi and Bose systems. Even for equilibrium systems at $T\neq 0$
Schwinger-Kadanoff-Baym-Keldysh approach is in many cases more convenient than the standard
Matsubara technique (applicable only for equilibrium systems) since it does not involve the Wick
rotation and the obtained results can be continuously transformed
to those computed in the standard Feynman-diagram technique at zero temperature.

The importance of coherence time effects on the production and absorption of field quanta from the
motion of source particles in matter has been first discussed by L.D.~Landau and
I.Ya.~Pomeranchuk~\cite{LandauP}.  In Ref.~\cite{LPM} A.B.~Migdal developed the complete
theoretical framework for the description of the bremsstrahlung radiation of ultra-relativistic electrons in
the process of multiple rescatterings on Coulomb centers. Successful measurements of such a
suppression of the bremsstrahlung radiation have been recently carried out at the Stanford Linear
Accelerator Center~\cite{eSLAC}, see also the review in Ref.~\cite{SKlein}. Now this effect is
named the Landau-Pomeranchuk-Migdal effect.

In the framework of his theory of finite Fermi systems A.B.~Migdal developed the description of the
soft pion degree of freedom in nuclear matter in application to atomic nuclei and neutron stars.
In vacuum, pions are the lightest quanta of the strong interactions between baryons.
In medium, pionic modes are softened even further due to the coupling to nucleon
particle-hole modes and can be easily excited even at low excitation energies, similar to
phonons in solids. As an intriguing
consequence of the pion softening A.B. Migdal suggested a possibility of the pion
condensation\footnote{Independently pion condensation was also suggested by D.J. Scalapino and R.
Sawyer \cite{SS72}.} at the increase of the baryon density, see Refs.~\cite{M71,M78,MSTV90}. Latter on, in
analogy to the pion condensation, the ideas of the kaon condensation \cite{BLRT94} and the charged
$\rho$ meson condensation \cite{V97,KV05} in the interiors of neutron stars were explored. Softening
of the pionic mode at finite temperature~\cite{VM78,VM82,D82} and at
non-equilibrium~\cite{VS87,V93,VBRS95} may manifest in neutron stars~\cite{V01} and heavy ion
collisions \cite{V93,RW}.

A.B. Migdal rose question on a possibility of existence of superdense abnormal nuclei glued by
the pion condensate \cite{M71,MMMS77}. Also a possibility of nuclei-stars was considered in
Ref.~\cite{VSC77}. The similar ideas on a possibility of quark nuclei, quark stars and hybrid stars
\cite{W84,AFO} are continued to be extensively explored nowadays, see \cite{Bombaci,BKV00}.

\subsection{White body radiation}

Below we consider the white-body radiation from a piece of a superfluid fermion matter. To be
specific we focus on the neutrino radiation from a piece of superfluid nucleon matter. Standard
Feynman technique of summation of squared matrix elements of reactions fails to calculate reaction
rates in the medium, since in general case there are no asymptotic states for source particles in
matter. Indeed, source particles continue to collide before and after radiation of a quantum.
This gives rise to finite imaginary parts of the self-energy
functions (particle widths). If one naively replaced the summation of all perturbative Feynman
diagrams (with free Green's functions)  by the summation of corresponding diagrams with dressed
Green's functions, it would lead to a double counting due to multiple repetitions of some
processes (for an extensive discussion of how one can treat this defect see
\cite{VS86,VS87,KV95,KV99}). This calls for a formalism dealing with closed diagrams (integrated over
all possible in-medium particle states) with full non-equilibrium Green's functions. Such a
general formalism was developed in Ref. \cite{KV95}. It treats on equal footing one-fermion and
multi-fermion processes as well as resonance reaction contributions of the boson origin, such as
processes with participation of zero sounds and reactions on the boson condensates. Decomposition
of  diagrams is done in terms of the full $G^{-+}$ Green's functions (Wigner densities). Each
diagram in the series with full Green's functions is free from the infrared divergencies. In such
a way one generalizes Landau-Pomeranchuk-Migdal treatment of the multiple scattering on external
centers to the treatment of the multiple scattering in matter. Both, the correct quasi-particle and
quasi-classical limits are recovered.

The formulation of the radiation problem in terms of closed diagrams calculated within the
non-equilibrium Green's function in quasi-particle approximation was performed in Ref.~\cite{VS87}
This approach was called the {\em "optical theorem formalism"}. In Refs.~\cite{VS86,VS87} it was
demonstrated that the standard calculations of reaction rates via integration of squared reaction
matrix elements and the results of the optical theorem formalism match exactly, provided
conditions for the quasi-particle approximation for fermions are fulfilled. Formally the matching
is done by cutting the closed diagrams. In general case considered in Ref.~\cite{KV95} going
beyond the quasi-particle approximation, the series of closed diagrams is constructed with respect
to the number of the $G^{-+}$ Green's functions. For low temperatures each $G^{-+}$ line brings
extra $(T/\epsilon_{\rm F})^2$ factor in the production rate of the radiating quanta,
$\epsilon_{\rm F}$ is the Fermi energy. In Ref.~\cite{KV95} the relations between reaction rates
at finite widths and the quasi-particle rates were found.

All real calculations of fermion superfluids were performed within quasi-particle approximation for fermions
(when fermion width is much less than all other relevant energetic scales). Below we focus on the
Larkin-Migdal approach to the cold fermion superfluids and formulate it in terms of the
Schwinger-Kadanoff-Baym-Keldysh technique to describe
fermion superfluids in equilibrium at  $T\neq 0$ and out of equilibrium.

\subsection{Neutrino cooling of neutron stars}

Physics of neutron star cooling is based on a number of ingredients, among which the neutrino
emissivity of the high-density hadronic matter in the star core is the important one.

After the first tens of seconds (at most hours), the typical temperature of a neutron star
decreases below  the so-called neutrino-opacity temperature $T_{\rm opac}\sim (1 - 2)$~MeV.
At these conditions neutrinos and anti-neutrinos can be radiated directly from the star interiors without
subsequent rescattering, since their mean-free path is much longer than the star
radius~\cite{VS86}. Hence, the star can be considered as a piece of a warm ``white''
body for neutrinos. Typical averaged lepton energy ($\sim$ several $T$) is much larger than the
nucleon particle width $\Gamma_N \sim T^2/\varepsilon_{{\rm F}}$.  Therefore, the nucleons can be
treated within the quasi-particle approximation. This observation simplifies consideration
essentially. One usually follows an intuitive way for the  separation of the processes according
to their phase spaces. The one-nucleon processes (if they are not forbidden by the energy-momentum
conservation) have the largest emissivity, $\epsilon_{\nu}\propto T^6$ for non-superfluid systems,
then two-nucleon processes come into play, $\epsilon_{\nu}\propto T^8$, and so on. In the optical
theorem formalism one-nucleon processes are determined by the self-energy $\Sigma^{-+}$ of virtual
$W$ and $Z$ bosons expanded in the series with respect to the number, $N$, of $G^{-+}G^{+-}$ loops
with full $"++"$ and $"--"$ vertices. The $N=1$ diagrams correspond to one-nucleon processes, the
$N=2$ diagrams to the two-nucleon processes, etc.

In the so-called {\em "standard scenario"} of the neutron star cooling, the processes were
calculated without taking into account in-medium effects. It was argued that the most important
channel at temperatures up to $T\sim 10^{8}$--$10^{9}$~K is the modified Urca (MU) process $n \, n
\rightarrow n \, p\, e\, \bar \nu$. First estimates of the MU emissivity were done in
\cite{BW65,TC65}. In Ref.~\cite{FM79,M79} B.~Friman and O.V.~Maxwell recalculated the emissivity of this
process in the model, where the nucleon-nucleon (NN) interaction was approximated by a free
one-pion exchange (FOPE).  The expression of the neutrino emissivity obtained by them was used in
various computer simulations, e.g., in Refs.~\cite{T79,NT81,SWWG96}. Besides the MU process, the
{\em "standard scenario"} includes also the processes of the nucleon (neutron and proton)
bremsstrahlung (NB) $n\, n\rightarrow n\, n \nu \bar{\nu}$ and $n\, p\rightarrow n\, p \nu
\bar{\nu}$, which contributions to the emissivity is smaller than those of the MU processes, see
Refs.~\cite{FSB75,FM79}. The density dependence of the reaction rates calculated with the FOPE is
rather weak and the neutrino radiation from a neutron star depends very weakly on the star mass.

There exists another class of so-called ``exotic'' processes, which occur only if some special
condition is fulfilled, i.e. when the nucleon density exceeds some critical values.
These are the direct Urca (DU)
processes on nucleons (e.g., $n\rightarrow pe\bar{\nu}$) and hyperons \cite{LPPH91}, pion Urca
reactions on a pion  condensate \cite{MBCDM77,VS84,VS86}, kaon
Urca processes on a kaon condensate \cite{BKPP88,T88}, $\rho$-Urca processes on charged
$\rho$-condensates \cite{V97,KV05}, DU processes on quarks \cite{Iwamoto}, DU processes on fermion
condensates \cite{VKZC00}. The values of critical densities are different for various processes
and are model dependent. For example, some relativistic mean-field models
produce the critical density of the DU reaction, $n_{c}^{\rm DU}$ as low as the nuclear saturation
density $n_0 \simeq 0.16$~fm$^{-3}$. However, the realistic, microscopically-based Urbana-Argone
equation of state~\cite{APR98} yields $n_{c}^{\rm DU}\simeq 5n_0$. The simulations of the neutron star
cooling history in Refs.~\cite{BGV,GV} have shown that the occurrence of the DU processes in the
neutron star with masses $M<1.5 M_{\odot}$ would lead to problems with the explanation of soft $X$-ray
data. The constraint on the equation of state of the dense nuclear matter, requiring a sufficiently
high value of $n_{c}^{\rm DU}$, was proposed in Ref.~\cite{KV05} and explored in details in
Ref.~\cite{Army}.

It was shown in Refs.~\cite{VS84,VS86,VS87,SV87,MSTV90,V01} that the neutrino emission from dense
hadronic component in neutron stars is subject of strong modifications due to collective effects
in the nuclear matter. Many new reaction channels open up in medium in comparison to the vacuum.
In Refs.~\cite{VS84,VS86,VS87,SV87,MSTV90,V01} the nucleon-nucleon interaction was considered
within the Landau--Migdal approach to Fermi liquids. The softening of the in-medium one-pion
exchange (MOPE) mode and other medium polarization effects, like nucleon-nucleon correlations in
the vertices, renormalization of the local part of $NN$ interaction due to loop effects, as well as a
possibility of the neutrino emission from intermediate reaction states and DU-like reactions
involving  zero sounds and boson condensates were incorporated. It was demonstrated in
Refs.~\cite{VS86,SV87,MSTV90,V01}, that for $n \gsim n_0$ the neutrino emissivity is mainly
determined by the medium modified Urca (MMU) process, in which the neutrino is radiated from the
intermediate reaction states. This fact changes significantly the absolute value and the density
dependence of the $nn\rightarrow npe\bar{\nu}$ process rate. The latter becomes very strong.
Therefore, for neutron stars with larger masses the resulting emissivity of the MMU process proves
to be substantially higher than the corresponding value (MU) calculated in the FOPE model of
Ref.~\cite{FM79}. For $n\gsim n_0$, the medium-modified nucleon bremsstrahlung (MNB) processes
yield a smaller contribution than MMU ones since the former does not include the neutrino
radiation from intermediate states, e.g. from intermediate pion. However, the MNB processes are
more efficient than the NB ones for such densities.

Oppositely, for $n<n_0$ the in-medium effects can moderately suppress the two-nucleon reaction
rates compared with those given by the FOPE model~\cite{BRSSV,HPR,Schwenk04}. The pion softening effect
disappears at $n\lsim 0.5$--$0.7\,n_0$, see Refs.~\cite{MSTV90,V01}, but the nucleon-nucleon
short-range repulsion effect remains. Inclusion of the nucleon-nulceon correlations without the
pion softening~\cite{RPLP} yields a suppression effect also at $n\gsim n_0$. Obviously, this
effect also follows from general consideration in Refs.~\cite{VS86,MSTV90}, if one artificially
suppresses the pion softening effect.

After the seminal work of A.B.~Migdal~\cite{M59}, various aspects of the nucleon superfluidity in
neutron stars were studied in the literature: The presence of a nucleon superfluid interacting with
the normal component is needed for explanation of glitches in pulsar periods and neutron
star quakes~\cite{ST83}. Explanation of pulsar cooling curves also requires an inclusion of
superfluid phases~\cite{GV}. Several superfluid phases are found possible. Phase transitions
between different phases may take place~\cite{KCZ}.

It is commonly accepted that most important are the superfluid phases with the spin-singlet pairing
of neutrons and protons, in the $1S_0$ state, and the spin-triplet pairing of neutrons in the
$3P_2$ states. The latter is believed to occur in neutron star interior at $n\gsim n_0$ in the
state with $m_J =0$, where $m_J$ is the projection of the total pair momentum onto a quantization axis.
In case $|m_J| =2$ the exponential suppression of the specific heat and the neutrino emissivity is
replaced by a power-law suppression since the gap vanishes at the poles of the Fermi sphere. This
possibility was mentioned for the first time in Ref.~\cite{VS87}, the corresponding reaction rates
were calculated in Ref.~\cite{YKL99}. However, a mechanism to realize  this interesting possibility
in neutron star cooling was not elaborated yet.

Many papers are devoted to the calculation of pairing gaps within different
approaches~\cite{APW,Tamagaki70,Amundsen85,Takatsuka93,KKC96,Schulze96,Elgaroy98,Khodel01,SF,KCTZ,Hebeler07,Chen08}.
The obtained results can be essentially different depending on a model for the nucleon-nucleon
interaction and a calculation scheme. The predictions of the neutron $3P_2$ gaps are especially
uncertain, e.g., compare Refs.~\cite{SF} and ~\cite{KCTZ}. For review see Ref.~\cite{SC06} and references
there in. Ref.~\cite{SF} argues that $3P_2$ gap should be
strongly
suppressed whereas Ref.~\cite{KCTZ} argues for its strong
enhancement. Reference~\cite{GV} calculated cooling curves using both
these assumptions and concluded that the cooling history is
naturally explained within assumption on  the suppressed  $3P_2$ gap.
Recently Ref.~\cite{PPLS10} studied the new data on the cooling of Cas A
object. Their conclusion is in favor of a suppressed $3P_2$ gap.
 At temperatures below the critical  temperatures of the neutron, $T_{cn}$, and proton,
$T_{cp}$, pairing, the reaction rates, considered above, are suppressed because of a decrease of the
available phase space. Initially, the suppression effects were included simply by multiplying the
rate of a two-nucleon process by the factors $e^{-2\Delta/T}$~\cite{M79}. Later, the phase-space
suppression factors (so called $R$ factors) have been treated more accurately in
Ref.~\cite{YLS99}.

In nucleon superfluids, there exist new neutrino-production mechanisms, which are forbidden for
$T>T_c$. These are the processes, suggested in Ref.~\cite{FRS76,VS87,SV87}, in which the creation of a
neutrino--anti-neutrino pair is associated with the breaking and formation of a Cooper pair -- the
so-called nucleon  pair breaking and formation (PBF) processes.
The emissivities of the nucleon PBF processes are suppressed at $T<T_c$ by the same
factor $\sim \exp (-2\Delta/T)$ as for the MU, NB, MMU and MNB processes. However,
in comparison to the all latter processes, the nucleon PBF processes have
the large one-nucleon phase-space volume~\cite{VS87,SV87}. The
existence of this new cooling mechanism demonstrates that influence of the nucleon pairing on
the neutrino production rates cannot be reduced just to an introduction of a simple phase-space
suppression factor.

Early works~\cite{FRS76,VS87,SV87,V01,KHY,YKL99,YLS99} which studied the PBF processes, did not care
about the conservation of the weak vector current. The latter is fulfilled only if the in-medium
renormalization of weak vertex functions is performed in accord with the renormalization of Green's
functions. This problem was tackled in Refs.~\cite{LP,KV1,KV2,KR,LPpairing,SMS,SR09}.
Reference~\cite{LP} argued that the emissivity of the $1S_0$ PBF processes should be dramatically
suppressed as $\propto v_{\rm F}^4$, where $v_{\rm F}$ is the Fermi velocity of non-relativistic
nucleons, provided the vector current conservation constraint is taken into account. The consistent
calculation of the PBF emissivity induced by the vector and axial-vector currents was performed in
Ref.~\cite{KV1,KV2} within the Larkin-Migdal-Leggett Fermi-liquid approach. The latter takes
properly into account correlation effects in both particle-particle and particle-hole channels. It
was demonstrated that the neutrino emissivity is actually controlled by the axial-vector current
and is suppressed only by the factor $\propto v_{\rm F}^2$, rather than $\propto v_{\rm F}^4$. Both
neutron PBF and proton PBF processes yield contributions of the same order of magnitude provided
strong and electromagnetic renormalizations of the proton weak vertices \cite{VS87,VKK,L00} are
included. In Ref.~\cite{LPpairing} one argues that for the $3P_2$ neutron pairing the vector
current conservation changes moderately the result obtained without its inclusion. As pointed out in
Ref.~\cite{GBSMK}, the suppression of the PBF processes at low densities might be served as a
possible explanation of the superburst ignition.

As we have mentioned, the convenient Nambu-Gorkov formalism developed for the description of
metallic superconductors, cf. Refs.~\cite{Nam60,Gor58,Schriffer}, does not distinguish interactions
in particle-particle and particle-hole channels. These interactions can be, however, essentially
different in a strongly interacting system, like in the nuclear matter and in the liquid He$^3$.
The adequate methods for Fermi liquids with pairing were developed for zero temperature by A.I.~Larkin
and A.B~Migdal in Ref.~\cite{LM63} (see also \cite{M67}) and for a finite temperature by A.J.~Leggett in
Ref.~\cite{Leg65a,Leg65}. The problem of calculation of a response function of a Fermi system to an
external interaction becomes tractable at cost of introduction of a set of Landau-Migdal parameters
for quasi-particle interactions. Parameters can be either evaluated microscopically or extracted
from analysis of experimental data, see Ref.~\cite{M67}. The technical difference of the
Larkin-Migdal and Leggett approaches is that the former approach works out equations for full
in-medium vertices, whereas the latter one calculates directly a response function. The former
approach was aimed at the study of transitions in nuclei, and the latter on the analyzes of
collective modes in superfluid Fermi liquid. The principal equivalence of both approaches was
emphasized already by A.J.~Leggett in Refs.~\cite{Leg65a,Leg65}. Reference \cite{KV2} demonstrates how
one may use both approaches in calculations of the PBF rates.

Reference~\cite{SVSWW97} was the first one, in which the most important in-medium effects were
incorporated in the numerical code for neutron star cooling. Among them neutron PBF and proton PBF
processes were treated as equally important. The PBF processes (but with free vertices)  were
incorporated also in the ``standard'' cooling scenario \cite{P98,YLS99} that led the authors of
Ref.~\cite{Page} to the suggestion of the minimal cooling paradigm. Detailed simulations of
different medium effects have been done in \cite{BGV,GV}. In contrast to the minimal cooling
paradigm, the medium modifications of all reaction rates lead to their pronounced density
dependence. For the PBF processes it is mainly due to the dependence of pairing gaps and
nucleon-nucleon correlation factors on the density. For MMU processes the reaction matrix elements
are strongly density dependent due to the softening of the exchanged pion and the dependence of
nucleon-nucleon correlation factors on the density. It establishes the strong
link between the cooling behavior of a neutron star and its mass
\cite{VS84,VS86,MSTV90,SVSWW97,BGV, GV}. The density dependence of the reaction rates provides a
smooth transition from {\emph ``standard'' } to {\emph ``non-standard''} cooling for the increasing
star-center density, i.e., for increasing the star mass. Thus, the inclusion of the most important in-medium
effects within the {\em{ ``nuclear medium cooling scenario''}} enables us to describe appropriately
both high and low surface temperatures obtained from analyzes of soft X-ray pulsar data. The
mentioned above moderate suppression of the PBF emissivity ($\propto v_{\rm F}^2$) at $1S_0$
pairing should not significantly affect general conclusions on the neutron star cooling history
done in previous works where it was not incorporated.

The paper is organized as follows. In Section~\ref{sec:noneq} we formulate description of normal
Fermi liquids at non-equilibrium. Softening of pionic degrees of freedom is taken into account.
In Section~\ref{sec:LM} we perform generalizations to the fermion superfluids at
non-equilibrium. The Larkin-Migdal equations are formulated on the Schwinger-Keldysh contour. A possibility
of the pairing in an arbitrary momentum state is considered. In Section~\ref{sec:White} we
introduce optical theorem formalism for normal and superfluid fermion systems out of equilibrium.
Cutting rules for closed diagrams expanded in series with respect to the number of $G^{-+}$ full
Green's functions are formulated. Important role of multi-piece diagrams is shown. Fermi liquid renormalizations are performed in
Section~\ref{sec:Fermi-liquid}. Equilibrium $T\neq 0$ systems with pairing are considered in
Section~\ref{sec:eqpairing}. In Section~\ref{sec:neutrino}, as an example, we find the
current-current correlator and in Section \ref{sec:PBF},
the neutrino emissivity from the PBF processes on neutrons paired in
$1S_0$ state. Technical details are given in Appendices.

\section{Description of non-equilibrium normal Fermi liquids}\label{sec:noneq}
\subsection{Dyson equations on Schwinger-Keldysh contour}

\begin{figure}
\includegraphics[width=7cm]{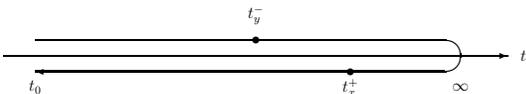}
\caption{Closed real-time contour with two external points $x,y$ on the contour}
\label{fig:contour}
\end{figure}
The non-equilibrium theory can  be entirely formulated on the {\em closed real-time Schwinger-Keldysh contour} (see
Fig.~\ref{fig:contour}) with the time argument running from $t_{0}$ to $\infty$ along {\em
time-ordered} branch and back to $t_{0}$ along {\em anti-time-ordered}
branch~\cite{Kel64}. We assume the reader to be familiar with this
real time formulation of the non-equilibrium theory.
Details can be found in Refs. \cite{Lif81,Dan84,IKV99,IKV00, KIV01} and in Appendix A.

In absence of pairing
one deals only with the ``normal'' contour Green's function defined as the expectation value of
contour-ordered products of operators
%
\begin{eqnarray}\label{Ga1}
i\,[\widehat{G}_n (x,y)]_{ab} &=& ^b\includegraphics[width=1.2cm]{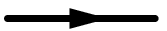}^a
\nonumber\\
&=& <N|\Tc \widehat{\Psi}_a (x) \widehat{\Psi}_{b}^{\dagger}(y)|N>.
\end{eqnarray}
The time-ordering, $\Tc$, goes here according to the time parameter running along the time contour
${\mathcal C}$. The averaging $<N|\dots|N>$  is taken over the $N$-particle non-equilibrium state;
$a ,b$ are spin indices. In the absence of the spin-orbital interaction the Green's function is
diagonal in the spin space
\be
[\widehat{G}_n (x,y)]_{ab}=G_n (x,y) \delta_{ab},
\nonumber
\ee
where $\delta_{ab}$ is the Kronecker symbol.

Because of the two contour branches one actually deals with four Green's functions unified in
matrices (so called Schwinger-Keldysh space). Further explanations are given in Appendix A. There
one can also find the helpful relations between Green's functions and self-energies.

The typical interaction time $\tau_{\rm int}$ for the change of the higher-order correlation
functions is usually much less than the typical relaxation time $\tau_{\rm rel}$, which determines
the kinetic stage of the system evolution. Describing the system at times
$t-t_0 \gg \tau_{\rm int}$, one can neglect initial correlations. This is in accordance with the
Bogolyubov-Klimontovich principle of the weakening of higher order correlations
\cite{Bogolyubov,Klimontovich}. The coarse-graining leads to time-irreversibility. Alternatively,
one also could suppose that the initial state is uncorrelated, like an in equilibrium ideal gas. This
corresponds to a situation when an information loss occurs right from the beginning, cf. Refs.
\cite{KB62,Dan84}. Assuming that we describe the system for  $t\gg\tau_{\rm int}$ we may use the
Wick decomposition that leads to the Dyson equation  formulated on the real time contour
\begin{eqnarray}\label{Dyson}
\hat{G}_{0x}^{-1}\widehat{G}_n (x,y) = \delta_{\mathcal{C}}(x,y)+
\int_{\mathcal{C}} \rmd  z \widehat{\Sigma}_n(x,z) \widehat{G}_n
(z,y),\end{eqnarray}
where
\begin{eqnarray}
\hat{G}_{0x}^{-1}=
 \left(i\,\partial_t+\frac{1}{2m}{\partial_{\vec{x}}^2}\right)\delta^a_b
\end{eqnarray}
in non-relativistic kinematics that we use in this work.  Here $\delta_{\mathcal{C}}$ is
$\delta$-function on the contour, $\widehat{G}_{0}$ is the free Green's function (thin line)
\begin{eqnarray}\label{S-def}
\hat{G}_{0x}^{-1} \widehat{G}^{0}(x,y) =\delta_{\mathcal{C}}(x,y),
\end{eqnarray}
and the self-energy $\widehat{\Sigma}_n$ is a functional of the
Green's functions. Being formulated with the standard diagrammatic rules the Dyson equation reads
\begin{eqnarray}\label{YaF-dsen}
\parbox{7cm}{\includegraphics[width=7cm]{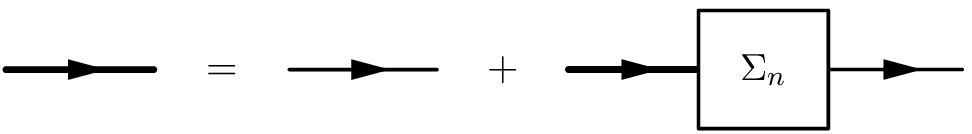}},
\end{eqnarray}
\be\label{Gn}
\widehat{G}_n =\widehat{G}_0 +\widehat{G}_0 \odot \widehat{\Sigma}_n \odot \widehat{G}_n \,.
\ee
The sign $\odot$ stands here for the contour coordinate folding, i.e. the integration,
$\int_{\mathcal{C}} \rmd  z$, over the spatial coordinate and the time coordinate  integrated along
the Schwinger-Keldysh time contour, cf. Eq.~(\ref{H=FG1}).
Alternatively it can be represented as
the usual four-dimensional coordinate integration, if  all quantities are treated as matrices in
the Schwinger-Keldysh space, see Eq.~(\ref{H=FG2}) in Appendix~\ref{Contour}. Thus, Eq.~(\ref{Gn})
is the symbolic equation for four ($G^{ij}$, $i,j=-,+$) Green's functions. The spin-index
contractions go in the standard manner: in the direction opposite to the fermion arrows. Throughout
this paper we shall use notations of Refs.~\cite{Lif81,KV95}, in which $\Sigma_n^{-+}$ and
$\Sigma_n^{+-}$  differ by the sign from the corresponding quantities introduced in
Ref.~\cite{IKV00}. In these notations Eq.~(\ref{Gn}) is rewritten as
\be
\label{Gn1}
\widehat{G}_n^{ij} =\widehat{G}_0^{ij} +\widehat{G}_0^{ik} \odot
\widehat{\Sigma}_n^{kl} \odot \widehat{G}_n^{lj}\,.
\ee
The equation for the retarded Green's function decouples from other equations
\be
\label{retD} \widehat{G}^R_n =\widehat{G}^R_0 +\widehat{G}^R_0
\odot \widehat{\Sigma}^R_n \odot \widehat{G}^R_n\,, \ee
and reads in diagrams as Eq.~(\ref{YaF-dsen}) (above, this equation was
formulated for the contour or matrix quantities).
Here $\widehat{\Sigma}^R
=\widehat{\Sigma}^{--}+\widehat{\Sigma}^{-+}$ is the retarded self-energy, see (\ref{Fretarded1}).
Similar equation exists for the advanced Green function $\widehat{G}^A_n$.

We also need to define two-particle Green's function
\begin{multline}
\widehat{K}_n (x,y;x',y')
\nonumber\\
=<N|\Tc \widehat{\Psi}^c (x')\widehat{\Psi}^d (y')
\widehat{\Psi}_{a}^{\dagger}(x)\widehat{\Psi}_{b}^{\dagger}(y)|N>
\end{multline}
and the two-particle interaction amplitude as a contracted part of $\widehat{K}$.
The system of equations for the non-equilibrium two-particle Green's functions was studied, e.g.,
in Refs.~\cite{Koshelkin99,Bornath99}.

As has been shown by Kadanoff and Baym, in case of smooth time-space changes of the system the
quasiclassical approximation can be applied to the non-equilibrium Dyson equations. Using  the
first-order gradient expansion for the quantities in the Wigner representation one obtains
Kadanoff-Baym kinetic equation for the $G^{-+}$ Green's function, which generalizes the standard
Boltzmann kinetic equation for quasi-free particle and the Landau kinetic equation for
quasi-particle to the case of particles with finite mass-widths. This generalized kinetic equation
is supplemented by the  equation for the retarded Green's function which is algebraic equation up
to second gradients. Self-consistent approximations \cite{Baym} to this kinetic scheme were
developed only recently, see \cite{IKV00,KIV01,IV09} and references therein.

\subsection{$NN$ interaction and pion degrees of freedom in nucleon matter}

Consider the nucleon-nucleon interaction formulated within the Fermi-liquid
approach with the explicit incorporation of the in-medium pion exchange.
For zero temperature it was done by A.B.~Migdal in Ref.~\cite{M78},
and, then, the approach was generalized for finite temperatures and
non-equilibrium systems in Refs.~\cite{VM78,VM82,D82,MSTV90,V93,V01}.
At excitation energies of our interest ($\epsilon^{*}\ll
\epsilon_{\rm F}$, $\epsilon^{*}\sim T$ in equilibrium) nucleons
are only slightly excited above their Fermi surfaces and all
processes occur in a narrow vicinity of the Fermi energy $\epsilon_{\rm F}$.
Within this approach the long-range processes
are treated explicitly, whereas short-range processes are described
by  local quantities approximated by phenomenological, so-called Landau-Migdal,  parameters.
At low excitation energies the $NN$ interaction amplitude is
presented on the Schwinger-Keldysh contour (or in matrix notation)
as follows
\be\label{TKFS}
\parbox{7.3cm}{\includegraphics[width=7.3cm]{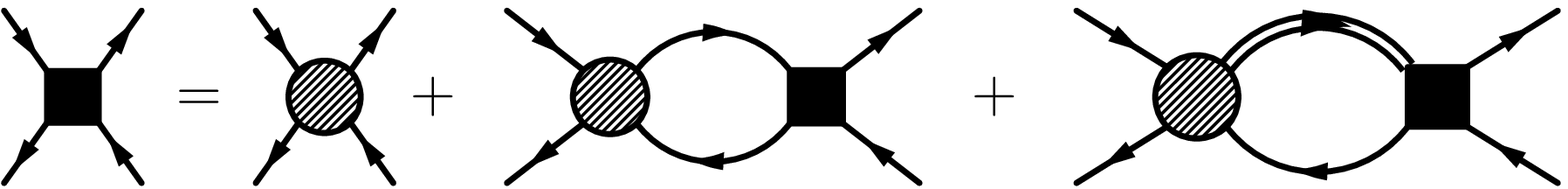}}\,,
 \ee
where
\be
\parbox{4.5cm}{\includegraphics[width=4.5cm]{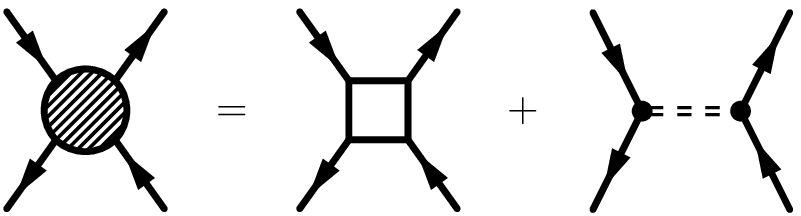}}\,.
\label{irred}
\ee
The solid line stands for a nucleon, the
double-line stands for a $\Delta$ isobar. Although the mass difference between the $\Delta$
and $N$, $m_{\Delta}-m_N \simeq 2.1 m_{\pi}\gg \epsilon_{\rm F}$ ($m_{\pi}$ is the pion mass) the
delta-nucleon hole term is numerically rather large, since the $\pi N\Delta$ coupling constant is
twice larger than the $\pi NN$ one, and the $\Delta$ spin-isospin degeneracy factor is 4 times
larger than that for nucleons. The doubly-dashed line corresponds to the exchange of the free pion
with inclusion of the contributions of the residual s-wave $\pi NN$ interaction and $\pi\pi$
scattering, i.e. the residual irreducible interaction to the nucleon particle-hole and
delta-nucleon hole insertions. The block in Eq.~(\ref{irred}), depicted by the empty square,
is irreducible with respect to particle-hole, delta-nucleon hole and pion states and is, by
construction, essentially more local than the contributions given by explicitly presented graphs.
Thus the empty block term should be much smoother function of its variables than the
terms (particle-hole, delta-hole, pion) presented explicitly  in Eqs. (\ref{TKFS}) and ~(\ref{irred}).
In principle, the short-range interaction should be calculated as function of the density, neutron
and proton concentrations, energy and momentum, temperature, etc. However, instead of doing
complicated calculations one often reduces it to the set of Landau-Migdal parameters, which one
extracts from analysis of experimental data on atomic nuclei.

The irreducible part of the interaction involving $\Delta$ isobar
is  constructed similarly to (\ref{irred})
\be
\parbox{4.5cm}{\includegraphics[width=4.5cm]{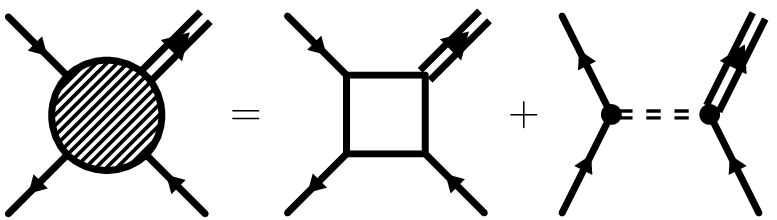}}\,.
\label{gam1-d}
\ee
The main part of the $N\Delta$ interaction is due to the pion exchange. Although information on
local part of the $N\Delta$ interaction is rather scarce, one can conclude \cite{MSTV90,SST99}
that the corresponding Landau-Migdal parameters are essentially smaller then those for $NN$
interaction. Therefore, and also for the sake of simplicity we will, in further, neglect the first graph on the
right-hand side of  Eq.~(\ref{gam1-d}).

The spectrum of the particle excitations is determined
by the spectral function given by the imaginary part of the retarded Green's function
($A=-2\Im G^R$). Resummation of
diagrams shown in (\ref{TKFS})  yields the following Dyson
equation for the retarded pion Green's function
\be
\includegraphics[width=7cm]{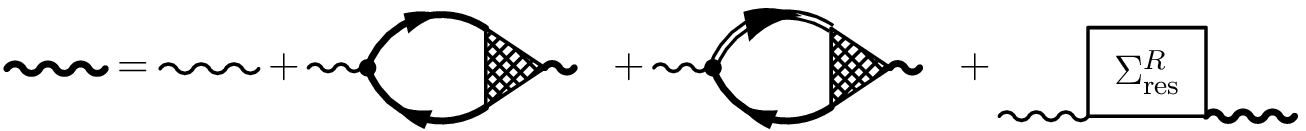}\,.
\nonumber\\
\label{pion-l}
\ee
Here $\Sigma_{res}^{R}$ is the residual retarded pion self-energy that includes the contribution
of all diagrams, which are not presented explicitly in (\ref{pion-l}), like   s-wave $\pi N$ and
$\pi\pi$ scatterings (included by doubly-dashed line in (\ref{irred})). The full vertex takes into
account $NN$ correlations
\be
\includegraphics[width=6cm]{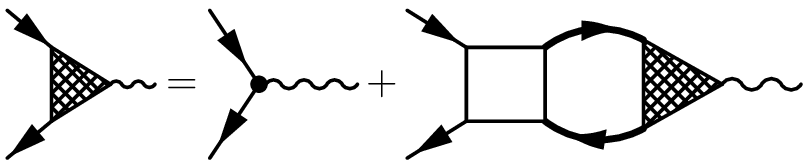}\,.
\label{dressed}
\ee

\begin{figure}
\centerline{
\includegraphics[width=7cm]{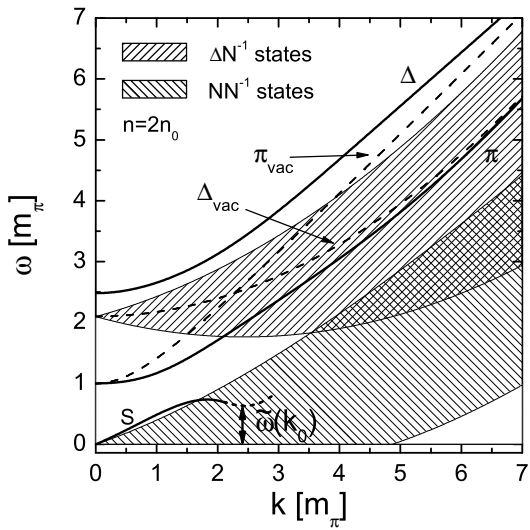}
} \caption{The typical behavior of the $\pi^{\pm,0}$ spectra
in  symmetric nuclear matter and of  $\pi^0$ spectrum in
asymmetric matter. The chosen typical value of the density
$n=2n_0$ is supposed to be smaller than the critical density for
the pion condensation.}
\label{Spectrum}
\end{figure}

In some regions of the $(\om,k)$ plane the pion spectral function
$A_{\pi}(\om,k)$ has sharp peaks along some lines $\om_i (k)$ which we call the spectral branches.
Nearby these lines one can use the quasi-particle approximation\footnote{The quasi-particle approximation
for the given particle
species is understood as putting the imaginary part of the
retarded self-energy in the  Green's function to zero. The quasi-particle
width $\Gamma =-2\Im \Sigma^R$ is then calculated with so-defined Green's functions.} and write the spectral
function as
\be
A_{\pi}(\om,k)\simeq \sum_{i}
\frac{2\pi\, \delta\big(\om -\om_i(k)\big)\phantom{_{\om_i (k)}}}
{\Big[2\om -\frac{\partial \Re\Sigma^R_{\pi} (\om ,k)}{\partial \om }\Big]_{\om_i (k)}}.
\ee
The spectrum of pionic quasi-particles possesses three branches for $\pi^{\pm,0}$ in the symmetric
nuclear matter ($N=Z$) and for $\pi^{0}$ in asymmetric matter ($N\neq Z$), e.g.
in  the neutron star matter. The typical spectrum is shown in
Fig.~\ref{Spectrum}. In the region $\omega\gsim m_\pi$ there are two branches: the $\Delta$ branch
and the pion branch. For $\om < m_{\pi}$ there is the spin sound branch (with $\omega \rightarrow
0$ when $k\rightarrow 0$). The hatches show the regions on the $(\om,k)$ plane with a
non-vanishing pion width, calculated within the quasi-particle approximation for nucleons and
$\Delta$ isobars. In the lower hatched region,  at $\omega < kp_{{\rm F},N}$ and $k\sim p_{{\rm
F},N}$, there are no quasi-particle branches and the pion width cannot be neglected. This is the
region of the Landau damping in the nucleon particle-hole channel. The pion spectral function is
enhanced in this region of $\omega$ and $k$ for $n>n_{c1}\sim (0.5$--$0.7)n_0$. We stress that the
pion spectral function calculated beyond the quasi-particle approximation for nucleons and
$\Delta$'s is much more involved, see Ref.~\cite{LKR}.

To specify  the enhancement of the $\pi^{\pm,0}$ spectral density for $N=Z$ and
of the $\pi^0$ for $N\neq Z$ in the Landau damping region it is
convenient to introduce the function
$$
\widetilde{\omega}^2(k)=-[\Re G^R_{\pi}(\omega =0, k,\mu_\pi )]^{-1}\,.
$$
Note that momenta passing through  the $NN$ interaction in the MU and MMU processes are $k\sim
p_{{\rm F},n}$, where $p_{{\rm F},n}$ is the nucleon Fermi momentum, and for the MNB process
$k\sim(0.9\mbox{--}1)\,p_{{\rm F},n}$~\cite{VS86}. Remarkably, the minimum on the function
$\widetilde{\omega}^2 (k)$ is realized at the similar momentum $k_0 \simeq
(0.9\mbox{--}1)\,p_{{\rm F},n}$.
Thus, the quantity $\widetilde{\omega}(k_0)$ called the {\em effective pion gap} controls the strength of the $NN$ interaction.
The $NN$ cross-section is $\propto 1/\widetilde{\om}^4$
for $\widetilde{\om}^2 < m_\pi^2$, provided the MOPE dominates, see Eq.~(\ref{MOPER}) below.
Note that for the  asymmetric nucleon
matter the pion gap is different for $\pi^0$ and for $\pi^{\pm}$ since neutral and charged
channels are characterized by different diagrams permitted by the charge conservation.

The pion chemical potentials  ($\mu_{\pi^{+}}\neq \mu_{\pi^{-}}\neq 0$, $\mu_{\pi^{0}}=0$) are
determined from equilibrium conditions for the reactions involving the corresponding pions. In the
neutron star matter $\mu_{\pi^-}$ follows from the condition of the chemical quasi-equilibrium
with respect to the reactions $n\rightarrow p\pi^-$ and $n \rightarrow pe\bar{\nu}$:
$\mu_{\pi^{-}}=\mu_e =\varepsilon_{{\rm F},n}-\varepsilon_{{\rm F},p}$, where $\epsilon_{{\rm
F},n}$, $\epsilon_{{\rm F},p}$ are Fermi energies of the neutron and proton. For a small-size
systems like atomic nucleus one should put $\mu_{\pi^{+}}=\mu_{\pi^{-}}=\mu_{\pi^{0}}= 0$.

At low pion energies (for $\pi^{\pm,0}$ for $N=Z$ and for $\pi^0$ for $N\neq Z$) the lowest-energy state
determining  by the pole of the pion Green's function is $i\beta \om \simeq \widetilde{\om}^2 (k_0)$ with $\beta
>0$ appeared due to the Landau damping, see Ref.~\cite{MSTV90}. Thus for $\widetilde{\om}^2 (k_0)>0$ the
pion excitations die out with time exponentially $\propto \exp(-\widetilde{\om}^2 (k_0)\, t/\beta)$.

\subsection{ Pion softening and pion condensation}

For $\widetilde{\om}^2 (k_0)<0$ the pion field grows exponentially with time as
$\exp(|\widetilde{\om}^2 (k_0)|\,t /\beta)$. Thus, the change in the sign of
$\widetilde{\omega}^2$ at $n=n_{c,\pi}$ marks the critical point of a
phase
transition to a state with a classical pion field (a pion condensate). The critical density
$n_{c,\pi}$ depends on the values of Landau-Migdal parameters, which are badly known for
asymmetric matter and for densities significantly larger than $n_0$. Nevertheless, some estimations
can be given. Various experiments have shown that the pion condensation does not manifest itself
in atomic nuclei as a volume effect, see Refs.~\cite{M78,MSTV90}\footnote{In the surface layer of
the nucleus a classical pion condensate field may appear due to a coupling of the gradient of the
$\sigma$ mean field to the pion, see \cite{Ripka}.}. Different model-dependent estimations
indicate that $n_{c,\pi}\sim (1.5\mbox{---}3)\, n_0$, depending on the pion species, the
proton-to-neutron ratio and the model used, see Refs.~\cite{MSTV90,Nakano}. Variational
calculations~\cite{APR} yield $n_{c,\pi}\simeq 2 n_0$ for isotopically symmetric nuclear matter
and $n_{c,\pi}\simeq 1.3 n_0$ for $\pi^0$ mesons in the neutron star matter.

\begin{figure}
\centerline{\includegraphics[height=4.5cm,clip=true]{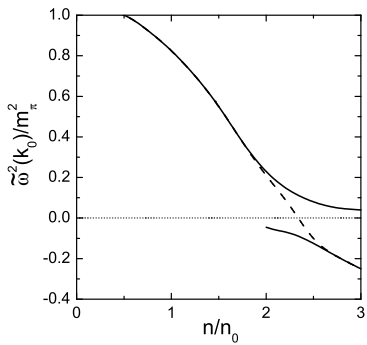}}
\caption{Effective pion gap (for $\pi^{\pm,0}$ for $N=Z$ and for $\pi^0$ for $N\neq Z$) versus the baryon
density as calculated in Refs.~\cite{MSTV90,V01}.
Compare with the pion spectrum presented in Fig.~\ref{Spectrum}.}
\label{piongap}
\end{figure}

Typical density behavior of $\widetilde{\omega}^2 (k_0)$ (for $\pi^{\pm},\pi^{0}$ at $N=Z$ and
for $\pi^0$ at $N\gg Z$)  is shown in Fig. \ref{piongap}.
At $n <n_{c1}$, $\widetilde{\omega}^2 (k)$ has the minimum for $k_0 =0$, i.e.
$\widetilde{\omega}^2 (k_0 =0) =m^2_{\pi}-\mu_{\pi}^2$. For such densities the value
$\widetilde{\omega}^2 (p_{{\rm F},n})$ essentially deviates from $m^2_{\pi}-\mu_{\pi}^2$ tending
to $m^2_{\pi}+p_{{\rm F},n}^2 -\mu_{\pi}^2$ in the low density limit.

At the critical point of the pion condensation ($n =n_{c,\pi}$) the value $\widetilde{\omega}^2
(k_0)$ with artificially neglected $\pi\pi$ fluctuations changes its sign (dashed line in
Fig.~\ref{piongap}). It symbolizes the occurrence of a second-order phase transition to an
inhomogeneous ($k_0\neq 0$) pion-condensate state. In reality, the $\pi\pi$ fluctuations are
significant in the vicinity of the critical point and the phase transition is of the first
order~\cite{D75,VM82,D82}. Therefore we depict two branches in Fig. \ref{piongap} (solid curves)
with positive and negative values of $\widetilde{\omega}^2 (k_0)$. Calculations in Ref.~\cite{D82}
demonstrated that at $n >n_{c,\pi}$ the free energy of the state with $\widetilde{\omega}^2
(k_0)>0$ and without the pion mean field becomes larger than the free energy of the state with
$\widetilde{\omega}^2 (k_0) <0$ and a finite pion mean field. Therefore at $n >n_{c,\pi}$ the
state with $\widetilde{\omega}^2 (k_0) >0$ is metastable and the state with $\widetilde{\omega}^2
(k_0)<0$ and the pion mean field $\varphi_\pi \neq 0$ becomes the ground state.

The quantity $\widetilde{\omega}^2 (k_0)$ demonstrates how much the virtual (particle-hole) mode
with pion quantum numbers is softened at the given density. For the symmetric nuclear matter at $n
=n_0$ the ratio $\alpha =G_\pi /G^0_\pi \simeq 6$ for  $\omega =0$, $k=p_{{\rm F},N}$. However,
this so-called {\em "pion softening"}~\cite{M78} does not significantly enhance the $NN$
scattering cross section because of the simultaneous essential suppression of the $\pi NN$ vertex
by nucleon-nucleon correlations. Indeed, the ratio of the $NN$ cross sections calculated with the
FOPE and MOPE  equals to
\begin{equation}\label{MOPER}
R=\frac{\sigma [\mbox{FOPE]}}{\sigma [\mbox{MOPE}]} \simeq
\frac{\gamma^4 (\omega\simeq 0,k\simeq p_{{\rm F},N} )
(m^2_{\pi}+p_{{\rm F},N}^2)^2}{\widetilde{\omega}^4 (p_{{\rm F},N}
)}\,,
\end{equation}
where $\gamma$ is the vertex dressing factor determined by Eq.~(\ref{dressed}),
$\gamma (n_0)\simeq 0.4$. For $n \lsim n_0$ one has $R\lsim 1$,
whereas already for $n =2n_0$ this estimate yields $R\sim 10$.
Thus, following Refs.~\cite{M78,MSTV90} one can evaluate the $NN$ interaction
for $n> n_0$ with the help of the MOPE, i.e.,
\be
\parbox{4cm}{\includegraphics[width=4cm]{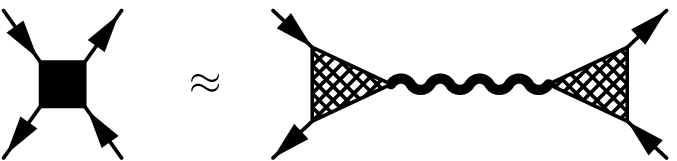}}\,.
\label{MOPE}
\ee
Here the bold wavy line relates to the in-medium pion. In the soft-pion approximation the same
one-pion exchange determines also interaction in the particle-particle channel
\be
\parbox{3cm}{\includegraphics[width=3cm]{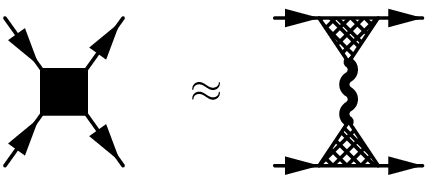}}\,.
\label{MOPE-pp}
\ee
Namely, this quantity determines the $NN$ interaction entering neutrino emissivities of the MMU
and MNB processes.

Often, one considers the softening of the one-pion exchange only,
neglecting the same effects for the two, three etc. pion exchanges, arguing
for their smallness because of extra integrations over the intermediate states, see
Ref.~\cite{M78}. At zero temperature these effects are numerically
small. Nevertheless, they result in the change of the order of the
phase transition (from the second order to the first order)~\cite{D75}.
Ignoring this small jump in the pion gap one may deal
with the MOPE interaction in the one (particle-hole or
particle-particle)  channel. The residual interactions are then
hidden in the values of Landau-Migdal parameters. In case of the
non-equilibrium and equilibrium $T\neq 0$ matter these
pion fluctuation effects contribute significantly to the pion
self-energy and should be taken into account, see Refs.~\cite{VM82,D82}.
Corresponding pion fluctuation contributions must be then extracted from the residual
interaction parameters.  Their self-consistent analysis in both
particle-hole and particle-particle channels has been performed
in Ref.~\cite{D82} within the Thomas-Fermi approximation ($k_0 \ll 2p_{{\rm
F},N})$.

\section{Description of non-equilibrium superfluid Fermi liquids}
\label{sec:LM}

Below, dealing with pairing phenomena in
non-equilibrium systems we assume that deviations from the equilibrium are rather small,
 bearing in mind that non-equilibrium
effects should not destroy the fermion pairing.

In a superfluid Fermi system a condensate of paired fermions is formed. It induces  non-vanishing
amplitudes for the transitions of a particle  to a hole state and vice versa. Thus, it is possible
to combine  one-particle state on top of the $N$-particle background  and one-hole state on top of
the background with $N+2$ particles. The one-particle--one-hole irreducible amplitudes of such
transitions are depicted as blocks
\be
-i[\hat{\Delta}^{(1)}(x)]_{ab}&=&
-i\,<N|\widehat{\Psi}_a(x)\,\widehat{\Psi}_b(x) |N+2>
\nonumber\\
&=&_{b}\includegraphics[width=1.5cm]{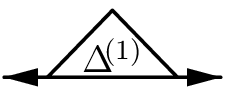}{}_a\,,
\nonumber\\
-i[\hat{\Delta}^{(2)}(x)]_{ab} &=&
-i\,<N+2|\widehat{\Psi}^\dagger_a(x)\,\widehat{\Psi}^\dagger_b(x)|N>
\nonumber\\
&=&_b\includegraphics[width=1.5cm]{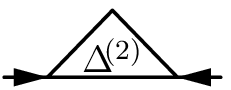}{}_a\,.
\label{Delta-block}
\ee
Their spin structure can be written in general case as
\be
\hat{\Delta}^{(1)}(x) &=&
\big(\Delta^{(1)}_0(x)\,\sigma_0+\vec{\Delta}^{(1)}_1(x)\,\vec{\sigma}\big)i\sigma_2\,,\quad
\nonumber\\
\label{Dspin}
\hat{\Delta}^{(2)}(x) &=&
i\sigma_2\,\big(\Delta^{(2)}_0(x)\, \sigma_0
+ \vec{\Delta}^{(2)}_1(x)\,\vec{\sigma}\big)\,. \ee
Here  $\vec{\sigma}=(\sigma_1,\sigma_2,\sigma_3)$ with $\sigma_i$ being the Pauli matrices;
$\sigma_0$ is the unit matrix in the spin space. The amplitudes $\Delta^{(1,2)}_0$ are induced by a
particle-particle interaction with even angular momenta $L=0,2,4,\dots$, and amplitudes
$\vec{\Delta}^{(1,2)}_1$,  by the interaction with odd angular momenta $L=1,3,5,\dots$.

Since $\widehat{\Delta}^{(1,2)}$ are functions of the only one contour coordinate, we have
$[\widehat{\Delta}^{(1,2)}]^{-+}=[\widehat{\Delta}^{(1,2)}]^{+-}=0$ and denote
$[\Delta^{(1,2)}]^{--}=[\Delta^{(1,2)}]^{R}=-[\Delta^{(1,2)}]^{++}$, as follows from Eq.~(\ref{Fretarded})
valid for any two-point functions.
From the definitions (\ref{Delta-block}) follows
\be
(\hat{\Delta}^{(1)}(x))^\dag=(\hat{\Delta}^{(2)}(x))^{\rm T}\,,
\ee and Eq.~(\ref{Dspin}) yields
\be
(\Delta^{(1)}_0)^*=\Delta^{(2)}_0\,,\quad
(\vec{\Delta}^{(1)}_1)^*=-\vec{\Delta}^{(2)}_1\,.
\ee

The transition amplitudes (\ref{Delta-block}) imply that a propagating particle
can be transformed in flight into a hole and vice versa. This is described by a new type of
propagators --- anamalous Green's functions --- which can be defined on the Schwinger-Keldysh
contour as
\be
&&[i\widehat{F}^{(1)}(x,y)]_{ab}=_b\parbox{15mm}{\includegraphics[width=1.5cm]{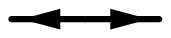}}_a
\nonumber\\ &&=<N|\Tc \widehat{\Psi}_a (x) \widehat{\Psi}_b (y)
|N+2> \, , \nonumber\\{} &&[i\widehat{F}^{(2)}(x,y)]_{ab}
=_b\parbox{15mm}{\includegraphics[width=1.5cm]{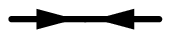}}_a
\nonumber\\ &&=<N+2|\Tc\widehat{\Psi}^{\dagger}_{a} (x)
\widehat{\Psi}^{\dagger}_{b} (y) |N>\,. \ee
For these two functions are valid relations
\be
[\widehat{F}^{(1,2)}(x,y)]_{ab}=-[\widehat{F}^{(1,2)}(y,x)]_{ba}\,
, \label{FT-F}\\
{}[\widehat{F}^{(1,2)}(x,y)]^\dag_{ab}=-[\widehat{F}^{(2,1)}(y,x)]_{ba}\,
. \label{Fdag-F} \ee
We also need the Green's function for a hole
\be
&&[i \widehat{G}^h (x,y)]_{ab}={}_a
\parbox{1cm}{\includegraphics[width=1cm,angle=180]{YaF-Gn.eps}}_b
\nonumber\\ &&=<N|\Tc\widehat{\Psi}_{{\rm C},a} (x)
\widehat{\Psi}^{{\dagger}}_{{\rm C},b} (y) |N> \label{Gh}\,, \ee
which is written in terms of the charge conjugated fermion operators $\widehat{\Psi}_{\rm
C}=\sigma_2\,\big(\widehat{\Psi}^\dagger\big)^{\rm T}$, where ${\rm T}$ stands for the
transposition operation in the spin space. The Green's functions of a particle and a hole are related as
\be
\widehat{G}^h (x,y)={\sigma}_2\,[\widehat{G}(y, x)]^{\rm T}\,\sigma_2=G(y, x)\, \delta_{ab}\,.
\ee

The Dyson equations for dressed normal and anomalous Green's functions in case of pairing (Gor'kov
equations) have  the form:
\begin{align}
\parbox{6.75cm}
{\includegraphics[height=.7cm]{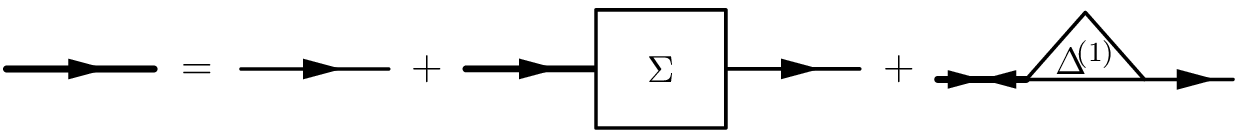}},
\label{dseG}\\
\parbox{5.6cm}
{\includegraphics[height=.7cm]{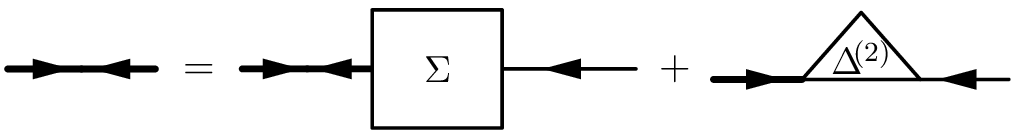}},
\label{dseF}
\\
\parbox{5.6cm}
{\includegraphics[height=.7cm]{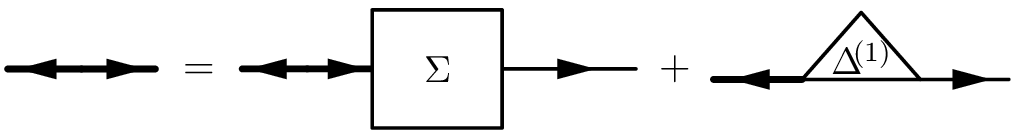}}.
\label{dseF1}
\end{align}
The thin line is the normal free Green's function of the fermion
particle
and the inverted thin line is the normal free Green's function of
the hole, $\hat{\Sigma}$ is the full self-energy of the particle
and $\hat{\Sigma}^h$, of the hole. Explicitly these Dyson
equations read as
\begin{align}
\hat{G} &= \hat{G}_0 + \hat{G}_0 \odot \hat{\Sigma} \odot \hat{G}
                       +\hat{G}_0 \odot \hat{\Delta}^{(1)} \odot \hat{F}^{(2)}\,,
\nonumber\\
\hat{F}^{(2)} &= \hat{G}^h_0 \odot \hat{\Sigma^{h}} \odot \hat{F}^{(2)}
                  + \hat{G}^h_0  \odot \hat{\Delta}^{(2)} \odot
\hat{G}\,,
\nonumber\\
\hat{F}^{(1)} &= \hat{G}_0 \odot \hat{\Sigma} \odot \hat{F}^{(1)}
               + \hat{G}_0  \odot \hat{\Delta}^{(1)} \odot \hat{G}^h\,.
\label{GEq-cont}
\end{align}
In terms of the dressed Green's functions for the system without pairing (\ref{Gn}) the Gor'kov
equations can be written shortly as \cite{LM63},
\begin{align}
\hat{G} &\simeq \hat{G}_n + \hat{G}_n  \odot \hat{\Delta}^{(1)} \odot \hat{F}^{(2)}\,,
\nonumber\\
\hat{F}^{(2)} &\simeq \hat{G}^h_n \odot \hat{\Delta}^{(2)} \odot \hat{G}\,,
\nonumber\\
\hat{F}^{(1)} &\simeq \hat{G}_n  \odot \hat{\Delta}^{(1)} \odot \hat{G}^h\,.
\label{Gorgap}
\end{align}
In these equations we dropped the terms containing the differences $(\hat{\Sigma}- \hat{\Sigma}_n)$ and
$(\hat{\Sigma}^h- \hat{\Sigma}_n^h)$. The self-energy $\Sigma$ includes the same set of diagrams as
the self-energy $\Sigma_n$, but constructed now from the Green's functions for the superfluid system, $G$
and $F$, instead of the normal Green's function $G_n$. Note that the anomalous Green's functions
$F$ can enter $\Sigma$ only in pairs to preserve the number of incoming and outgoing fermion lines in each vertex.
Hence, the terms containing $F$'s are $\propto\Delta^2$.
Since in the momentum representation
$G-G_n\propto \Delta^2$ and $G\to G_n$ for $|p_0-\epsilon_\rmF|/\Delta>>1$,
each integral over the energy $p_0$ in the self-energy is accumulated only in the vicinity of the Fermi
surface. Thus the neglected terms are small as $\Delta^2/\epsilon_{\rm F}^2$\,, cf. also Ref.~\cite{LM63}.

Hereafter the thin line in diagrams will stand for the full
Green's function of the system without pairing. From
Eq.~(\ref{Gorgap}) it follows immediately that the Green's
function $G$ is diagonal in the spin space. Then the Green's
functions $F^{(1)}$ and $F^{(2)}$ have the same spin structure as
the blocks (\ref{Delta-block}):
\be
\hat{F}^{(1)}(x,y) &=&
\big(F^{(1)}_0(x,y)\,\sigma_0+\vec{F}^{(1)}_1(x,y)\,\vec{\sigma}\big)i\sigma_2\,,\quad
\nonumber\\
\label{Fspin}
\hat{F}^{(2)}(x,y) &=&
i\sigma_2 \big(F^{(2)}_0(x,y)\, \sigma_0 + \vec{F}^{(2)}_1(x,y)\,\vec{\sigma}\big)\, .
\ee
In further, we will assume that only one type of pairing may occur: either in a state with an even
angular momentum or in a state with an odd angular momentum.\footnote{It is however not excluded
that at some conditions a part of fermions can be paired in one state whereas other part, in
another state.} Then we have either only $\Delta^{(1,2)}_0\neq 0$ or only
$\vec{\Delta}^{(1,2)}_1\neq 0$, and correspondingly, either $F^{(1,2)}_0\neq 0$ or
$\vec{F}^{\,\,(1,2)}_1\neq 0$, but not both simultaneously. In this case the Green's function
$\hat{G}$ remains diagonal in the spin space. Then we may  follow derivations of \cite{LM63} with
a little difference that all  our expressions are valid on the Schwinger-Keldysh contour.

The graphical equations for $\Delta^{(1)}$ and $\Delta^{(2)}$ are
\be
\includegraphics[width=6cm]{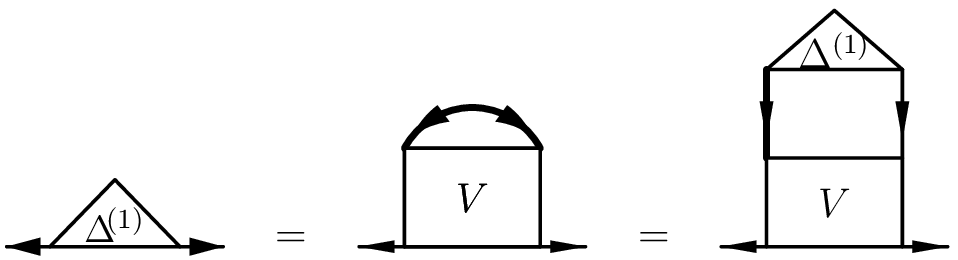}\,,
\label{Gapdiag1}
\\
\includegraphics[width=6cm]{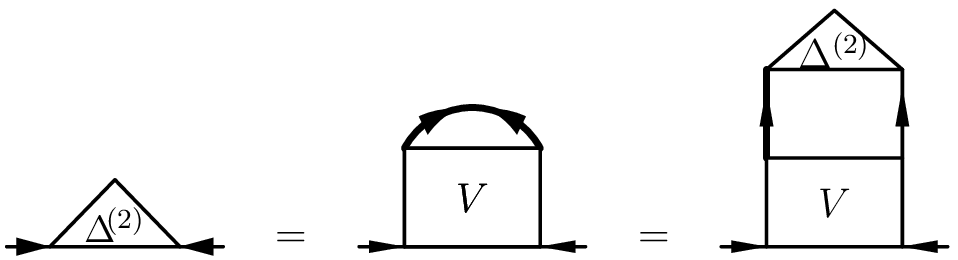}\,.
\label{Gapdiag2}
\ee
Here $\widehat{V}$ is a two-particle irreducible interaction, that  determines the full in-medium
particle-particle (``pp'') scattering amplitude via graphical equation
\be
\parbox{6cm}{\includegraphics[width=6cm]{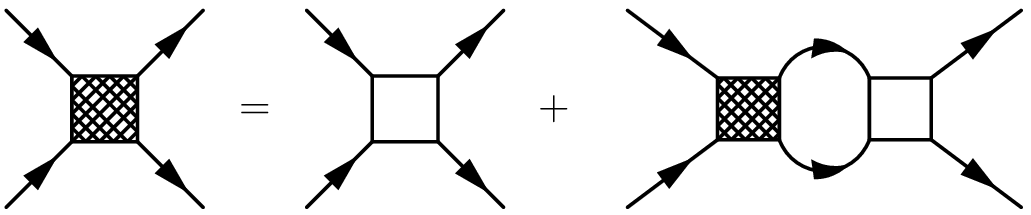}}\,,
\label{Tpp-diag}
\ee
which in terms of contour foldings becomes
\be
\widehat{T}_{\rm pp}=\widehat{V}-i\,\widehat{V}\boxdot
\big(\widehat{G}_n\,\widehat{G}_n\big)\boxdot\widehat{T}_{\rm pp}
\,.
\label{Tpp}
\ee
Both $\widehat{V}$ and $\widehat{T}$ are to be understood as
formulated on the Schwinger-Keldysh contour,
\be
&&\widehat{T}_{\rm pp}(x',y';x,y)=\widehat{V}(x',y';x,y)
\nonumber\\
&&
-i\,\int_{\mathcal{C}}\rmd z_1\int_{\mathcal{C}}\rmd z_2
    \int_{\mathcal{C}}\rmd z_3\int_{\mathcal{C}}\rmd z_4
\widehat{V}(x',y';z_1,z_2)
\nonumber\\
&&
\times
\widehat{G}_n(z_1,z_3)\,\widehat{G}_n(z_2,z_4)
\widehat{T}_{\rm pp}(z_3,z_4;x,y),
\ee
or being matrices in the Schwinger-Keldysh space. The sign $\boxdot$ in (\ref{Tpp}) stands for
integration over two four-dimensional coordinates with the time running over the Schwinger-Keldysh
contour. The quantities $\hat{V}$ and $\hat{T}$ are also matrices in the spin space, and we
parameterize $\hat{V}$ as
\be
&&\widehat{V}_{cd,ab} =
i\parbox{2cm}{\includegraphics[width=2cm]{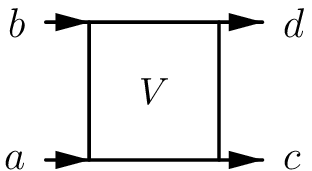}}
\label{Vdef}\\
&&\quad
= V_0 (i\sigma_2)_{dc} (i\sigma_2)_{ab}  +
  V^{\alpha\beta}_1 (\vec{\sigma}^\alpha i\sigma_2\,)_{dc} (i\sigma_2\vec{\sigma}^\beta)_{ab}\,.
\nonumber
\ee
With this definition the interactions $V_0$ and $V_1$ correspond to the scattering of two fermions
with the total spin zero and one, respectively. The interaction block entering
Eq.~(\ref{Gapdiag1}) is given by
\be
\parbox{2cm}{\includegraphics[width=2cm]{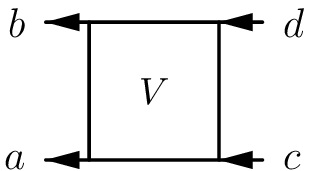}}=
\widehat{V}_{ab,cd}\,. \label{VTdef}
\ee
Using this definition we can analyze the spin structure of Eq.~(\ref{Gapdiag2})
\be
&&[\hat{\Delta}^{(2)}]_{ab} = -i[\hat{F}^{(2)}]_{cd}\boxdot
\widehat{V}_{cd,ab} \nonumber\\ &&= -i
[\hat{G}^h_n]_{c\uc}
\odot [\hat{\Delta}^{(2)}]_{\uc \ud}\odot \hat{G}_{\ud
d}
\boxdot \widehat{V}_{cd,ab} \,.
\label{d2} \ee
Substituting here Eqs.~(\ref{Dspin},\ref{Fspin}) and (\ref{Vdef}) and taking into account that the
full Green's functions $G$ and $G_n$ are diagonal in the spin-space,  we obtain
\be
&&[i\sigma_2\big(\Delta^{(2)}_0\, \sigma_0 + \vec{\Delta}^{(2)}_1\,\vec{\sigma}\big)]_{ab}
\nonumber\\
&&
=-i\, \big[G_n^h\odot i\sigma_2(\Delta^{(2)}_0\, \sigma_0 +
\vec{\Delta}^{(2)}_1\,\vec{\sigma}\big) \odot G \big]_{cd}
\nonumber\\ &&
\boxdot \big[V_0 (i\sigma_2)_{dc}
(i\sigma_2)_{ab}  +
  V^{\alpha\beta}_1 (\vec{\sigma}^\alpha i\sigma_2 )_{dc}
  (i\sigma_2\vec{\sigma}^\beta)_{ab}\big].
\nonumber
\ee
Separating terms with $\sigma_0$ and $\vec{\sigma}$ Pauli matrices and making use of the relations
$(i\sigma_2)^2=-1$ and ${\rm Tr}\{\sigma^\alpha\sigma^\beta\}=2\, \delta^{\alpha\beta}$  we find
\begin{align}
&\Delta^{(2)\phantom{\alpha}}_0=2\,i\,\big(G_n^h\odot\Delta^{(2)\phantom{\delta}}_0\odot
G\big) \boxdot V_0 \,,
\nonumber\\
&\vec{\Delta}^{(2)\alpha}_1=2\,i\,\big(G_n^h\odot\Delta^{(2)\beta}_1\odot
G\big) \boxdot V_1^{\beta\alpha}\,.
\label{D2}
\end{align}
For $\Delta_0^{(1)}$ and $\vec{\Delta}_1^{(1)}$ we obtain exactly the same equations as
(\ref{D2}) but with the  replacement $G_n^h\leftrightarrow G$.

An external field $V^{\rm ext}$ acting on a superfluid  Fermi system can induce four different
effects determined by the four vertex functions related to the creation of particle and hole $\tau$,
anti-particle and anti-hole $\widehat{\tau}^h$, two particles $\widehat{\tau}^{(2)}$, and two
holes $\widehat{\tau}^{(1)}$. The vertices can be graphically depicted as
\be
{\includegraphics[width=1.6cm]{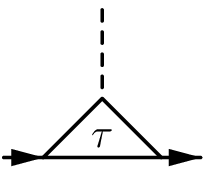}}=-i\tau,\quad
{\includegraphics[width=1.6cm]{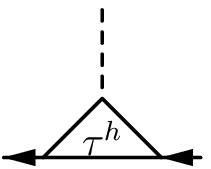}}=-i\tau^h,
\nonumber\\
{\includegraphics[width=1.6cm]{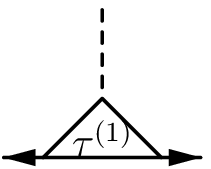}}=-i\tau^{(1)},\quad
{\includegraphics[width=1.6cm]{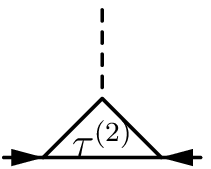}}=-i\tau^{(2)}\,.
\label{vert0}
\ee
In matrix notations, vertices have three indices, $\widehat{\tau}^{ij}_{k}$\,, where the
lower index relates to the external dash line. The couplings of the external field to the particle
and to the hole are related as
\be
\widehat{\tau}^h(x,z,y)= [\widehat{\tau}(y,z,x)]^{\rm T}\,.
\label{tauh-c} \ee
The coupling  of an external field to the non-relativistic fermion
is described by the $2\times 2$ matrix acting in the fermion spin
space. Any rank-2 matrix can be decomposed into a unit matrix
$\sigma_0$ and Pauli matrices $\vec{\sigma}$. Thus, we have
\begin{widetext}
 \be
\hat{\tau} &=& t_0\, \sigma_0 + \vec{\sigma}\,\vec{t}_1\,,\,\,
\hat{\tau}^h = t^h_0\, \sigma_0 + \vec{\sigma}^{\rm T}\,\vec{t}^{\,h}_1\,,
\nonumber\\
\hat{\tau}^{(1)}&=&
\big(t_0^{(1)}\, \sigma_0 + \vec{\sigma}\,\vec{t}_1^{\,\,(1)}\big)\, i\sigma_2,
\nonumber\\
\hat{\tau}^{(2)}&=&
i\sigma_2\,\big(t_0^{(2)}\, \sigma_0+
\vec{\sigma}\,\vec{t}_1^{\,\,(2)}\big)\,,
\label{tau-gen-struc}
 \ee
and similarly for the bare vertices
\be
\hat{\tau}_0 = t_{0,0}\, \sigma_0 +
\vec{\sigma}\,\vec{t}_{0,1}\,,\,\, \hat{\tau}_0^h = t_{0,0}^h\,
\sigma_0 + \vec{\sigma}^{\rm T}\,\vec{t}^{\,h}_{0,1}\,.
\label{bare-tau}
\ee The
vertices obey the  following  equations defined on the
Schwinger-Keldysh contour:
\be
\includegraphics[width=7.8cm]{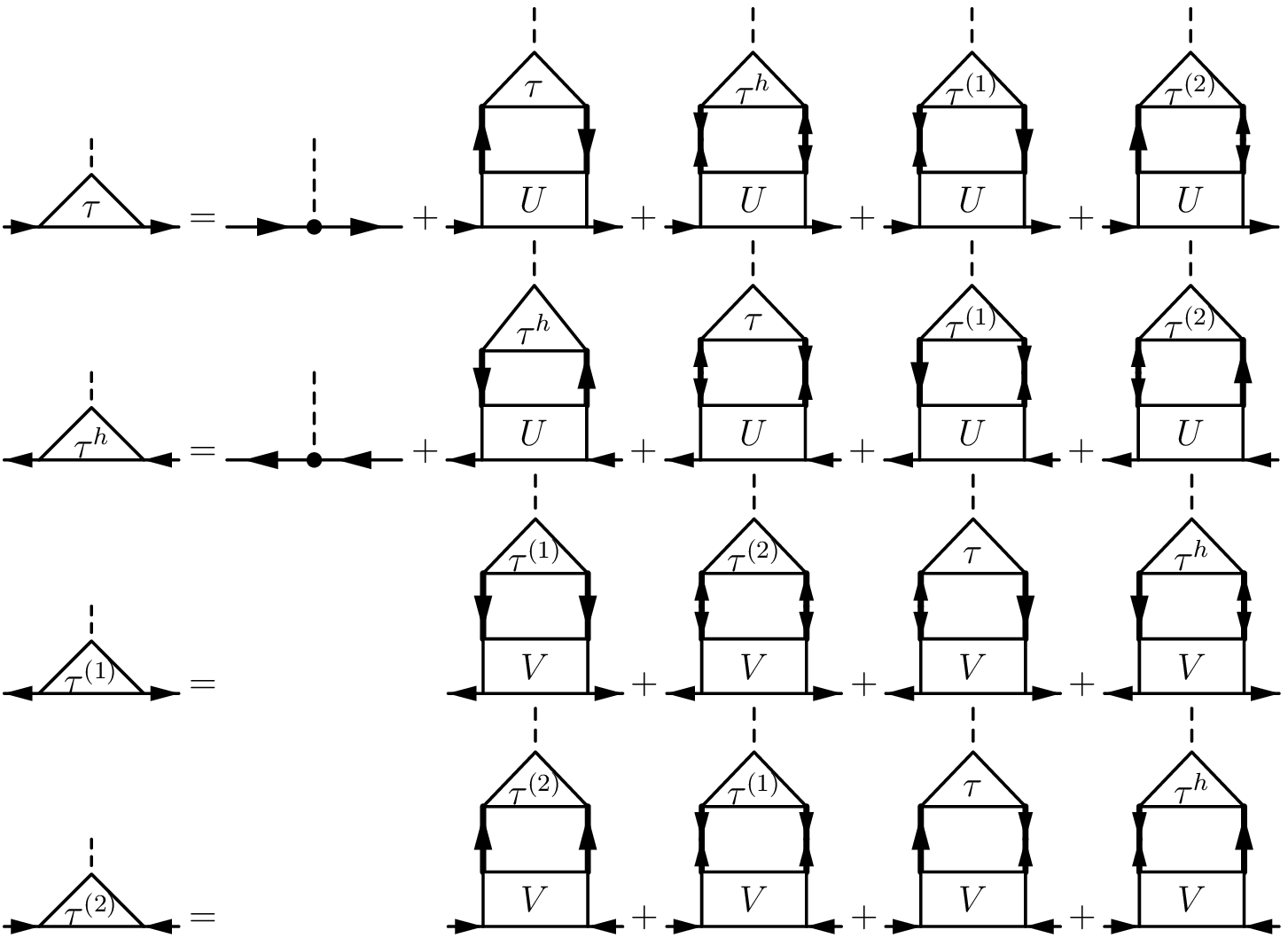}
\nonumber\\
\label{dr-vert}
\ee
Intermediate lines in (\ref{dr-vert}) are of all possible signs in
the Schwinger-Keldysh space. In operator form equations for vertices
read
\begin{subequations}
\be
\hat{\tau}&=&\hat{\tau}_0 - i
\Big[
    \hat{G}\odot \hat{\tau}\odot\hat{G}
   +\hat{F}^{(1)} \odot\hat{\tau}^h \odot\hat{F}^{(2)}
   +\hat{G}\odot\hat{\tau}^{(1)}\odot\hat{F}^{(2)}
   +\hat{F}^{(1)}\odot\hat{\tau}^{(2)}\odot \hat{G}
\Big]\boxdot\hat{U}\,,
\label{taun}\\
\hat{\tau}^h &=&\hat{\tau}_0^h -i
\Big[
  \hat{G}^h \odot \hat{\tau}^h \odot\hat{G}^h
+ \hat{F}^{(2)}\odot\hat{\tau}\odot \hat{F}^{(1)}
+ \hat{F}^{(2)}\odot\hat{\tau}^{(1)}\odot\hat{G}^h
+ \hat{G}^h\odot\hat{\tau}^{(2)}\odot \hat{F}^{(1)}
\Big]\boxdot\hat{U}\,,
\label{tauh} \\
\hat{\tau}^{(1)}&=&\phantom{\hat{\tau}_0^h} -i\Big[
 \hat{G}\odot\hat{\tau}^{(1)}\odot \hat{G}^h
+\hat{F}^{(1)}\odot\hat{\tau}^{(2)}\odot\hat{F}^{(1)}
+\hat{G}\odot\hat{\tau}\odot\hat{F}^{(1)}
+\hat{F}^{(1)}\odot \hat{\tau}^h\odot\hat{G}^h
\Big]\boxdot\hat{V}\,,
\label{tau1}\\
\hat{\tau}^{(2)}&=&\phantom{\hat{\tau}_0^h} -i\Big[
 \hat{G}^h\odot\hat{\tau}^{(2)}\odot \hat{G}
+\hat{F}^{(2)}\odot\hat{\tau}^{(1)}\odot\hat{F}^{(2)}
+\hat{F}^{(2)}\odot\hat{\tau}\odot\hat{G}
+\hat{G}^h \odot\hat{\tau}^h \odot \hat{F}^{(2)}
\Big]\boxdot\hat{V}\, .
\label{tau2}
\ee
\label{tau-eq}
\end{subequations}
\end{widetext}
Here $U$ is the particle-hole irreducible interaction, which
determines the full particle-hole scattering amplitude via the
equation
\be
\widehat{T}_{\rm ph}=\widehat{U}
-i\,\widehat{U}\boxdot \big(\widehat{G}\,\widehat{G}^h\big)\boxdot\widehat{T}_{\rm ph}\,.
\label{Tph}
\ee
In diagrams this equation has the form
\be
\parbox{6cm}{\includegraphics[width=6cm]{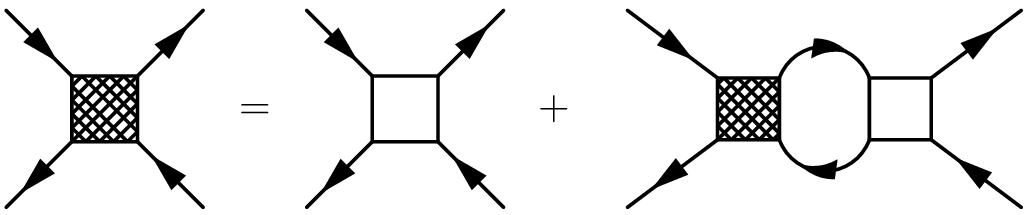}}\,,
\label{Tph-diag} \ee
which differs from (\ref{Tpp-diag}) by inversion of one of the nucleon lines. In the spin space
the matrix $\widehat{U}$ is defined as
\be
\widehat{U}_{dc,ab} &=& i\parbox{2cm}{\includegraphics[width=2cm]{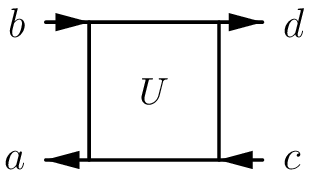}}
\nonumber\\
&=& U_0\,  \delta_{dc} \delta_{ab} +U_1^{\alpha\beta}  \,\vec{\sigma}^\beta_{dc} \,\vec{\sigma}^\alpha_{ab}.
\label{Udef}\ee
The amplitudes $U_0$ and $U_1$ are determined on the contour or they are matrices in the
Schwinger-Keldysh space.

Having at our disposal the compete spin structure of all elements
we can work out the spin algebra in Eqs.~(\ref{tau-eq}). As an
illustration we consider the second and third terms in
Eq.~(\ref{taun}):
\begin{widetext}
\begin{multline}
[(t_0-t_{0,0})\, \sigma_0 + \vec{\sigma}\,(\vec{t}_1-\vec{t}_{0,1})]_{ab}=
-i\big[
\big(F^{(1)}_0\,\sigma_0+\vec{F}^{(1)}_1\,\vec{\sigma}\big)i\sigma_2 \odot
\big(t^h_0\, \sigma_0 + \vec{\sigma}^{\rm T}\,\vec{t}^{\,h}_1\big) \odot
i\sigma_2\, \big(F^{(2)}_0\, \sigma_0 + \vec{F}^{(2)}_1\,\vec{\sigma}\big)
\nonumber\\
+
G\sigma_0\odot\big(t_0^{(1)}\, \sigma_0 + \vec{\sigma}\,\vec{t}_1^{\,\,(1)}\big)\, i\sigma_2
\odot
i\sigma_2\, \big(F^{(2)}_0\, \sigma_0 + \vec{F}^{(2)}_1\,\vec{\sigma}\big)+\dots\big]_{cd}
\nonumber
\boxdot (U_0\, \delta_{dc}  \delta_{ab} +U_1^{\alpha\beta}
\,\vec{\sigma}^\beta_{dc} \,\vec{\sigma}^\alpha_{ab} )\,.
\end{multline}
For the spin-scalar $t_0$ component we obtain
\begin{align}
t_0-t_{0,0}=-2\,i\,\big[&
-\big(F^{(1)}_0\odot t^h_0 \odot F^{(2)}_0
+ \vec{F}^{(1)}_1 \odot t^h_0 \odot \vec{F}^{(2)}_1
\big)
-G\odot \big(t_0^{(1)}\odot F^{(2)}_0 +\vec{t}_1^{\,\,(1)}\odot \vec{F}^{\,(2)}_1\big)
+\dots\big]\boxdot U_0\,.
\label{examp-t0}
\end{align}
We dropped here all terms containing simultaneously $F^{(1,2)}_0$ and $\vec{F}^{\,\,(1,2)}_1$
Green's functions, since we assumed only one type of pairing in our system. Using the trace ${\rm
Tr}\{\sigma^{\alpha}\sigma^\beta\sigma^\gamma\}=i\,2\,\epsilon_{\alpha\beta\gamma} $ we obtain
the term $i\epsilon_{\alpha\beta\gamma} \vec{F}^{\,(1)\alpha}_1\odot\vec{t}_1^{\,\,h(1)\beta}
\odot \vec{F}^{\,(2)\gamma}_1$, which vanishes since the vectors $\vec{F}^{\,(1)}$ and
$\vec{F}^{\,(2)}$ are colinear and may differ only by a phase, see Eq.~(\ref{Fdag-F}).

 The
equation for the spin-vector vertex $\vec{t}_1$ is more involved
\begin{align}
\vec{t}_1^{\,\,\alpha}-\vec{t}_{0,1}^{\,\,\alpha}=-2\,i\,\big[&
-\big(
-F^{(1)}_0\odot\vec{t}_1^{\,\,h\beta} \odot F^{(2)}_0
-(\delta_{\gamma\delta}\delta_{\lambda\beta} -
\delta_{\gamma\lambda} \delta_{\delta\beta}+\delta_{\gamma\beta} \delta_{\delta\lambda})
\vec{F}^{\,\,(1)\gamma}_1\odot\vec{t}_1^{\,\,h\delta}\odot\vec{F}^{\,\,(2)\lambda}_1\big)
\nonumber\\
&
-G\odot \big(\vec{t}_1^{\,\,(1)\beta}\odot F^{(2)}_0 +
\big(t_0^{(1)}\delta_{\beta\delta} +i\epsilon_{\beta\gamma\delta} \vec{t}^{\,\,(1)\gamma}_1\big)
\odot\vec{F}_1^{\,\,(2)\delta}\big)
+\dots\big]\boxdot U_1^{\alpha\beta}\,.
\label{examp-t1}
\end{align}
We used here the trace
${\rm Tr}\{\sigma^{\alpha}\sigma^\beta\sigma^\gamma\sigma^\delta\}=2\, (
\delta_{\alpha\beta}\delta_{\gamma\delta} - \delta_{\alpha\gamma} \delta_{\beta\delta}
+\delta_{\alpha\delta} \delta_{\beta\gamma})\,.
$
 The different
signs of terms in the first bracket appear because
$\sigma_2\,\vec{\sigma}^{\,\rm T}\, \sigma_2=-\vec{\sigma}$. We
elaborate
the spin structure of $V$
at hand of the
second and fourth terms in Eq.~(\ref{tau1}):
\begin{multline}
[\big(t_0^{(1)}\, \sigma_0 + \vec{\sigma}\,\vec{t}_1^{\,\,(1)}\big)\, i\sigma_2]_{ab}=
-i\big[
\big(F^{(1)}_0\,\sigma_0+\vec{F}^{(1)}_1\,\vec{\sigma}\big)i\sigma_2
\odot i\sigma_2\,\big(t_0^{(2)}\, \sigma_0+ \vec{\sigma}\,\vec{t}_1^{\,\,(2)}\big)
\odot \big(F^{(1)}_0\,\sigma_0+\vec{F}^{(1)}_1\,\vec{\sigma}\big)i\sigma_2
\nonumber\\
+
\big(F^{(1)}_0\,\sigma_0+\vec{F}^{(1)}_1\,\vec{\sigma}\big)i\sigma_2
\odot (t^h_0\, \sigma_0 + \vec{\sigma}^{\rm T}\,\vec{t}^{\,h}_1)\odot G^h\sigma_0
+\dots
\big]_{cd}
(V_0 (i\sigma_2)_{dc} (i\sigma_2)_{ab}  +
  V^{\alpha\beta}_1  (i\sigma_2\vec{\sigma}^\beta)_{dc}(\vec{\sigma}^\alpha i\sigma_2\,)_{ab}
  )\,.
\nonumber
\end{multline}
This equation reduces to
\begin{align}
t_0^{(1)}=-2\, i\,\big[&
+\big(F^{(1)}_0\odot t_0^{(2)} \odot F^{(1)}_0
     +\vec{F}^{(1)}_1 \odot t_0^{(2)} \odot \vec{F}^{(1)}_1
\big)
-(F^{(1)}_0\odot t^h_0 - \vec{F}^{\,\,(1)}_1\odot\vec{t}^{\,\,h}_1) \odot G^h+\dots\big]\boxdot V_0\,,
\label{examp-tt0}\\
\vec{t}_1^{\,\,(1)\alpha}=-2\,i\big[&
+\big(F^{(1)}_0\odot \vec{t}_1^{\,\,(2)\beta} \odot F^{(1)}_0
+(\delta_{\gamma\delta}\delta_{\lambda\beta} - \delta_{\gamma\lambda}
\delta_{\delta\beta}+\delta_{\gamma\beta} \delta_{\delta\lambda})
\vec{F}^{(1)\gamma}_1 \odot \vec{t}_1^{\,\,(2)\delta} \odot \vec{F}^{(1)\lambda}_1
\big)
\nonumber\\
&
-\big(
\vec{F}^{\,\,(1)\gamma}_1\odot
\big(t^{h}_0\delta_{\gamma\beta} - i\epsilon_{\beta\gamma\delta} \vec{t}^{\,\,h\delta}_1 \big)
- F^{(1)}_0\odot \vec{t}^{\,\,h\beta}_1
\big) \odot G^h+\dots\big]\boxdot V_1^{\alpha\beta}\,.
\label{examp-tt1}
\end{align}
Proceeding this way and collecting all terms  we rewrite Eqs.~(\ref{tau-eq}) for the scalar
vertices as
\begin{subequations}
\begin{align}
t_0-t_{0,0}= - 2\,i
\Big[&
    G\odot \tau_0\odot G
-F^{(1)}_0\odot t^h_0 \odot F^{(2)}_0
- G\odot t_0^{(1)}\odot F^{(2)}_0
-F^{(1)}_0\odot t_0^{(2)}\odot G
\nonumber\\
&- \vec{F}^{(1)}_1 \odot t^h_0 \odot \vec{F}^{(2)}_1
 - G\odot\vec{t}_1^{\,\,(1)}\odot \vec{F}^{\,(2)}_1
 -\vec{F}^{\,(2)}_1\odot \vec{t}_1^{\,\,(1)}\odot G
\Big]\boxdot U_0\,,
\label{tsol:t0}\\
t^h_0-t^h_{0,0}= - 2\,i
\Big[&
    G^h\odot \tau^h_0\odot G^h
-F^{(2)}_0\odot t_0 \odot F^{(1)}_0
-  F^{(2)}_0\odot t_0^{(1)}\odot G^h
-G^h\odot t_0^{(2)}\odot F^{(1)}_0
\nonumber\\
&- \vec{F}^{(2)}_1 \odot t_0 \odot \vec{F}^{(1)}_1
 - \vec{F}^{\,(2)}_1\odot\vec{t}_1^{\,\,(1)}\odot G^h
 - G^h\odot \vec{t}_1^{\,\,(2)}\odot \vec{F}^{\,(1)}_1
\Big]\boxdot U_0\,,
\label{tsol:t0h}
\end{align}
\begin{align}
t_0^{(1)}=+2\, i\,\big[&
G\odot t^{(1)}_0\odot G^h - F^{(1)}_0\odot t_0^{(2)} \odot F^{(1)}_0
+ G \odot   t_0\odot F^{(1)}_0 + F^{(1)}_0\odot t^h_0 \odot G^h
\nonumber\\
&-\vec{F}^{(1)}_1 \odot t_0^{(2)} \odot \vec{F}^{(1)}_1
 + G \odot  \vec{t}_1 \odot \vec{F}^{\,\,(1)}_1
 - \vec{F}^{\,\,(1)}_1\odot\vec{t}^{\,\,h}_1 \odot G^h\big]\boxdot V_0\,,
\label{tsol:t01}\\
t_0^{(2)}=+2\, i\,\big[&
G^h\odot t^{(2)}_0\odot G - F^{(2)}_0\odot t_0^{(1)} \odot F^{(2)}_0
+ F^{(2)}_0 \odot   t_0\odot G  + G^h\odot t^h_0 \odot F^{(2)}_0
\nonumber\\
&-\vec{F}^{(2)}_1 \odot t_0^{(1)} \odot \vec{F}^{(2)}_1
 +  \vec{F}^{\,\,(2)}_1\odot  \vec{t}_1 \odot G
 -  G^h\odot\vec{t}^{\,\,h}_1 \odot \vec{F}^{\,\,(1)}_1 \big]\boxdot V_0\,.
\label{tsol:t02}
\end{align}
\label{tsol:0}
\end{subequations}
For the vector vertices we have
\begin{subequations}
\begin{align}
\vec{t}_1^{\,\,\alpha}-\vec{t}_{0,1}^{\,\,\alpha}=&-2\,i\,\big[
G\odot \vec{t}_1^{\,\,\alpha}\odot G
+F^{(1)}_0\odot\vec{t}_1^{\,\,h\beta} \odot F^{(2)}_0
-G\odot \vec{t}_1^{\,\,(1)\beta}\odot F^{(2)}_0
- F^{(1)}_0\odot\vec{t}_1^{\,\,(2)\beta}\odot G
\nonumber\\
&
+(\delta_{\gamma\delta}\delta_{\lambda\beta} -
\delta_{\gamma\lambda} \delta_{\delta\beta}+\delta_{\gamma\beta} \delta_{\delta\lambda})
\vec{F}^{\,\,(1)\gamma}_1\odot\vec{t}_1^{\,\,h\delta}\odot\vec{F}^{\,\,(2)\lambda}_1
\nonumber\\
& -G\odot\big(  t_0^{(1)}\delta_{\beta\delta} +
 i\epsilon_{\beta\gamma\delta}\vec{t}_1^{\,\,(1)\gamma}\big)\odot \vec{F}^{\,\,(2)\delta}_1
 - \vec{F}^{\,(1)\gamma}_1  \odot
 \big( t_0^{(2)}\delta_{\gamma\beta} +i\epsilon_{\beta\gamma\delta} \vec{t}_1^{\,\,(2)\delta}\big)\odot G
\big]\boxdot U_1^{\alpha\beta}\,,
\label{tsol:t1}\\
\vec{t}^{\,\,h\alpha}_1-\vec{t}^{\,\,h\alpha}_{0,1}=& - 2\,i
\Big[
G^h\odot \vec{t}^{\,\,h\beta}_1\odot G^h
+F^{(2)}_0\odot \vec{t}^{\,\beta}_1 \odot F^{(1)}_0
+F^{(2)}_0\odot \vec{t}_0^{\,(1)\beta}\odot G^h
+G^h\odot \vec{t}_0^{\,\,(2)\beta}\odot F^{(1)}_0
\nonumber\\
&
+(\delta_{\gamma\delta}\delta_{\lambda\beta} -
\delta_{\gamma\lambda} \delta_{\delta\beta}+\delta_{\gamma\beta} \delta_{\delta\lambda})
\vec{F}^{\,\,(2)\gamma}_1\odot\vec{t}_1^{\,\,\delta}\odot\vec{F}^{\,\,(1)\lambda}_1
\nonumber\\
&
+ \vec{F}^{\,(2)\gamma}_1\odot
 \big( t_0^{\,\,(1)}\delta_{\gamma\beta} +i\epsilon_{\beta\gamma\delta} \vec{t}_1^{\,\,(1)\delta}
 \big)\odot G^h
+ G^h\odot
 \big( t_0^{(2)}\delta_{\beta\delta} +i\epsilon_{\beta\gamma\delta}\vec{t}_1^{\,\,(2)\gamma} \big)
 \odot \vec{F}^{\,(1)\delta}_1
\Big]\boxdot U_1^{\alpha\beta}\,,
\label{tsol:t1h}\\
\vec{t}_1^{\,\,(1)\alpha}=&+2\,i\big[
G\odot \vec{t}^{\,\,(1)\beta}\odot G^h - F^{(1)}_0\odot \vec{t}_1^{\,\,(2)\beta} \odot F^{(1)}_0
+ G\odot \vec{t}^{\,\,\beta}_1 \odot F^{(1)}_0
- F^{(1)}_0\odot \vec{t}^{\,\,h\beta}_1 \odot G^h
\nonumber\\
&
-(\delta_{\gamma\delta}\delta_{\lambda\beta} - \delta_{\gamma\lambda}
\delta_{\delta\beta}+\delta_{\gamma\beta} \delta_{\delta\lambda})
\vec{F}^{(1)\gamma}_1 \odot \vec{t}_1^{\,\,(2)\delta} \odot \vec{F}^{(1)\lambda}_1
\nonumber\\
&
+G\odot\big( t_0\delta_{\beta\delta}+i\epsilon_{\beta\gamma\delta}\vec{t}_1^{\,\gamma}\big)
\odot\vec{F}^{\,\,(1)\delta}_1
+\vec{F}^{\,\,(1)\gamma}_1\odot \big(t^{h}_0\delta_{\gamma\beta}
-i\epsilon_{\beta\gamma\delta}\vec{t}_1^{\,\,h\delta}\big)\odot G^h
\big]\boxdot V_1^{\alpha\beta}\,,
\label{tsol:t11}\\
\vec{t}_1^{\,\,(2)\alpha}=&+2\,i\big[
G^h\odot \vec{t}^{\,\,(2)\beta}\odot G
- F^{(2)}_0\odot \vec{t}_1^{\,\,(1)\beta} \odot F^{(2)}_0
+ F^{(2)}_0\odot \vec{t}^{\,\,\beta}_1 \odot G
- G^h\odot \vec{t}^{\,\,h\beta}_1 \odot F^{(2)}_0
\nonumber\\
&
-(\delta_{\gamma\delta}\delta_{\lambda\beta} - \delta_{\gamma\lambda} \delta_{\delta\beta}+\delta_{\gamma\beta} \delta_{\delta\lambda})
\vec{F}^{(2)\gamma}_1 \odot \vec{t}_1^{\,\,(1)\delta} \odot \vec{F}^{(2)\lambda}_1
\nonumber\\
&
+\vec{F}^{\,\,(2)\gamma}_1\odot
\big( t_0\delta_{\gamma\beta}+i\epsilon_{\beta\gamma\delta}\vec{t}_1^{\,\delta}\big)\odot G
+  G^h \odot \big(t^{h}_0\delta_{\beta\delta}
-i\epsilon_{\beta\gamma\delta}\vec{t}_1^{\,\,h\gamma}\big)\odot \vec{F}^{\,\,(1)\delta}_1
\big]\boxdot V_1^{\alpha\beta}\,.
\label{tsol:t12}
\end{align}
\label{tsol:1}
\end{subequations}
Recall that Eqs.~(\ref{tsol:0}) and (\ref{tsol:1}) are written
here simultaneously  for both types of pairing with even and odd
angular momenta. For the former case we must retain only the terms
with $F^{(1,2)}_0$, in the latter one, the terms with
$\vec{F}^{(1,2)}_1$.

Response  of the system to the external field is described by the self-energy
\begin{align}
-i\, \widehat{\Sigma}&=\parbox{10cm}{\includegraphics[width=10cm]{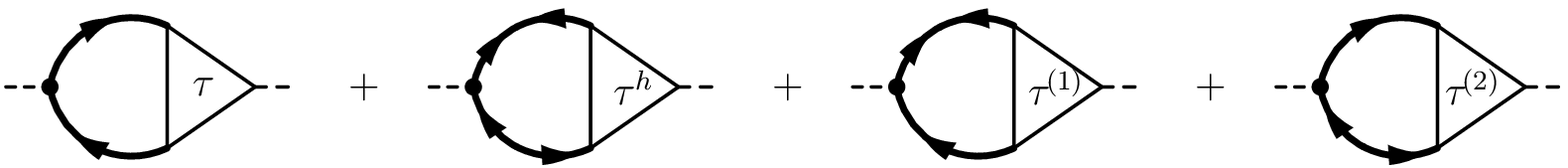}}
\label{chi-graph}\\
&= - \widehat{\tau}_0\boxdot \Big[
\widehat{G}\odot \widehat{\tau} \odot \widehat{G} +
\widehat{F}^{(1)}\odot \widehat{\tau}^h \odot \widehat{F}^{(2)}
+ \widehat{G}\odot \widehat{\tau}^{(1)} \odot \widehat{F}^{(2)}
+ \widehat{F}^{(1)}\odot \widehat{\tau}^{(2)} \odot \widehat{G}\Big]\,.
\label{chi-ss}
\end{align}
After taking the spin trace we obtain
\begin{align}
\Sigma=-2\,i\,t_{0,0}\boxdot
\Big[& G \odot t_{0} \odot G - F_0^{(1)} \odot t_0^h\odot F_0^{(2)}
     - G\odot t_0^{(1)}\odot F_0^{(2)} -  F_0^{(1)}\odot t_0^{(2)}\odot G
\nonumber\\
&-\vec{F}_1^{\,\,(1)}\odot t_0^h\odot \vec{F}_1^{\,\,(2)} - G\odot \vec{t}_1^{\,\,(1)}\odot \vec{F}_1^{\,\,(2)} -
 \vec{F}_1^{\,\,(1)}\odot \vec{t}_1^{\,\,(2)}\odot G \Big]
\nonumber\\
-2\,i\,\vec{t}_{0,1}^{\,\alpha}\boxdot
\Big[& G\odot \vec{t}_1^{\,\,\alpha}\odot G + F_0^{(1)}\odot \vec{t}_1^{\,\,h\alpha}\odot F_0^{(2)}
     - G\odot \vec{t}_1^{\,\,(1)\alpha}\odot F_0^{(2)} - F_0^{(1)}\odot \vec{t}_1^{\,\,(2)\alpha}\odot G
\nonumber\\
&+\big(\delta_{\alpha\beta}\delta_{\gamma\delta}-\delta_{\alpha\gamma}\delta_{\beta\delta}
 +\delta_{\alpha\delta}\delta_{\beta\gamma}\big)
\, \vec{F}_1^{\,\,(1)\beta}\odot \vec{t}_1^{\,\,h \gamma}\odot
\vec{F}_1^{\,\,(2)\delta} \nonumber\\ &- G\odot
\big(t_0^{(1)}\delta_{\alpha\gamma}+i\epsilon_{\alpha\beta\gamma}
\vec{t}_1^{\,\,(1)\beta}\big)\odot\vec{F}_1^{\,\,(2)\gamma}
-\vec{F}_1^{\,\,(1)\beta}\odot
\big(t_0^{(2)}\delta_{\alpha\beta}+i\epsilon_{\alpha\beta\gamma}
\vec{t}_1^{\,\,(2)\gamma}\big)\odot G \Big]\,. \label{chi-full}
\end{align}
\end{widetext}
If external field couples to a conserved current the self-energy
(\ref{chi-ss}) must support the current conservation obeying the
relations
\be
\partial_{x}^\mu\, \Sigma_{\mu\nu}(x,y)=\partial_y^\nu\,
\Sigma_{\mu\nu}(x,y)=0\,. \label{conserv-ss}
 \ee
Following our notations the first Lorentz index, $\mu$, in $\Sigma_{\mu\nu}(x,y)$  is attached to the
right vertex (at the contour coordinate $x$) in diagrams
(\ref{chi-ss}) and the second index, $\nu$, is attached to the left vertex (at
the contour coordinate $y$). Relations (\ref{conserv-ss}) can be
fulfilled if the full vertex functions (\ref{vert0}) satisfy the
Ward identity
\be
&&\widehat{G}(x,z)-\widehat{G}(z,y)
\nonumber\\
&&
=   \widehat{G}(x,x')      \odot \partial_z^\mu\widehat{\tau}_\mu(x',z,y')
   \odot \widehat{G}(y',y)
\nonumber\\
&&
+  \widehat{F}^{(1)}(x,x')\odot
\partial_z^\mu\widehat{\tau}^h_\mu(x',z,y')     \odot
\widehat{F}^{(2)}(y',y)
\nonumber\\
&&
+  \widehat{G}(x,x') \odot
\partial_z^\mu\widehat{\tau}^{(1)}_\mu(x',z,y') \odot \widehat{F}^{(2)}(y',y)
\nonumber\\
&&
+   \widehat{F}^{(1)}(x,x')\odot
\partial_z^\mu\widehat{\tau}^{(2)}_\mu(x',z,y') \odot
\widehat{G}(y',y) \label{Ward-contour} \ee
derived first in Ref.~\cite{LM63} and generalized here for
coordinates on the Schwinger-Keldysh contour or for matrices in
the Schwinger-Keldysh space.


\section{Optical theorem formalism}
\label{sec:White}

\subsection{Radiation from a piece of non-equilibrium matter}

Optical theorem formalism is an efficient tool to calculate reaction rates including finite
particle widths and other in-medium effects~\cite{VS87,KV95,KV99}.  Assume that we deal with a
 system of a finite size (white body) transparent for radiating quanta. To be specific let us consider
anti-neutrino--lepton ($\bar{\nu}l$) production. By the lepton we mean the electron $e$, muon $\mu^{-}$, or neutrino
$\nu$. We assume that the system is opaque for $e$ and $\mu^{-}$ but
transparent for $\nu$ and $\bar{\nu}$. Then it is convenient to express all quantities in the
Wigner representation doing the transformation (\ref{Wigner-def}) from coordinates $x=X-\xi /2$,
$y=X+\xi /2 $ to the corresponding Wigner coordinates  $(X;q) =(t,\vec{X};\om,\vec{q})$. The
probability of the  anti-neutrino--lepton production can be expressed in terms of
the evolution operator $S$,
\begin{gather}
\frac{\rmd\mathcal{W}^{\rm tot}_{X^{*}\rightarrow \bar\nu l}}{\rmd^3 x\rmd t \rmd^4q}
= \intop\rmd \Phi_l
(-i\Sigma^{-+})\delta^{(4)}(q-q_l-q_{\bar\nu})\,,
\label{optfirst}\\
-i\Sigma^{-+}(x,q)=\sum_{\{X^{*}\}} \langle \bar\nu l|\delta S^\dagger |X^{*}\rangle \langle X^{*}|\delta
S|\bar\nu l\rangle\,, \label{Sigpm-gen}
\end{gather}
where $\delta S =S-1$, and
\be
\rmd\Phi_l=\frac{(1-n_l) d q_l^3 dq_{\bar{\nu}}^3}{(2\pi)^6\,4\,
\omega_l \,\omega_{\bar{\nu}}}
\label{lept-phase-space}
\ee
is the phase-space volume of an antineutrino with the
four-momentum $q_{\bar\nu}=(\om_{\bar{\nu}},\vec{q}_{\bar{\nu}})$
and a lepton with the four-momentum $q_l=(\om_l,\vec{q}_l)$.
If $l=\nu$,  the occupation number $n_l$ is to be put zero.
The summation goes over the complete set of all possible intermediate
states $\{X^{*}\}$. In Eq.~(\ref{Sigpm-gen}) we suppose that electrons or muons can be
treated in the quasi-particle approximation. Then there is no need (although
possible) to consider them as intermediate states.

Making use of  smallness of the weak coupling, we expand the evolution operator as
\begin{equation}\label{Smatrix}
\delta S\approx - i \, \intop_{-\infty}^{+\infty} \mathcal{T}
\,\bigl\{V_W(x_0 )\, S_{\rm nucl} (x_0 )\bigr\} d x_0 \,,
\end{equation}
where $V_W$ is the Hamiltonian of the weak interactions in the interaction representation, $S_{\rm
nucl}$ is the part of the $S$ matrix corresponding to strong nuclear interactions, and
$\mathcal{T}\{...\}$ is the chronological time ordering operator. After substitution of $\delta S$
into (\ref{Sigpm-gen}) and averaging over a non-equilibrium initial state of the nuclear
system, there appear chronologically ordered ($G^{--}$, $F^{--}$), anti-chronologically ordered
($G^{++}$, $F^{++}$) and disordered ($G^{+-}$, $F^{+-}$ and $G^{-+}$, $F^{-+}$) exact Green's
functions.

Once the reaction probability is evaluated according to Eq.~(\ref{optfirst}), the neutrino emissivity
in the neutral channel, i.e. with $\bar{\nu}\nu$ production,
is given by \cite{VS87}
\be
\epsilon_{\bar\nu\nu} =\int \frac{\rmd \mathcal{W}^{\rm tot}_{X^{*}\rightarrow
\bar\nu \nu}} {\rmd^3 x \rmd t\rmd^4q}\, \om\, \rmd \om\,\rmd^3 q
\,. \label{emis} \ee
The emissivity in the charged channel, i.e. with $\bar{\nu} l$ production ($l=e^-\,,\mu^-$),
is as follows
\be
\epsilon_{\bar\nu l} =\int \frac{\rmd \mathcal{W}^{\rm tot}_{X^{*}\rightarrow
\bar\nu l}} {\rmd^3 x \rmd t\rmd^4q\rmd^3 q_{\bar\nu}}\, \om_\nu\,\rmd^3 q_{\bar\nu}\,\rmd^4 q
\,. \label{emis-nul}
\ee

The Lagrangian density for the lepton-nucleon interactions is
\be
L=\frac{G}{\sqrt{2}}\big( j_{{\rm ch},\mu} l_{\rm ch}^{\mu}
+j_{{\rm neut},\mu} l_{\rm neut}^{\mu}+\mbox{h.c.}\big),
\label{weak-Lag}
\ee
where $G=1.166\times 10^{-5}$~GeV$^{-1}$ is the Fermi coupling constant and there are two
contributions from charged (ch) and neutral (neut) currents; ``h.c.'' stands for hermitian
conjugated terms. The lepton currents are
\be
&&l_{\rm neut}^\mu =\bar{\psi}_{\nu}\gamma^{\mu}(1-\gamma_5 )\psi_{\nu} ,
\nonumber\\
&&l_{\rm ch}^\mu =\bar{\psi}_e\gamma^{\mu}(1-\gamma_5 ) \psi_\nu\,.
\label{lept-curr}
\ee

The nucleon currents $j_{\rm ch}$ and $j_{\rm neut}$ have  vector and axial-vector components
\begin{align}
&j_{{\rm ch},\mu}=j^{np}_{\mu}\,,\quad j_{{\rm neut},\mu}=j^{pp}_{\mu}+j^{nn}_{\mu} ,
\nonumber\\
&j^{np}_{\mu}=\Psi_p^{\dagger}(p')( J^{V}_{\mu} - g_A\, J^{A}_{\mu} )\Psi_n(p),
\nonumber\\
&j^{pp}_\mu =\phantom{-}\frac{1}{2}\Psi_{p}^{\dagger}(p') (c_v J^V_\mu -g_A\, J^A_\mu) \Psi_{p}(p) ,
\nonumber\\
&j^{nn}_\mu =-\frac{1}{2}\Psi_{n}^{\dagger}(p') (J^V_\mu-g_A\, J^A_{\mu})\Psi_{n}(p),
\label{currents}
\end{align}
where the four vectors $J^{V,A}$ can be written for
non-relativistic nucleons as, cf. Ref.~\cite{EW},
\be
J^{V} =\big(1,{\textstyle\frac{\vec{p}\,'+\vec{p}}{2m_N}}\big)
\,,\quad J^{A}=\big({\textstyle
\frac{\vec{\sigma}(\vec{p}\,'+\vec{p})}{2m_N}},\vec{\sigma}\big),
\label{curr-vert} \ee
and $c_v =1-4\sin^2\theta_{\rm W} \simeq 0.08$, $\theta_{\rm W}$ is the Weinberg angle; $g_A \simeq
1.26$ is the axial-vector coupling constant, $p=(p_0,\vec{p})$, $\vec{p}$ and $\vec{p}{\,'}$ are momenta of incoming
and outgoing nucleons. Compared to the frequently used expression, that includes only leading terms in
the non-relativistic expansion, e.g., see Ref.~\cite{FM79}, we following \cite{KV1,KV2} retain here
sub-leading terms $\propto v_{\rm F}$. Although in many cases the terms $\propto v_{\rm F}$ yield
small corrections to leading-order results, in some particular cases the leading-term contribution
may vanish because of symmetry constraints and then sub-leading terms become dominant. Such an
example will be studied below. The bare current vertex involves the bare nucleon mass, see
Eq.~(\ref{curr-vert}). In medium, however, nucleon wave functions are normalized to one
quasi-particle rather than to one free particle, provided the nucleons are treated
in the quasi-particle approximation. Hence, the bare nucleon mass $m_N$ is to be replaced by the
in-medium nucleon mass $m^*_N$.

The structure of the weak-interaction Lagrangian (\ref{weak-Lag}) suggests that we can detach
leptonic currents
\be
\label{optpi}
\Sigma^{-+}=\frac{G^2}{2}\, \Sigma^{-+}_{{\rm nucl},\mu\nu}\, \sum_{\rm spin}\{l^\mu l^\nu\} \,,
\ee
and deal with the object determined by the strong interactions only
\begin{align}
&\Sigma^{-+}_{{\rm nucl},\mu\nu}(X,q) \nonumber\\ &\qquad=\int d^4
\xi e^{iq\xi} i\langle j_{\nu}^{\dagger}(X-\xi /2)j_{\mu}(X+\xi /2
)\rangle \,. \label{sigmamp}
\end{align}
$\Sigma^{-+}_{{\rm nucl},\mu\nu}$ is the full $(-+)$ self-energy for the nuclear processes;
the current $j$ stands here for charged or neutral nucleon currents defined in (\ref{currents}).
Quantum states and operators are taken in the Heisenberg picture. The sum in Eq.~(\ref{optpi})
runs over lepton spins.

In the graphical form, the general expression for the probability of the lepton (electron, muon,
neutrino) and anti-neutrino production  is depicted as
\be
\frac{G^2}{2}\, i\Sigma^{-+}_{{\rm nucl},\mu\nu}\,  \sum_{\rm spin}\{l^\mu l^\nu\}
=\parbox{2.5cm}{\includegraphics[width=2.5cm]{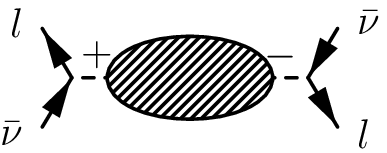}}\,\,\, .
\label{Smattot}
\ee
The hatched block has the meaning of the $(-+)$ self-energy of the virtual $Z$ or $W$ bosons which
convert in a lepton $l$ and an anti-neutrino $\bar{\nu}$. The block determines the gain term in the
generalized kinetic equation for the full $(-+)$ Green's functions of the virtual boson. If one
integrates over $e$/$\mu$ (closes the $e$/$\mu$ line in diagrams), one recovers the gain term for
the $\bar{\nu}$ in the charged current processes, see Refs.~\cite{KV95,IKV00,Fauser}.
This circumstance allows  to use the given method in
different non-equilibrium problems, like in description of the neutrino/antineutrino transport.
Note that the generalized kinetic equation for $\bar{\nu}$ is greatly simplified, if conditions for
the quasi-particle approximation are fulfilled, see \cite{KV99,IKV00,SD99}.

The integration over the lepton phase-space can be performed separately from the calculation of
$\Sigma^{-+}_{{\rm nucl},\mu\nu}$. If we introduce the leptonic tensor
\be
T_{\rm lept}^{\mu\nu}(q)=\intop \rmd  \Phi_l\, \sum_{\rm spin}\{l^\mu l^\nu \}\, \delta^{(4)}(q-q_l-q_{\bar\nu})\,,
\label{Lmunu}
\ee
the reaction probability (\ref{optfirst}) takes the form
\be
\frac{\rmd\mathcal{W}^{\rm tot}_{X^{*}\rightarrow \bar\nu \nu}}{\rmd^3 x\rmd t \rmd^4q}
=-i\frac{G^2}{2}\, \Sigma^{-+}_{{\rm nucl},\mu\nu}(X,q)\,T_{\rm lept}^{\mu\nu}(q)\,.
\label{Wneut}
\ee
For evaluation of the emissivity in Eq.~(\ref{emis-nul}) we also need the following expression
\begin{align}
\intop\frac{\rmd\mathcal{W}^{\rm tot}_{X^{*}\rightarrow \bar\nu l }}
{\rmd^3 x\rmd t \rmd^4q \rmd^3 q_{\bar\nu}}\om_{\bar\nu}\rmd^3 q_{\bar\nu}
=&-i\frac{G^2}{2}\, \Sigma^{-+}_{{\rm nucl},\mu\nu}(X,q)\,
\nonumber\\
&\times\widetilde{T}_{\rm lept}^{\mu\nu}(q)\,,
\label{Wcharg}
\end{align}
where $l=e^-,\mu^-$ and
\be
\widetilde{T}_{\rm lept}^{\mu\nu}(q)=\!\!
\intop \rmd  \Phi_l\,\om_{\bar\nu} \sum_{\rm spin}\{l^\mu l^\nu \}\,
\delta^{(4)}(q-q_l-q_{\bar\nu})\,.
\label{tLmunu}
\ee
The tensors $T_{\rm lept}^{\mu\nu}$ and $\widetilde{T}_{\rm lept}^{\mu\nu}$ are calculated in
Appendix~\ref{app:lept-tensor}.
Note that the self-energies, $\Sigma_{\rm nucl}^{-+}$, in Eqs.~(\ref{Wneut}) and (\ref{Wcharg})
are to be constructed with the neutral and charged nucleon currents from Eq.~(\ref{currents}),
respectively.

As we have discussed in the Introduction, the standard calculation of the reaction rates is done
with the help of summation of the squared matrix elements of reactions, see \cite{YLS99}. This is
fully correct procedure, if one treats the processes perturbatively, i.e. provided there is a
small expansion parameter. One nucleon processes are related to perturbative $\Sigma^{-+}$
diagrams with only one the nucleon $G_0^{-+}$ Green's function in expansion of (\ref{Smattot}),
two-nucleon processes are related to the diagrams with two nucleon $G_0^{-+}$ Green functions,
etc. However, this procedure fails when applied to strongly interacting systems. The description
of even a one-nucleon process includes infinite number of perturbative diagrams with the
$NN$-interactions, since the coupling is not small. Nevertheless, one is able to separate
processes using the quasi-particle approximation provided excitation energies are sufficiently low
$\om\ll \epsilon_{\rm F}$ (when the fermion width is small). Then, diagrams with one
quasi-particle nucleon $G^{-+}/F^{-+}$ Green's function describe  the one-nucleon reactions, with
two $G^{-+}/F^{-+}$ Green's functions describe two-nucleon reactions, etc. The calculations of the
reaction rates based on application of the optic theorem formalism and calculations using the
ordinary formalism of computing squared reaction-matrix elements yield the same
results~\cite{VS86,VS87}. In general case, when particle widths are not small, situation becomes
much more involved. Then  calculations using the squared matrix elements become invalid, and the
only possibility to calculate the emissivity from the piece of matter is to use the closed-diagram
technique. Below we formulate a general method and then discuss the quasi-particle approximation.

\subsection{
Diagrammatic decomposition in terms of full ($-+$) Green functions
}
\label{subsec:pm-decomp}

\subsubsection{Fermions with finite width}

The hatched block in Eq. (\ref{Smattot}) is the sum of all closed diagrams written in terms of
full Green's functions. External ($-+$) signs mean that each diagram in the series contains at
least one ($-+$) nucleon Green's function ($G^{-+}$ and additionally $F^{-+}$, for a system with pairing). The
latter function is especially important. It obeys the Kadanoff-Baym kinetic equation. Various
contributions from $\{X\}$ can be classified according to the number $N$ of exact $G^{-+}/F^{-+}$
nucleon Green's functions (lines in the diagram):
\begin{align}
&\frac{d{\cal W}^{\rm  tot}_{X^{*}\rightarrow\bar\nu l}}{d^3 x d t}
= \intop \rmd\Phi_l\, \delta^{(4)}(q-q_l-q_{\bar\nu})
\nonumber\\
&\quad\times \left[
\parbox{25mm}{
\includegraphics[width=2.5cm]{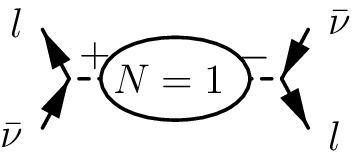}}
+
\parbox{25mm}{
\includegraphics[width=2.5cm]{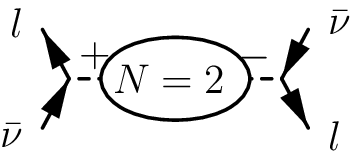}}+
\dots\right] .
\label{qp}
\end{align}
The quasi-particle approximation for fermions can be utilized only if the energy of radiating
quanta $\om$ is larger than the nucleon width ($\om \sim \epsilon^* \sim T \gg \Gamma_N$), i.e.,
inequality $\om\sim T \ll \epsilon_{\rm F}$ must hold, see Refs.~\cite{VS87,KV95}. In this case
the contributions of various processes encoded in a closed diagram can be made visible by cutting
the diagrams through the ($+-$) and ($-+$) lines. The cut means taking off the energy integral
provided the spectral functions of fermionic quasi-particles can be reduced to the
$\delta$-functions. This restricts the fermion energy to an in-medium mass shell. The $N=1$ term
describes the DU process, and $N=2$, the MMU and MNB processes.

In general case, when the fermion width cannot be neglected, the cut through the ($-+$), ($+-$) lines has
only a symbolic meaning. Nevertheless, the separation of the diagrams according to the number of
the full ($-+$)  Green's functions proves to be helpful also in this case \cite{KV95}. Note that now each
diagram in (\ref{qp}) represents a whole class of perturbative diagrams of any order in the
interaction strength and in the number of loops.

The full set of topologically distinct skeleton diagrams for $\Sigma^{-+}$ written in terms of
full ($-+$) Green's functions can be explicitly presented as a series in
$N$~\cite{KV95}. For $N=1$ and $N=2$ we have
\begin{align}
&\parbox{2cm}{\includegraphics[width=2cm]{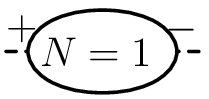}}=\,
\parbox{2cm}{\includegraphics[width=2cm]{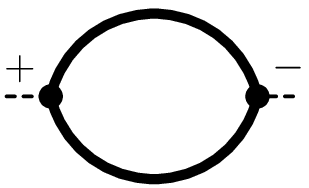}}\,,
\label{N=1-diag}
\end{align}
\begin{align}
&\parbox{2cm}{\includegraphics[width=2cm]{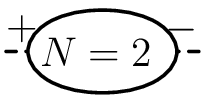}}=\,
\parbox{2.1cm}
{\includegraphics[height=1.8cm]{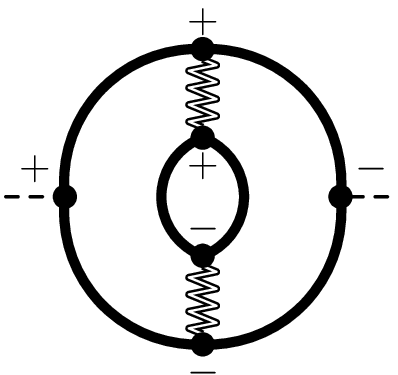}}
+
\parbox{2.4cm}
{\includegraphics[height=1.6cm]{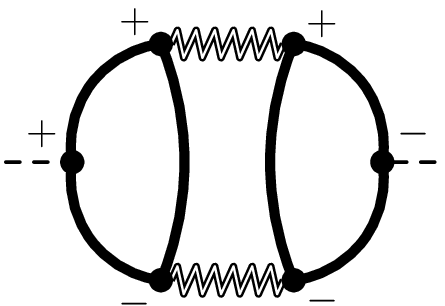}}
\nonumber\\
&
 +
\parbox{2.4cm}
{\includegraphics[height=1.6cm]{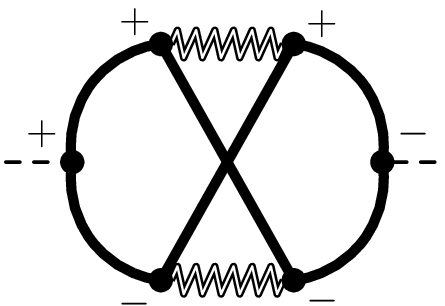}}
+
\parbox{3cm}
{\includegraphics[height=1.6cm]{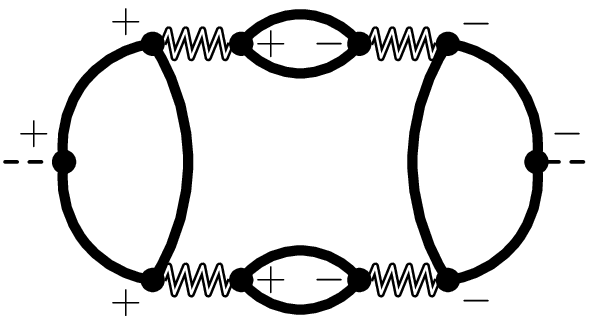}}\,.
\label{N=2-diag}
\end{align}
For $N=3$ we have, for example, contributions of the type
\be
\parbox{2cm}{\includegraphics[width=2cm]{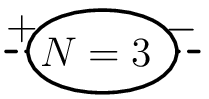}}=
\parbox{4cm}{\includegraphics[width=4cm]{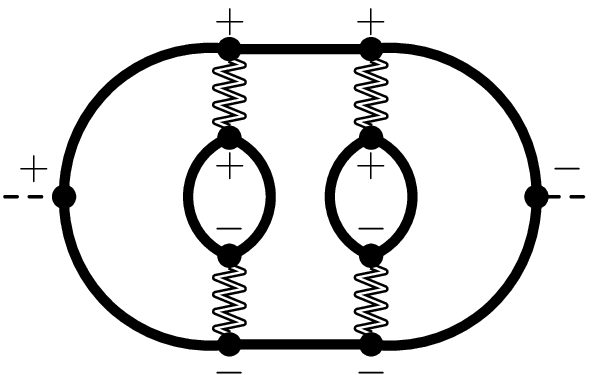}}
\nonumber\\
+
\parbox{5cm}{\includegraphics[width=5cm]{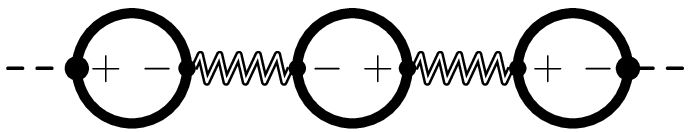}}+\dots\,.
\label{N=3-diag}
\ee
Note that for $N\geq 3$ there appear multi-cut diagrams (see the
last explicitly presented diagram in (\ref{N=3-diag})).
The $NN$ interaction block in Eqs.~(\ref{N=1-diag},\ref{N=2-diag}) and (\ref{N=3-diag}) is the full block
containing the vertices of one particular sign, e.g.,
\begin{equation}
\label{zigzag}
\parbox{6cm}{\includegraphics[width=6cm]{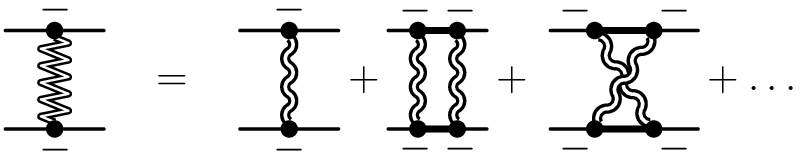}}
\end{equation}
and the analogous equation for the $(++)$ block. Since only the same-sign vertices are permitted
in Eq.~(\ref{zigzag}), no $(\pm,\mp)$ [i.e., $(+,-)$ and $(-,+)$] lines appear in these diagrams. The thick double-wavy lines stand
here for the  exact boson ($--$) Green's functions or an iterated two-body potentials:
\be
\parbox{3.5cm}{\includegraphics[width=3.5cm]{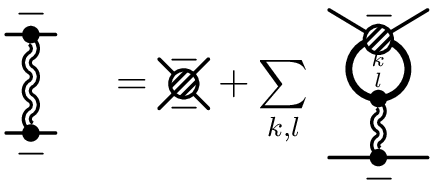}}\,.
\label{wavyline}
\ee
The full dot in the vertex is the renormalized in-medium vertex, which includes all  diagrams with
one sign, i.e. it is irreducible with respect to the full ($\pm\mp$) nucleon Green's
functions. This means that it contains only $(--)$ or $(++)$ Green's functions. We denote such
vertices as $\tau^{-}$ and $\tau^{+}$, e.g.,
\begin{equation}\label{3point}
\parbox{6cm}{\includegraphics[width=6cm]{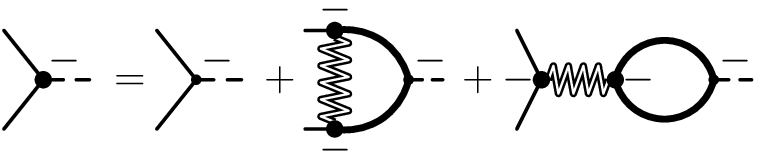}}\,.
\end{equation}
Here we assume that the bare vertex is time-local, i.e. it carries on only one Keldysh index
instead of three indices in general case. Note that full one-sign Green's functions entering Eqs.
(\ref{zigzag},\ref{wavyline},\ref{3point}), would contain alternative sign diagrams, if they were
expanded in perturbative series with respect to the bare Green's functions. To simplify discussion, we do
not include in Eq.~(\ref{3point})  the processes when the weak interaction (dashed line) is
coupled directly to the intermediate pion line due to the $\pi\pi\to \bar{\nu} l$ processes, see
the first diagram in Eq.~(\ref{MMU123}) below.

We did not show the direction of fermion lines in the diagrams, since it can be picked out at will
in closed-loop diagrams. Once some direction is chosen, the arrows in the diagrams
(\ref{zigzag},\ref{wavyline},\ref{3point}) follow.

For a theory of fermions interacting with bosons the contribution with the fewest number of
external particles is three (rather than four as for processes described with the Boltzmann
kinetics). Indeed, the cut through the one-loop diagram in Eq.~(\ref{N=1-diag}) shows that in
dense matter an off-shell fermion can decay into a fermion plus a boson an {\it vice versa}. For
these processes it is important that all particles have a finite width in dense matter. The
formation and decay processes which are forbidden by the energy-momentum conservation in the free
space, can occur in the dense matter without principal restrictions. Therefore the most important
term in the series (\ref{qp}) is the first (one-loop) diagram (\ref{N=1-diag}), which is
positively definite, and corresponds to the first term of the classical Langevin equation, for
details see~\cite{KV95}.

The classical Langevin process deals with the propagation of a single charge (say a proton) in a
neutral medium (e.g in the neutron matter). Therefore for this case only those diagrams occur,
where both photon vertices attach to the same proton line. In the quasiclassical limit for
fermions (with small fermion occupations $n_f\ll 1$;  in case of
 equilibrium matter  $n_f (p_0) =[\exp((p_0 -\mu)/T)+ 1]^{-1}$) all the diagrams
\begin{align}
&-i\Sigma^{-+}_{\mathrm{cl}}=
\parbox{2cm}{\includegraphics[width=2cm]{YaF-oneloop-noarr.eps}}
+
\parbox{2.0cm}
{\includegraphics[height=1.8cm]{YaF-mexhatinsert.eps}}
\nonumber\\
&\quad+
\parbox{2.8cm}
{\includegraphics[height=1.8cm]{YaF-2mexhatinsert.eps}}
+
\parbox{3.5cm}
{\includegraphics[height=1.8cm]{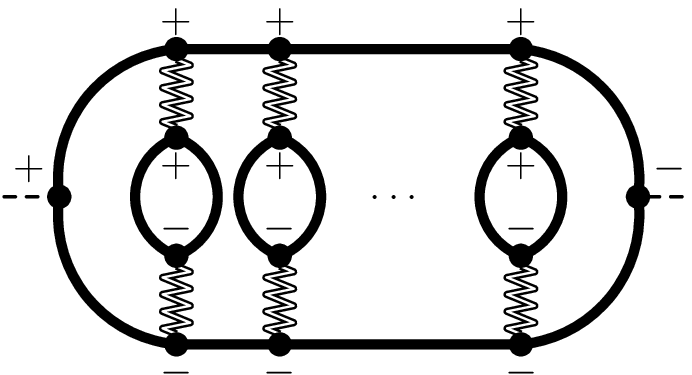}}\,
\label{classicaldiagram}
\end{align}
with arbitrary number of $(-+)$ lines are summed up leading to the diffusion result, see
Ref.~\cite{KV95} for detailed discussion. Each of these diagrams with the $n$  vertical insertions
corresponds to the $n$-th term in the Langevin process, where hard scatterings occur at random
with a constant mean collision rate $\Gamma_f =-2\Im\Sigma^R_f$.

For particle propagation in an external field, e.g., for the scattering on infinitely heavy
centers (proper Landau-Pomeranchuk-Migdal effect), only the one-loop diagram remains, where
the fermion line is given by
\be
\includegraphics[width=4cm]{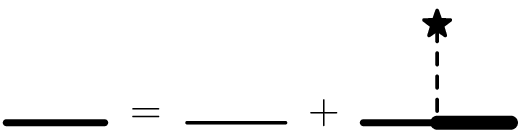}\,,
\ee
since one deals then with a genuine one-body problem.

In the general case of a non-equilibrium system all above  equations, being presented in the
Wigner representation,  are  very cumbersome. To simplify the problem one may use the gradient
expansion in the convolutions of two-point functions, see  Eq.~(\ref{grad-exp}). In general case,
the Wigner transformation will produce an infinite tower of nested gradient terms. Assuming that
a piece of non-equilibrium matter under consideration evolves very slowly in time and space, one may
keep only first-order gradient terms. First-order gradient terms in the expansion of the collision
term $C_{\rm col}=\Sigma^{-+}G^{+-}-\Sigma^{+-}G^{-+}$ are attributed to the memory effects in the
generalized kinetic equation \cite{IKV00}. In the standard derivation of the kinetic Kadanoff-Baym
equation one simplifying usually drops these effects \cite{KB62,Lif81}. As has been shown in
\cite{IKV00} the memory terms are of the same gradient order as other terms in the Kadanoff-Baym
equation and should be kept, because of the local part of the collision term is also of the first
gradient order (since $C_{\rm col}$ being zero in the thermal equilibrium state). However, in the
given paper we are interested  only in the calculation of the production rates in  direct
reactions from a piece of the non-equilibrium matter, which are fully determined by the quantity
$\Sigma^{-+}$. Since $\Sigma^{-+}\neq 0$ in the thermal equilibrium, the gradient corrections to
it are small and can be neglected provided the given non-equilibrium piece of matter  evolves very
slowly in time and space, that we further assume. Therefore, in further we will calculate only  the
local part of the $\Sigma^{-+}$ term.

\subsubsection{Quasiparticle approximation for fermions}\label{subsec:resummation}

The one-loop diagram (\ref{N=1-diag}) calculated  with the quasi-particle fermion propagators determines the
one-nucleon reactions: the DU reactions and the PBF (in case of the superfluid matter)
\cite{VS87,SV87,V01}. The contribution to the DU process vanishes for $n <n_{c}^{\rm DU}$.

The two-nucleon processes are encoded in the $N=2$ term in Eq.~(\ref{qp}) and are given
in the quasi-particle approximation by the diagrams
\begin{align}
&\parbox{25mm}
{\includegraphics[width=2.5cm]{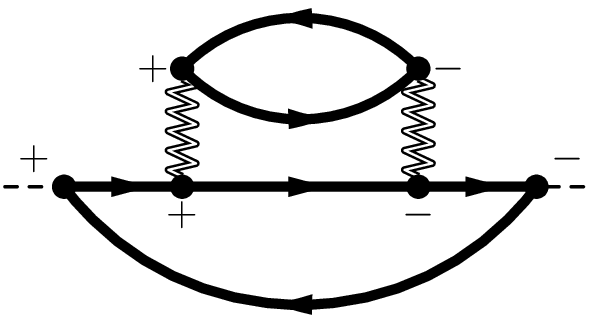}}
+\parbox{2.1cm}
{\includegraphics[height=1.8cm]{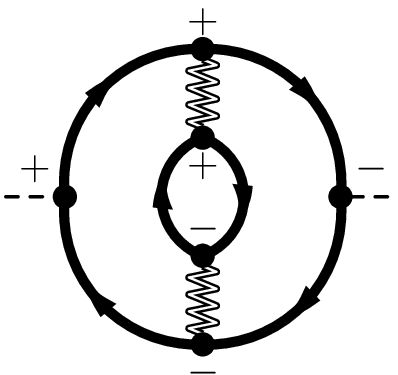}}
\nonumber\\
&+
\parbox{2.4cm}
{\includegraphics[height=1.6cm]{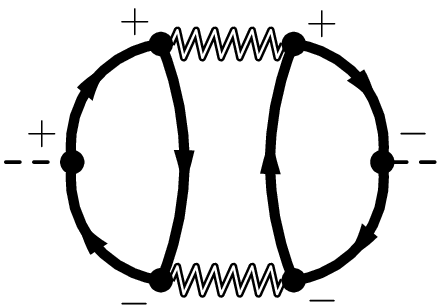}}
+
\parbox{2.4cm}
{\includegraphics[height=1.6cm]{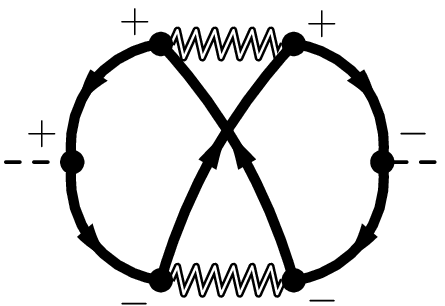}}
\nonumber\\
&+
\parbox{3cm}
{\includegraphics[height=1.6cm]{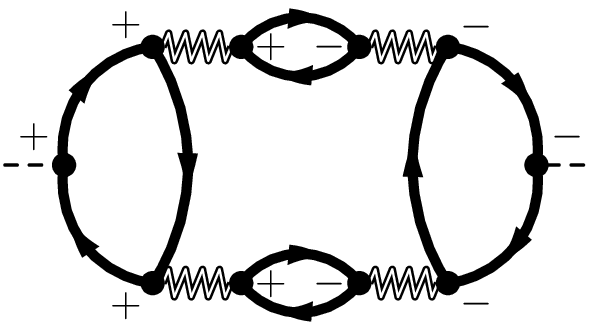}}\,.
\label{two}
\end{align}
Note that the first diagram in (\ref{two}) is not allowed in terms of the full Green's functions
with the width (compare with (\ref{N=2-diag})) but it should be explicitly presented in the
quasi-particle picture, see \cite{VS87,KV95}. After the cut over ($-+$), ($+-$) lines the diagrams
(\ref{two}) are separated by two pieces and correspond to the processes
 \be\label{pairneutr}
 \parbox{15mm}
 {\includegraphics[width=1.5cm]{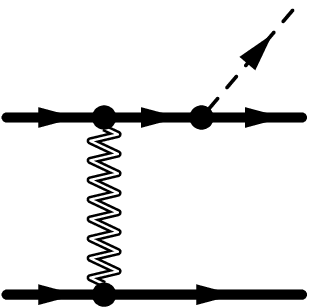}}
 \quad , \quad
 \parbox{15mm}
 {\includegraphics[width=1.5cm]{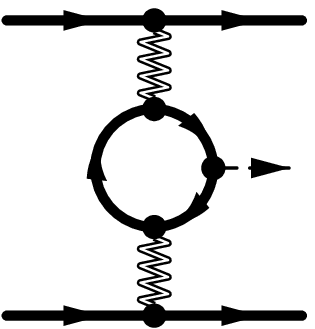}}\,
 \ee
shown here for the paired potential interaction.

In case when the $NN$ interaction amplitude is mainly controlled by the soft pion exchange in the
reaction channel under consideration (for $n\gsim n_0$), instead of (\ref{pairneutr}) one has
\be
\parbox{2cm}{\includegraphics[width=2cm]{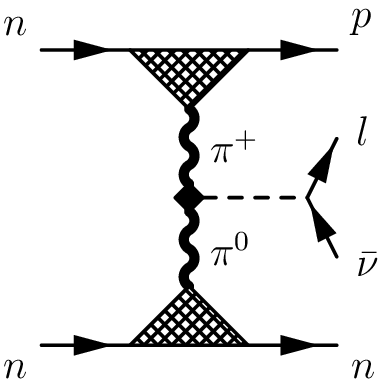}},\,\,
\parbox{2cm}{\includegraphics[width=2cm]{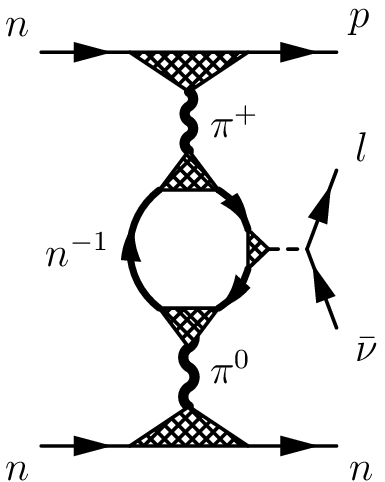}},\,\,
\parbox{2cm}{\includegraphics[width=2cm]{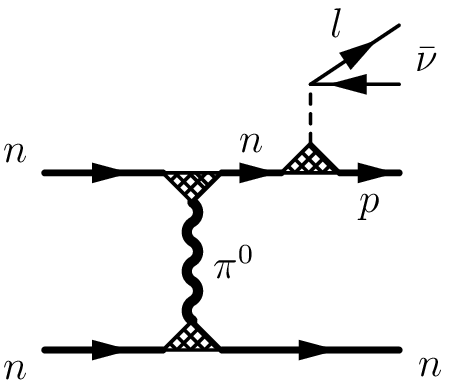}}\,...
\label{MMU123}
\ee
Ref.~\cite{VS86} calculated the rate of MMU and MNB processes taking into account in-medium
effects for the case of non-superfluid nucleon matter. Evaluations \cite{VS86,MSTV90,V01} have
shown that the dominating contribution to MMU rate comes from the first two diagrams of the series
(\ref{MMU123}), whereas the third diagram, which naturally generalizes the corresponding MU (FOPE)
contribution, gives only a small correction for $n \gsim n_0$. As is seen from comparison of Eqs.
(\ref{pairneutr}) and (\ref{MMU123}), the first diagram (\ref{MMU123}) is absent, if one
approximates the nucleon-nucleon interaction  by a two-body potential.

The diagrams that can be cut into more than two pieces (e.g., see the last explicitly presented
diagram in Eq.~(\ref{N=3-diag})) are proportional to powers of independent
$({{L}}^{-+})^{2+n}(L^{+-})^{m}$  loops, $m,n$ are positive integer numbers,  whereas the
diagrams for a two-nucleon process have only two $L^{-+}$ loops, and they decay after the cut into two parts.

In Ref.~\cite{KV95} it was shown how for the processes with the radiation of soft quanta one can
simply incorporate the effects of a finite fermion width into the results calculated within the
quasi-particle approximation for fermions (i.e., the fermion width $\Gamma_{f}=-2\Im\Sigma^R_f$ is
put to zero).  For this purpose it is sufficient to multiply the quasi-particle result by a
pre-factor. For example, comparing one-loop result at a non-zero value of the nucleon width
$\Gamma_f$ with the first non-zero diagram in the quasi-particle approximation (when one puts
$\Gamma_f \rightarrow 0$ in the expressions for the Green's functions) one gets
\begin{align}\label{corr0}
\parbox{1.5cm}{\includegraphics[width=1.5cm]{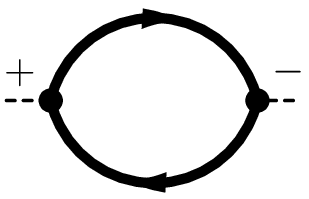}}=
C_0(\omega)
\left\{
\parbox{28mm}{\includegraphics[width=28mm]{YaF-selfinsert-arrows.eps}}
\right\}_{\mbox{QP}}\,. &
\end{align}
QP means here the quasi-particle result.
At a small momentum  $\vec{q}$ of the radiated quantum  the correction factor is equal to
\begin{equation}
\label{C}
C_0(\omega)=\frac{\omega^2}{\omega^2+\Gamma^2_N}\,,
\end{equation}
which removes the singularity of the quasi-particle production rate for small $\omega$. This factor
complies with the replacement $\omega\rightarrow \omega+i\,\Gamma_N$ in the retarded Green's
function. Correction factors for the higher order diagrams also can be derived. Here we  quote
corresponding results for the next lowest order diagrams:
\begin{align}
\parbox{22mm}{\includegraphics[width=22mm]{YaF-mexhatinsert-arrows.eps}}
&=C_1(\omega)\left\{\;
\parbox{22mm}{\includegraphics[width=22mm]{YaF-mexhatinsert-arrows.eps}}
  \right\}_{\mbox{QP}},\label{corr1}\\
\parbox{22mm}{\includegraphics[width=22mm]{YaF-glassesloop-arrows.eps}}
  &=C_0(\omega)
  \left\{
\parbox{22mm}{\includegraphics[width=22mm]{YaF-glassesloop-arrows.eps}}
  \right\}_{\mbox{QP}}\label{corr2}
\end{align}
with $C_0(\omega)$ from (\ref{C}) and
\begin{equation}
\label{C1}
  C_1(\omega)=\omega^2\frac{\omega^2-\Gamma^2_N}{(\omega^2+\Gamma^2_N)^2}\,.
\end{equation}
In case $T \ll \epsilon_{\rm F}$ for typical energy of the radiation $\om \sim T$ one has $\om \gg
\Gamma_N$, since $\Gamma_N \propto T^2/\epsilon_{\rm F}$,
 and thus $C_0 \simeq C_1 \simeq 1$, see ~\cite{KV95}. In general case the full radiation rate is obtained by
summation of all diagrams in the series (\ref{qp}).


\subsection{Resummation of the two-fermion interaction  out of
equlibrium.
Bosonization of the interaction
}\label{subsec:boson}

In this section we work out resummation of the two-fermion interaction amplitude starting from a
bare interaction, which is local in time but not necessarily  local in space. This is the
generalization of the procedure performed in Ref.~\cite{KV95} for the  point-like interaction,
local both in space and in time. We construct the compact expression for $\Sigma^{-+}$ self-energy
in equilibrium and non-equilibrium cases and illustrate how the decomposition with respect to the
number of $(+-)$ and $(-+)$ lines works. In order to simplify the consideration we first study the
normal matter and then perform generalizations to the superfluid matter.

Consider the particle-hole channel with the full two-fermion interaction amplitude determined by
\begin{align}
\parbox{7cm}{
\includegraphics[width=7cm]{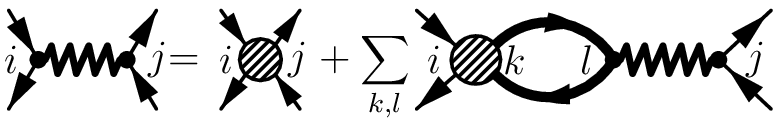}
}
\nonumber\\
\label{diag-calG-iter}
\end{align}
with some particle-hole irreducible bare time-local interaction $\mathcal{G}_0$.
Without the first-order gradient terms included, the diagrams in Eq.~(\ref{diag-calG-iter}) correspond to the
following expression in the Wigner representation
\begin{align}\label{Gcal-eq}
&\mathcal{G}^{ji}(p',p;q)= \mathcal{G}^{ji}_0(p',p)
+\sum_{kl}\intop\frac{\rmd p''^4}{(2\pi)^4\, i} \\
&\times\mathcal{G}^{jl}(p',p'';q)\, G^{lk}(p''+\frac{q}{2})\,
G^{kl}(p''-\frac{q}{2})\,\mathcal{G}^{ki}_0(p'',p;q)\nonumber
\end{align}
where each element is to be understood as depending additionally on the Wigner's $X$ variable.
Since in the local approximation exploited here
this variable will enter  all expressions as a common parameter, we will not write it
explicitly in the expressions below. In Eq.~(\ref{Gcal-eq}), $p$, $p'$ and $p''$ are the relative
momenta in incoming, outgoing and intermediate channels, respectively; $q$ is the exchanged
momentum in the particle-hole channel. Note that in case under consideration the bare interaction
is diagonal in the Schwinger-Keldysh space, i.e
\be
\mathcal{G}_0^{\pm \mp}=0\,,\quad
\mathcal{G}^{--}_0=-\mathcal{G}^{++}_0=\mathcal{G}_0\,.
\label{Gcal0}
\ee
We introduce the products of the Green's functions
\be
\widetilde{L}^{jk}(p;q)=-i\, G^{jk}(p+q/2)\, G^{kj}(p-q/2)\,,
\label{Ldef} \ee
which are related to the bare self-energies as
\be
\Sigma_0^{ij}=\intop\frac{\rmd p^4}{(2\pi)^4}
\tau^{i}_0(p;q)\widetilde{L}^{ij}(p;q)\, \tau_0^{j}(p;q)\,.
\label{Sigma0}
\ee
Since here the bare vertices are assumed to be
time local, they carry only one Keldysh index, $\tau^{ij}_k=\tau^i\,
\delta_{ij}\delta_{ik}$\,. For the vertex independent on the fermion momentum $p$ the self-energy
reads
\be
\Sigma_0^{ij}=\tau_0^2\, L^{ij}(q) \,,\,\,
L^{ij}(q)=\intop\frac{\rmd p^4}{(2\pi)^4}
\widetilde{L}^{ij}(p;q)\,, \label{Linteg}
 \ee
where we introduced the loop-functions $L^{ij}$.

If we formally extend the products (\ref{Ldef}) as
$$ \widetilde{L}^{jk}(p;q)\to \widetilde{L}^{jk}(p,p';q)=
\widetilde{L}^{jk}(p;q)(2\pi)^4\delta^{(4)}(p-p')\,, $$
Eq.~(\ref{diag-calG-iter}) can be presented as
\begin{align}
&\mathcal{G}^{ji}(p',p;q)= \mathcal{G}^{ji}_0(p',p)
+\sum_{kl}\intop\frac{\rmd p''^4}{(2\pi)^4\, i} \intop\frac{\rmd
p'''^4}{(2\pi)^4}
\nonumber\\
&\quad\times\mathcal{G}^{jl}(p',p'';q)
\widetilde{L}^{lk}(p'',p''';q) \mathcal{G}^{ki}_0(p''',p;q)\,.
\label{Gcal-eq-II}
\end{align}
The integral equation~(\ref{Gcal-eq-II}) can be interpreted as a matrix equation in the
discretized momentum space~\cite{RLandau83}. The integration turns into a summation over the grid
points with the appropriate weights. Then in terms of matrices $\mathcal{G}$, $\mathcal{G}_0$ and
$\widetilde{L}$, Eq. (\ref{diag-calG-iter}) takes the form
\be
\mathcal{G}^{ji}= \mathcal{G}_0^{ji} +
\sum_{kl}\mathcal{G}^{jl}\cdot \widetilde{L}^{lk} \cdot
\mathcal{G}_0^{ki}\,,
\label{calG-iter}
\ee
where dot products emphasize the matrix multiplication. For example, the bare self-energy reads in
these notations as  $\Sigma_0^{ij}=\tau_0^i\cdot \widetilde{L}^{ij}\cdot \tau_0^j$. Working with
matrices we can proceed with the solution of Eq.~(\ref{calG-iter}) which we rewrite as
\be
&&\mathcal{G}^{--}=\mathcal{G}^{--}_0
+\mathcal{G}^{--}\cdot\widetilde{L}^{--}\cdot\mathcal{G}^{--}_0
+\mathcal{G}^{-+}\cdot\widetilde{L}^{+-}\cdot\mathcal{G}^{--}_0,
\nonumber
\\
&&\mathcal{G}^{++}=\mathcal{G}^{++}_0
+\mathcal{G}^{++}\cdot\widetilde{L}^{++}\cdot\mathcal{G}^{++}_0
+\mathcal{G}^{+-}\cdot\widetilde{L}^{-+}\cdot\mathcal{G}^{++}_0,
\nonumber
\\
&&\mathcal{G}^{+-}=
 \mathcal{G}^{+-}\cdot\widetilde{L}^{--}\cdot\mathcal{G}^{--}_0
+\mathcal{G}^{++}\cdot\widetilde{L}^{+-}\cdot\mathcal{G}^{--}_0,
\nonumber
\\
&&\mathcal{G}^{-+}=
 \mathcal{G}^{-+}\cdot\widetilde{L}^{++}\cdot\mathcal{G}^{++}_0
+\mathcal{G}^{--}\cdot\widetilde{L}^{-+}\cdot\mathcal{G}^{++}_0.
\label{G-equation}
 \ee
Introducing the quantity, called in~\cite{KV95} the residual
interaction,
\be
\mathcal{G}^{\pm\pm}_{\rm res} &=&  \mathcal{G}^{\pm\pm}_0\cdot \big[1-\widetilde{L}^{\pm\pm}\cdot
\mathcal{G}^{\pm\pm}_0\big]^{-1}
\nonumber\\
&=& \big[1-\mathcal{G}^{\pm\pm}_0\cdot \widetilde{L}^{\pm\pm} \big]^{-1}\cdot \mathcal{G}^{\pm\pm}_0\,,
\label{Gres}
\ee
we  rewrite the above set of equations as
\be
\mathcal{G}^{--}= \mathcal{G}^{--}_{\rm res}
+\mathcal{G}^{-+}\cdot\widetilde{L}^{+-}\cdot\mathcal{G}^{--}_{\rm res},
\nonumber\\
\mathcal{G}^{++}= \mathcal{G}^{++}_{\rm res}
+\mathcal{G}^{+-}\cdot\widetilde{L}^{-+}\cdot\mathcal{G}^{++}_{\rm res},
\nonumber\\
\mathcal{G}^{+-}=
\mathcal{G}^{++}\cdot\widetilde{L}^{+-}\cdot\mathcal{G}^{--}_{\rm res},
\nonumber\\
\mathcal{G}^{-+}=
\mathcal{G}^{--}\cdot\widetilde{L}^{-+}\cdot\mathcal{G}^{++}_{\rm res}.
\nonumber
\ee
Substituting $\mathcal{G}^{+-}$, $\mathcal{G}^{-+}$ from the last two equations into the first two equations
and using the notations
\be
Z_r^{\mp\mp}=\big[1-\widetilde{L}^{\mp\pm}\cdot\mathcal{G}^{\pm\pm}_{\rm res}
\cdot\widetilde{L}^{\pm\mp}\cdot\mathcal{G}^{\mp\mp}_{\rm res} \big]^{-1},
\nonumber\\
Z_l^{\mp\mp}=\big[1-\mathcal{G}^{\mp\mp}_{\rm res}\cdot \widetilde{L}^{\mp\pm}
\cdot\mathcal{G}^{\pm\pm}_{\rm res} \cdot\widetilde{L}^{\pm\mp} \big]^{-1}
\label{ZRL}
\ee
we arrive at the formal solution of Eq.~(\ref{calG-iter}):
\begin{subequations}
\be
\mathcal{G}^{--} &=& \mathcal{G}^{--}_{\rm res}\cdot Z^{--}_r = Z^{--}_l\cdot  \mathcal{G}^{--}_{\rm res},
\label{Gcalmm}\\
\mathcal{G}^{++} &=& \mathcal{G}^{++}_{\rm res}\cdot Z^{++}_r = Z^{++}_l\cdot  \mathcal{G}^{++}_{\rm res},
\label{Gcalpp}\\
\mathcal{G}^{+-} &=& \mathcal{G}^{++}_{\rm res}\cdot Z^{++}_r\cdot\widetilde{L}^{+-}
\cdot\mathcal{G}^{--}_{\rm res}
\nonumber\\
&=& \mathcal{G}^{++}_{\rm res}\cdot\widetilde{L}^{+-}\cdot Z^{--}_l\cdot\mathcal{G}^{--}_{\rm res},
\label{Gcalpm}\\
\mathcal{G}^{-+} &=& \mathcal{G}^{--}_{\rm res}\cdot Z^{--}_r\cdot\widetilde{L}^{-+}
                        \cdot\mathcal{G}^{++}_{\rm res}
\nonumber\\
&=& \mathcal{G}^{--}_{\rm res}\cdot\widetilde{L}^{-+}\cdot Z^{++}_l\cdot\mathcal{G}^{++}_{\rm res}.
\label{Gcalmp}\ee
\label{Gcalsol}
\end{subequations}
This solution for the bosonized interaction  describes propagation of effective boson, such as
phonon, plasmon etc., in non-equilibrium systems. These effective bosons can be treated on the same
footing as all other effective quanta. The phase space distribution of such bosons is given by the
$\mathcal{G}^{+-}$ and $\mathcal{G}^{-+}$ Wigner densities.

 Eq. (\ref{retD}) for the retarded Green's function  decouples from
the other Eqs. (\ref{Gn1}).
Let us demonstrate that the same occurs for the
resummed interaction (\ref{Gcalsol}).
We define the quantity
\be
\widetilde{{L}}^{R}\equiv \widetilde{{L}}^{--} -
\widetilde{{L}}^{-+}\,,
\label{LR}
\ee
which, being integrated over 4-momentum $p$ gives the retarded loop-function
$L^{R}= L^{--} - L^{-+}$ as follows from (\ref{Fretarded1}).
Similarly we define
\be
\widetilde{{L}}^{A}=\widetilde{{L}}^{--} - \widetilde{{L}}^{+-}\,.
\label{LA}
\ee
Using relations (\ref{Lconjugat}) we are able to prove that these quantities are connected by the
relation $\big[\widetilde{{L}}^{R}\big]^\dag=\widetilde{{L}}^{A}$ like the retarded and advanced
Green's functions and self-energies. For the case of an energy-independent bare interaction
$\mathcal{G}_0$, Eqs.~(\ref{completeL},\ref{RR-LL},\ref{Lcompl-int}) imply the useful relation
\be
\mathcal{G}_0\cdot \big(\widetilde{{L}}^{--} - \widetilde{{L}}^{-+}\big)\cdot \mathcal{G}_0
=\mathcal{G}_0\cdot \big(\widetilde{{L}}^{+-} - \widetilde{{L}}^{++}\big)\cdot \mathcal{G}_0\,.
\nonumber\\
\label{LcomplG0}
\ee
With the help of this relation we obtain
from Eqs.~(\ref{G-equation}) and Eq.~(\ref{Gcal0}):
\be
\mathcal{G}^{--}+\mathcal{G}^{-+}= \mathcal{G}_0+(\mathcal{G}^{--}+\mathcal{G}^{-+})\cdot \widetilde{L}^{R}\cdot
\mathcal{G}_0,
\nonumber\\
-\mathcal{G}^{+-} - \mathcal{G}^{++}=\mathcal{G}_0-
(\mathcal{G}^{+-}+\mathcal{G}^{++})\cdot\widetilde{L}^{R}\cdot
\mathcal{G}_0\,,
\nonumber \ee
where we used that
$\big(\mathcal{G}^{--}\cdot \widetilde{L}^{--}
-\mathcal{G}^{-+}\cdot \widetilde{L}^{++}
+\mathcal{G}^{-+}\cdot \widetilde{L}^{+-}
-\mathcal{G}^{--}\cdot \widetilde{L}^{-+}\big)\cdot\mathcal{G}_0=
(\mathcal{G}^{--}+\mathcal{G}^{-+})\cdot \widetilde{L}^{R}\cdot\mathcal{G}_0$
and analogously
$
\big(\mathcal{G}^{++}\cdot \widetilde{L}^{++}+
\mathcal{G}^{+-}\cdot \widetilde{L}^{-+}+
\mathcal{G}^{+-}\cdot \widetilde{L}^{--}+
\mathcal{G}^{++}\cdot \widetilde{L}^{+-}\big)\cdot\mathcal{G}_0=
-(\mathcal{G}^{++}+\mathcal{G}^{+-})\cdot \widetilde{L}^{R}\cdot\mathcal{G}_0
$\,.
Thus, we can introduce the retarded interaction amplitude
\be
\mathcal{G}^R &=&\mathcal{G}^{--}+\mathcal{G}^{-+}=
-\mathcal{G}^{+-} - \mathcal{G}^{++}
\label{GR-def}\\
&=&\mathcal{G}_0\cdot [1-\widetilde{L}^R\cdot \mathcal{G}_0]^{-1}
=[1-\mathcal{G}_0\cdot \widetilde{L}^R]^{-1} \cdot
\mathcal{G}_0\,,
\nonumber \ee
expressed only through the quantity $\widetilde{L}^R$, which
convolution with $\mathcal{G}_0$, like $\mathcal{G}_0\cdot\widetilde{L}^R\cdot\mathcal{G}_0$
possesses the retarded properties.

\subsection{Physical meaning of multi-piece diagrams}

In general case the total radiation rate is obtained by summation of all diagrams in (\ref{qp}).
Some of the diagrams shown, e.g., in the second line in Eq.~(\ref{N=3-diag}) give more than two
pieces, if being cut. Therefore, they do not reduce to the Feynman amplitudes. The role of these diagrams
will be illustrated in the given sub-section.

\subsubsection{Non-equilibrium systems}

Consider the RPA-like set of the self-energy diagrams
\begin{align}
-i\Sigma^{ji}_{\rm RPA} =\parbox{6.2cm}{
\includegraphics[width=6.2cm]{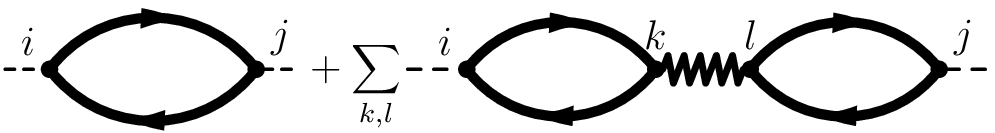}
}. \label{diag:sigji}
\end{align}
Note that this is only a RPA subset of all possible self-energy diagrams. In the Wigner
representation with the omitted gradient terms Eq.~(\ref{diag:sigji}) reads
\be
\Sigma^{ij}_{\rm RPA} =\tau_0^{i}\cdot \widetilde{L}^{ij}\cdot \tau_0^j
+\sum_{kl}\tau_0^i\cdot
\widetilde{L}^{ik}\cdot\mathcal{G}^{kl}\cdot
\widetilde{L}^{lj}\cdot\tau_0^j .
\label{sigij}
\ee
According to Eq.~(\ref{diag:sigji}) the quantity $\Sigma^{-+}_{\rm RPA}$, which determines the
production probability, includes now the following terms
\be\label{diag:sigji1}
\parbox{7.5cm}{\includegraphics[width=7.5cm]{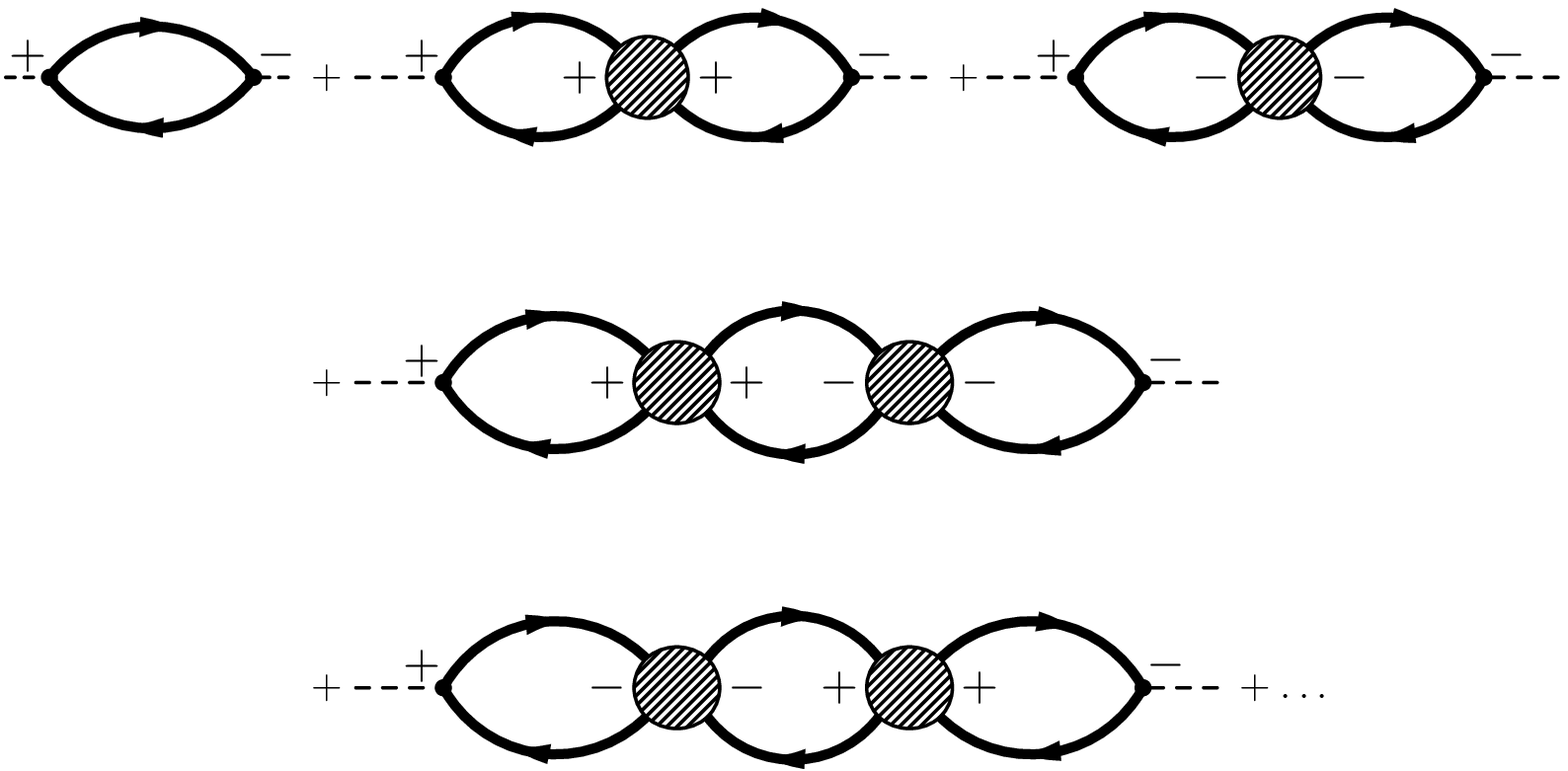}}
\ee
We see that if one  singles out infinite tower of terms with only one $(-+)$ loop, their summation
will lead to the renormalization of left and right vertices in the one-loop diagram, see the $N=1$
term in Eq.~(\ref{N=1-diag}). In addition Eq.~(\ref{sigij}) contains terms with many repeated
$(+-)$ and $(-+)$ loops, i.e. the multi-piece diagrams. However, the RPA series does not include
many other terms with $N\ge 2$, as it is seen from comparison with Eq.~(\ref{N=3-diag}).

From Eq.~(\ref{sigij}) we write now
\be
\Sigma^{-+}_{\rm RPA}  &=&\tau_0^{-}\cdot \widetilde{L}^{-+}\cdot \tau_0^+
\nonumber\\ &+& \tau_0^-\cdot
\widetilde{L}^{--}\cdot\mathcal{G}^{--}\cdot
\widetilde{L}^{-+}\cdot\tau_0^+ \nonumber\\ &+& \tau_0^-\cdot
\widetilde{L}^{-+}\cdot\mathcal{G}^{++}\cdot
\widetilde{L}^{++}\cdot\tau_0^+ \nonumber\\ &+& \tau_0^-\cdot
\widetilde{L}^{-+}\cdot\mathcal{G}^{+-}\cdot
\widetilde{L}^{-+}\cdot\tau_0^+ \nonumber\\ &+& \tau_0^-\cdot
\widetilde{L}^{--}\cdot\mathcal{G}^{-+}\cdot
\widetilde{L}^{++}\cdot\tau_0^+\,. \label{Sigmamp}
 \ee
Using another self-energy
\be
\Sigma^{--}_{\rm RPA}  &=&\tau_0^{-}\cdot \widetilde{L}^{--}\cdot \tau_0^-
\nonumber\\
&+& \tau_0^-\cdot \widetilde{L}^{--}\cdot\mathcal{G}^{--}\cdot
\widetilde{L}^{--}\cdot\tau_0^-
\nonumber\\
&+& \tau_0^-\cdot \widetilde{L}^{-+}\cdot\mathcal{G}^{++}\cdot
\widetilde{L}^{+-}\cdot\tau_0^-
\nonumber\\
&+& \tau_0^-\cdot \widetilde{L}^{-+}\cdot\mathcal{G}^{+-}\cdot
\widetilde{L}^{--}\cdot\tau_0^-
\nonumber\\
&+& \tau_0^-\cdot \widetilde{L}^{--}\cdot\mathcal{G}^{-+}\cdot
\widetilde{L}^{+-}\cdot\tau_0^-\,,
\label{Sigmamm}
\ee
we can determine the retarded combination $\Sigma^R_{\rm RPA}  = \Sigma^{--}_{\rm RPA} +\Sigma^{-+}_{\rm RPA}$.
The direct calculation yields
\begin{align}
&\Sigma^{--}_{\rm RPA} +\Sigma^{-+}_{\rm RPA} = \tau_0^- \cdot \widetilde{L}^R\cdot \tau_0^-
\nonumber\\
&+ \tau_0^- \cdot
\widetilde{L}^R\cdot(\mathcal{G}^{--}+\mathcal{G}^{-+})\cdot \widetilde{L}^R\cdot \tau_0^-
\nonumber\\
&+ \tau_0^- \!\cdot\!
\widetilde{L}^R\!\cdot\!(\mathcal{G}^{--}+\mathcal{G}^{++}+\mathcal{G}^{+-}+\mathcal{G}^{-+})
\!\cdot\!\widetilde{L}^R\cdot \tau_0^-\,.
\nonumber
\end{align}
Taking into account Eq.~(\ref{GR-def}) we obtain
\begin{align}
\Sigma^R_{\rm RPA} &=\Sigma^{--}_{\rm RPA} +\Sigma^{-+}_{\rm RPA}
\nonumber\\
&=\tau_0^- \cdot \widetilde{L}^R\cdot \tau_0^-+
\tau_0^- \cdot \widetilde{L}^R\cdot\mathcal{G}^R\cdot \widetilde{L}^R \tau_0^-
\nonumber\\
&=\tau_0^- \cdot \widetilde{L}^R\cdot[1-\mathcal{G}_0\cdot \widetilde{L}^R]^{-1} \tau_0^-.
\label{SigmaR}
\end{align}
Equations~(\ref{GR-def}) and~(\ref{SigmaR}) express the retarded interaction and the retarded
self-energy through the bare interaction $\mathcal{G}_0$ and the quantity $\widetilde{L}^R$
determined by Eq. (\ref{LR}).

It is possible to present  expression for $\Sigma^{-+}_{\rm RPA}$ in a more
compact form. Using  relations
\begin{subequations}
\begin{align}
\widetilde{L}^{\pm\pm}\cdot\mathcal{G}^{\pm\pm}_{\rm res}
&=\big[1-L^{\pm\pm}\cdot \mathcal{G}^{\pm\pm}_0\big]^{-1}- 1,
\\
\widetilde{L}^{-+}\cdot\mathcal{G}^{++}_{\rm res}&\cdot
Z^{++}_r\cdot \widetilde{L}^{+-}\cdot\mathcal{G}^{--}_{\rm res}
\cdot \widetilde{L}^{-+}\nonumber
\\
&=\widetilde{L}^{-+}\cdot\Big(Z^{++}_l -1\Big),
\\
\widetilde{L}^{-+}\cdot Z^{++}_l &=
Z^{--}_r\cdot\widetilde{L}^{-+},
\end{align}
\end{subequations}
we  rewrite Eq.~(\ref{Sigmamp}) as
\be
\Sigma^{-+}_{\rm RPA}  &=& \tau_{\rm res}^-\cdot Z^{--}_r\cdot\widetilde{L}^{-+}\cdot\tau_{\rm res}^+
\nonumber\\
&=& \tau_{\rm res}^-\cdot \widetilde{L}^{-+}\cdot Z^{++}_l\cdot\tau_{\rm res}^+,
\label{Sigmamp-res}
\ee
where  renormalized vertices
\be
\tau^\pm_{\rm res}&=&\tau^\pm_0\cdot \big[1-\widetilde{L}^{\pm\pm}\cdot \mathcal{G}^{\pm\pm}_0\big]^{-1}
\nonumber\\
&=& \big[1- \mathcal{G}^{\pm\pm}_0\cdot \widetilde{L}^{\pm\pm}\big]^{-1}\cdot \tau_0^\pm
\label{tau-res} \ee
are the solutions of Eq.~(\ref{3point}) (and of similar equation for the $(+)$ vertex)
with the omitted second term on the r.h.s. (witin the RPA
approximation). With the help of Eq.~(\ref{Sigmamp-res}) expressed in terms of
$\widetilde{L}^{ij}$ one can calculate the reaction rates  associated with the processes described
by Eq.~(\ref{diag:sigji}) (in case of non-equilibrium slowly evolving systems with small spatial
gradients).

\subsubsection{Equilibrium systems}

In equilibrium the expressions  derived above can be simplified considerably and expressed through
the real and imaginary parts of the function $\widetilde{L}^R$:
\begin{subequations}
\begin{align}
&\mathcal{G}_0\cdot\widetilde{L}^{-+}\cdot\mathcal{G}_0 =2i n_b(\om)
\Im \big(\mathcal{G}_0\cdot\widetilde{L}^{R}\cdot\mathcal{G}_0\big),
\label{loop-equil:mp}\\
&\mathcal{G}_0\cdot\widetilde{L}^{+-}\cdot\mathcal{G}_0
=2i[1+ n_b(\om)]\Im\big(\mathcal{G}_0\cdot\widetilde{L}^{R}\cdot\mathcal{G}_0\big)\,,
\label{loop-equil:pm}\\
&\mathcal{G}_0\cdot\widetilde{L}^{--}\cdot\mathcal{G}_0
=\phantom{-} \mathcal{G}_0\cdot\widetilde{L}^R\cdot\mathcal{G}_0
\nonumber\\
&\qquad\qquad\qquad
+2 i n_b(\om)\Im \big(\mathcal{G}_0\cdot\widetilde{L}^{R}\cdot\mathcal{G}_0\big)\,,
\label{loop-equil:mm}\\
&\mathcal{G}_0\cdot\widetilde{L}^{++}=-\mathcal{G}_0\cdot\widetilde{L}^R\cdot\mathcal{G}_0
\nonumber\\
&\qquad\qquad\qquad
+2i [n_b(\om) +1] \Im\big(\mathcal{G}_0\cdot\widetilde{L}^{R}\cdot\mathcal{G}_0\big)\,,
\label{loop-equil:pp}
\end{align}
\label{loop-equil}
\end{subequations}
where $n_b(\om) =[e^{\om/T}-1]^{-1}$ are the equilibrium boson occupations. These relations are
derived in Appendix \ref{eq.rel}.

We note that the self-energy (\ref{Sigmamp-res}) can be written as
\be
\Sigma^{-+}_{\rm RPA} &=&\tau_0^-\cdot [\mathcal{G}_0^{--}]^{-1}\cdot\mathcal{G}_{\rm res}^{--}\cdot
\widetilde{L}^{-+}
\nonumber\\
&&\cdot [1-\mathcal{G}_{\rm res}^{++}\cdot \widetilde{L}^{+-} \mathcal{G}_{\rm res}^{--}\cdot
\widetilde{L}^{-+}]^{-1}
\nonumber\\
&&\cdot [1-\mathcal{G}_{\rm res}^{++}\cdot \widetilde{L}^{++}]^{-1}\cdot \tau_0^+\,,
\label{Sigmamp-2}
\ee
where with the help of the equilibrium relations
(\ref{loop-equil}) we express
\begin{align}
\mathcal{G}_{\rm res}^{--}\cdot \widetilde{L}^{-+}\cdot \mathcal{G}_0=&
\frac{2\, i\, n_b}{1+i\, (2\, n_b+1)\, Y}\cdot Y\cdot \mathcal{G}_0\,,
\nonumber\\
\mathcal{G}_{\rm res}^{++}\cdot \widetilde{L}^{+-}\cdot \mathcal{G}_0=&
\frac{-2\, i\, (n_b+1)}{1-i\, (2\, n_b+1)\, Y}\cdot Y\cdot \mathcal{G}_0\,,
\nonumber\\
[1-\mathcal{G}_{\rm res}^{++}\cdot \widetilde{L}^{++}]^{-1}\cdot \mathcal{G}_0
=&[1-i\, (2\, n_b+1)\, Y]^{-1}
\nonumber\\
&\cdot [1-\mathcal{G}_0^{--}\cdot\Re \widetilde{L}^R]^{-1}\cdot \mathcal{G}_0
\label{Mintro}
\end{align}
through the common matrix
$$ Y=[1-\mathcal{G}_0^{--}\cdot \Re\widetilde{L}^R]^{-1}\cdot \mathcal{G}_0^{--}\cdot \Im
\widetilde{L}^R\,.
$$
Since functions of the matrix $Y$ commute with each other, we can simplify Eq.~(\ref{Sigmamp-2}) as
\begin{align}
\Sigma^{-+}_{\rm RPA}&= 2\,i\,n_b\,\tau_0^-\cdot \mathcal{G}^{-1}_0\cdot  Y\cdot [ 1+ Y\cdot  Y] ^{-1}
\nonumber\\
&\cdot [1-\mathcal{G}_0\cdot\Re\widetilde L^R]^{-1}\cdot \tau_0^+
\nonumber\\
&= 2\,i\,n_b\,\tau_0^-\cdot
\mathcal{G}_0^{-1}\cdot\big[ \mathcal{G}_0\cdot\Im \widetilde L^{R}+(1-\mathcal{G}_0\cdot\Re \widetilde L^R)
\nonumber\\
&\cdot [\mathcal{G}_0\cdot\Im \widetilde L^{R}]^{-1}\cdot(1-\mathcal{G}_0\cdot\Re L^R) \big]^{-1}\cdot \tau_0^+\,.
\label{Sigmamp-3}
\end{align}
It is known that in the equilibrium the production rate can be also calculated with the help of
the retarded self-enegy. Note that in this case the expression (\ref{SigmaR}) for $\Sigma^R_{\rm
RPA}$ can be also obtained from the direct summation of the series of diagrams
\be\label{nomult}
\parbox{6cm}{
\includegraphics[width=6cm]{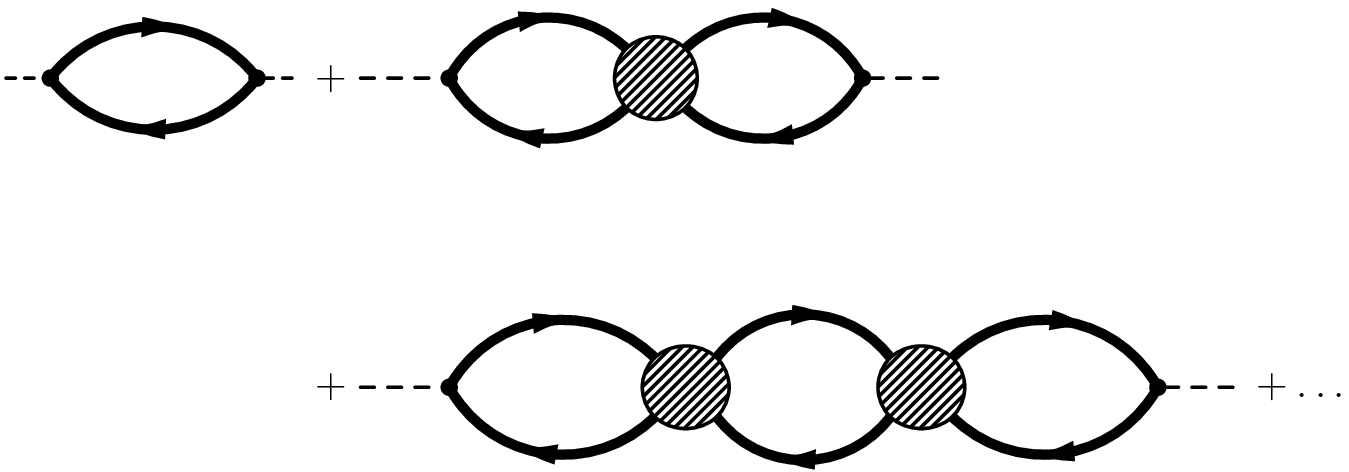}
}\ee
where all quantities are defined for $T\neq0$ within the standard Matsubara technique (using
discrete frequencies  $i\om_n$). After the common replacement $i\om_n\to \om+i0$ one obtains the
analytical continuation to the retarded function. Applying  standard rules to take the real and
imaginary parts of the inverse of a complex matrix\footnote{If for a complex matrix $M=\Re M +i
\Im M$, there exist an inverse matrix $M^{-1}$ then the real and imaginary parts of the inverse
matrix are equal to
\be
&&\Re \big(M^{-1}\big)= -\frac{1}{\Im M} \Re M\, \Im \big(M^{-1}\big)\,,
\label{Reinverse}
\\
&&\Im \big(M^{-1})=\frac{-1}{\Im M+ \Re M\frac{1}{\Im M}\Re M}\,.
\label{Iminverse}
\ee
 }
from Eq.~(\ref{SigmaR}) we obtain
 \begin{align}
\Im\Sigma^R_{\rm RPA}=&\tau_0\cdot \mathcal{G}_0^{-1}\cdot \Im[1-\mathcal{G}_0\cdot \widetilde L^R]^{-1}\cdot \tau_0
\nonumber\\
=&\tau_0\cdot \mathcal{G}_0^{-1} \cdot
\big[\mathcal{G}_0\cdot \Im\widetilde L^R+(1-\mathcal{G}_0\Re\widetilde L^R)
\nonumber\\
&\cdot
[\mathcal{G}_0\Im\widetilde{L}^R]^{-1}\cdot(1-\mathcal{G}_0\Re\widetilde L^R) \big]^{-1}\cdot
\tau_0\,.
\label{ImSigR}
 \end{align}
Comparing Eqs.~(\ref{Sigmamp-3}) and (\ref{ImSigR}) we recover the standard equilibrium relation
\begin{align}
\Sigma^{-+}=-2\,i\,\Im\Sigma^{R}\,n_b\,,
 \label{standr}
\end{align}
see (\ref{Seq}). The sign minus compared to the corresponding relation for $L$, appears here since
the self-energy includes vertices  $\tau_0^-=-\tau_0^+=\tau_0$\,.

Thus, we have shown that in case of equilibrium systems the reaction rates can be found either by using
Matsubara technique and then recovering $\Im\Sigma^R_{\rm RPA}$ given by diagrams (\ref{nomult}), or equivalently
by summing up series of diagrams (\ref{diag:sigji}) within the Schwinger-Keldysh formalism. {\em{We
emphasize that in the latter case not only the $N=1$ term but also corresponding multi-piece
diagrams must be included.}}

Let us now analyze the contribution of genuine one-nucleon processes separated according to the
$G^{\pm\mp}$ counting introduced in Sec.~\ref{subsec:pm-decomp}. They are represented by the only
diagram (\ref{N=1-diag})
\begin{equation}
\label{one-loops}
-i\Sigma^{-+}_{N=1}=\parbox{20mm}{\includegraphics[width=21mm]{YaF-oneloop-pm.eps}}\,.
\end{equation}
It produces the expression
\be
\Sigma^{-+}_{N=1}&=&\tau_{\rm res}^-\cdot \widetilde{L}^{-+}\cdot \tau_{\rm res}^+
=\tau_0^-\cdot [\mathcal{G}_0^{--}]^{-1}\cdot\mathcal{G}_{\rm res}^{--}
\nonumber\\
&&\cdot \widetilde{L}^{-+}
[1-\mathcal{G}_{\rm res}^{++}\cdot \widetilde{L}^{++}]^{-1}\cdot
\tau_0^+\,.
\label{Sigmamp-N=1}
\ee
Making use of Eq.~(\ref{Mintro}) we obtain
\begin{align}
\Sigma^{-+}_{N=1}=&
 2\,i\,n_b\,\tau_0^-\cdot \mathcal{G}_0^{-1}\cdot\big[(2\, n_b+1)^2
\mathcal{G}_0\cdot\Im \widetilde L^{R}
\nonumber\\
&+(1-\mathcal{G}_0\cdot\Re \widetilde L^R)\cdot[\mathcal{G}_0\cdot\Im \widetilde L^{R}]^{-1}
\nonumber\\
&\cdot(1-\mathcal{G}_0\cdot\Re \widetilde L^R)
\big]^{-1}\cdot \tau_0^+\,.
\label{Sigmamp-N=1-2}
\end{align}
We see that in contrast to Eq.~(\ref{Sigmamp-3}) there is an extra factor $(2\, n_b+1)^2$ in the
denominator of Eq.~(\ref{Sigmamp-N=1-2}). Thus the  relation~(\ref{standr}) holds for
$\Sigma^{-+}_{N=1}$ only approximately, e.g. for $n_b\ll 1$.

This example teaches us that multi-piece diagrams may yield a contribution to the total production
rate, beyond that is given by the purely one-nucleon diagram~(\ref{N=1-diag}).

Moreover, as we will show below, only for the rates calculated with Eqs.~(\ref{diag:sigji}),
(\ref{SigmaR}) (and Eqs.~(\ref{Sigmamp-3}), (\ref{ImSigR}), respectively) the condition of the
vector current conservation is exactly fulfilled. It would be fulfilled only approximately
(e.g., for $n_b \ll 1$) or even violated in general case, if the rates were calculated
according to Eq.~(\ref{Sigmamp-N=1}). Bearing  in mind that in many cases it is important to keep
conservation laws on exact level, provided it is possible, we may re-interpret which diagrams
correspond to the one-nucleon, two-nucleon, and other processes: We will ascribe a diagram to the
one-nucleon process, if after the cut through the $(\pm\mp)$ lines it decays into two pieces with
two fermion legs, supplemented by the corresponding multi-piece terms. Diagrams producing two
pieces with four fermion legs each plus the corresponding multi-piece terms, describe two-nucleon
processes, etc.
We stress that
only taking multi-piece diagrams into account one recovers exact conservation laws (like the vector
current conservation) in sub-sets of diagrams responsible to one-nucleon, two-nucleon  processes, etc.

\subsection{Extension to a superfluid system}

In the system with pairing  we have to deal with
the larger number of interaction amplitudes and loop-functions $\widetilde{L}$.
There are altogether 16 quantities corresponding to
the possible direction of the arrows. The
uniform description can be achieved in the so-called "arrow space" introduced by A.J.~Leggett in
Ref.~\cite{Leg65a}, where each element is labeled according to the direction of the arrows attached to
it. We use label 1  for the arrow to the left (``$+$'' in the Leggett's notations) and label 2  for
the arrow to the right (``$-$'' in the Leggett's notations). For instance the particle-hole interaction
$\mathcal{G}_0$ and $\mathcal{G}$ entering Eq.~(\ref{calG-iter}) will bring the indices
$(\mathcal{G}_0)^{22}_{11}$.
The 16 elements of the interaction amplitudes and the loop can be now arranged in the $4\times 4$ matrix
according to the basis
\be\textstyle
\left\{\left(\begin{array}{c}1\\2\end{array}\right),
\left(\begin{array}{c}2\\1\end{array}\right),
\left(\begin{array}{c}1\\1\end{array}\right),
\left(\begin{array}{c}2\\2\end{array}\right) \right\}\,.
\label{AS-basis} \ee
The bare interaction matrix combines the interactions in particle-hole $(\mathcal{G}_0)^{22}_{11}$,
hole-particle $(\mathcal{G}_0)^{11}_{22}$, particle-particle $(\mathcal{G}_0)^{22}_{22}$, and
hole-hole $(\mathcal{G}_0)^{11}_{11}$  channels arranged in the diagonal matrix
\be
(\op{\mathcal{G}}_0)_{\mu\nu} &=& \left[\begin{array}{cccc}
(\mathcal{G}_0)^{11}_{22} & & & \\ & (\mathcal{G}_0)^{22}_{11} & &
\\ &  & (\mathcal{G}_0)^{11}_{11}& \\ &  &  &
(\mathcal{G}_0)^{22}_{22}
\end{array}
\right]_{\mu\nu} \nonumber\\ &=&
\left[\parbox{3cm}{\includegraphics[width=3cm]{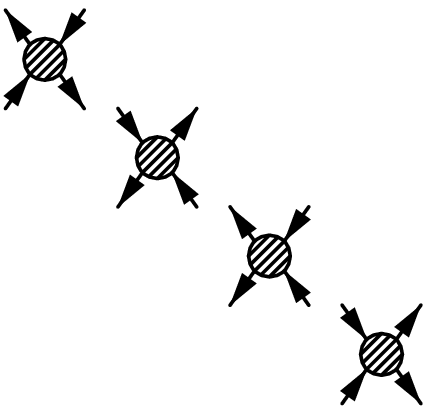}}\right]_{\mu\nu}\,.
\label{G0-AS}
\ee
The indices $\mu$ and $\nu$ run from 1 to 4.
Note that interaction in particle-hole and particle-particle channels must not be the same.
The matrix of the $\widetilde{L}$ functions reads
\be
\op{\widetilde L}_{\mu\nu} &=&
\left[\begin{array}{cccc}
\widetilde L^{11}_{22} & \widetilde L^{12}_{21} & \widetilde L^{11}_{21} & \widetilde L^{12}_{22}\\
\widetilde L^{21}_{12} & \widetilde L^{22}_{11} & \widetilde L^{21}_{11} & \widetilde L^{22}_{12} \\
\widetilde L^{11}_{12} & \widetilde L^{12}_{11} & \widetilde L^{11}_{11} & \widetilde L^{12}_{12} \\
\widetilde L^{21}_{22} & \widetilde L^{22}_{21} & \widetilde L^{21}_{21} & \widetilde L^{22}_{22}
\end{array}
\right]_{\mu\nu}
\nonumber\\
&=& \left[\parbox{4cm}{\includegraphics[width=4cm]{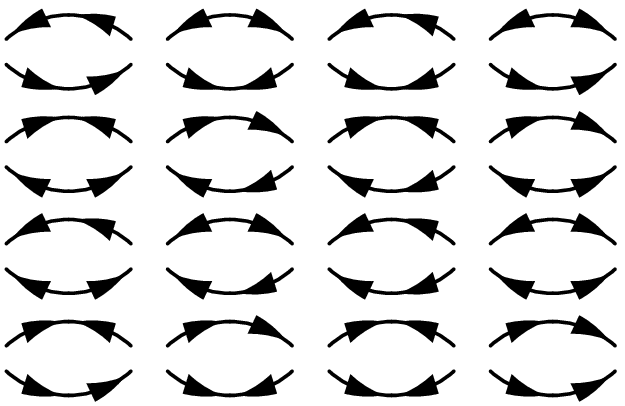}}\right]_{\mu\nu}\,.
\label{L-AS}
\ee
In terms of the matrices $\op{\mathcal{G}}_0$, $\op{\mathcal{G}}_0$ and $\op{\widetilde{L}}$
Eq.~(\ref{calG-iter}) can be written as
\be
\op{\mathcal{G}}^{ji}_{\mu\nu}= \op{\mathcal{G}}^{ji}_{\mu\nu} +
\sum_{\gamma\delta}\sum_{kl}\op{\mathcal{G}}^{jl}_{\mu\gamma}\cdot
\op{\widetilde{L}}^{lk}_{\gamma\delta}
 \cdot (\op{\mathcal{G}}_0)^{ki}_{\delta\nu}
\label{calG-iter-AS} \ee
in double-index notations, where the Latin
indices run in the Schwinger-Keldysh space, $j,i=+,-$ and the Greek indices
run over the basis (\ref{AS-basis}) in the arrow space, $\mu,\nu=1,2,3,4$.
Since all above derivations were performed for the matrix object, the generalization of the results
(\ref{Gcalsol}) and (\ref{GR-def}) is now given  in terms of the nested matrices (\ref{G0-AS}) and (\ref{L-AS}).

The equation for the self-energy (\ref{sigij}) takes  the following form
\be
\Sigma^{ij}=\op{\tau}_0^{\dag i}\cdot \op{\widetilde{L}}^{ij}\cdot \op{\tau}_0^j
+\sum_{kl}\op{\tau}_0^{\dag i}\cdot
\op{\widetilde{L}}^{ik}\cdot\op{\mathcal{G}}^{kl}\cdot
\op{\widetilde{L}}^{lj}\cdot\op{\tau}_0^j ,
\nonumber\\
\label{sigij-AS}
\ee
where  $\op{\tau}_0^j$ is the array in the arrow space
$\op{\tau}_0^j=(\tau_0^{h,j},\tau_0^j,0,0)$\,.
The result (\ref{Sigmamp-res}) will be expressed now through the renormalized vertex
$\op{\tau}_{\rm res}^{j}=(\tau^{h,j},\tau^j,\tau^{(1)j},\tau^{(2)j})$\,, obeying the equation
\be
\op{\tau}_{\rm res}^\pm\cdot
 \big[1- \op{\mathcal{G}}^{\pm\pm}_0\cdot
 \op{\widetilde{L}}^{\pm\pm}\big]^{-1}=\op{\tau}_0^\pm\,,
\ee
which has the diagramatic representation as in Eq.~(\ref{dr-vert})\,.

\subsection{Application to the point-like interactions}

To present the matrix relations derived above in a more transparent form, we  consider a model
case of a momentum independent bare interaction, $\mathcal{G}^{ij}_0$, and bare vertex,
$\tau^j_0$. The matrix Eqs. (\ref{Gcalsol}) turn into the algebraic ones with the replacement
$\mathcal{G}_0\cdot \widetilde{L}^{ij}\cdot\mathcal{G}_0\to \mathcal{G}_0^2\,L^{ij}$\,. The
expressions derived in this section become very compact in this case.

\subsubsection{Non-equilibrium systems}

The  series of diagrams (\ref{nomult}) for the $\Sigma^{R}$ is easily summed
up to the known result
\begin{align}
\Sigma^R_{\rm RPA} &=\frac{\Sigma_{0}^R}{1-\widetilde{\mathcal{G}}_0 \Sigma_0^R},
\label{F.l1}\\
\Im\Sigma^{R}_{\rm RPA} &=\frac{ \Im \Sigma_{0}^{R}} {\big(1-\widetilde{\mathcal{G}}_0\Re \Sigma_0^{R}\big)^2
+\big(\widetilde{\mathcal{G}}_0\Im \Sigma_0^{R}\big)^2}.
\label{F.l}
\end{align}
In order to simplify subsequent diagrammatic representation we
introduced the new object
$\widetilde{\mathcal{G}}=\mathcal{G}/\tau_0^2$.

On the other hand, from (\ref{Gcalsol}) we find
\begin{eqnarray}
\widetilde{{\mathcal G}}^{-+}
=\frac{-\widetilde{\mathcal{G}}_0^2\Sigma_0^{-+}}
{
[1-\widetilde{\mathcal{G}}_0\Sigma_0^{--}][1+\widetilde{\mathcal{G}}_0 \Sigma_0^{++}]
+\widetilde{\mathcal{G}}_0^2 \Sigma_0^{+-}\Sigma_0^{-+}
}.
\nonumber\\
\label{renormGmp}
\end{eqnarray}

It is convenient to re-group diagrams for $\Sigma^{-+}_{\rm RPA}$  in Eq. (\ref {diag:sigji}) as
\begin{align}
&-i\Sigma^{-+}_{\rm RPA}=
\label{diag:sigjimp}\\
&\parbox{6cm}{\includegraphics[width=6cm]{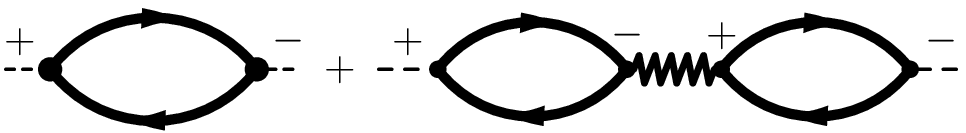} }
\nonumber
\end{align}
Thus, $\Sigma^{-+}$ satisfies the  equation
\begin{align}
\Sigma^{-+}_{\rm RPA}=
\frac{\Sigma_{0}^{-+}}{[1-\widetilde{\mathcal{G}}_0^{--} \Sigma_0^{--}]
[1-\widetilde{\mathcal{G}}_0^{++}\Sigma_0^{++}]}
+ \Sigma_0^{-+}\widetilde{\mathcal{G}}^{+-}\Sigma_0^{-+}.
\label{smp}
\end{align}
Using  (\ref{Gcalsol}) we find
\begin{align}
\Sigma^{-+}_{\rm RPA}=\frac{\Sigma_{0}^{-+}}{
[1-\widetilde{\mathcal{G}}_0\Sigma_0^{--}][1+\widetilde{\mathcal{G}}_0\Sigma_0^{++}]
+\widetilde{\mathcal{G}}_0^2\Sigma_0^{+-}\Sigma_0^{-+}}.
\label{renormSmp}
\end{align}
\begin{widetext}
Similarly, for $\Sigma^{--}_{\rm RPA}$ diagrams in Eq. (\ref {diag:sigji})
can be re-grouped as
\be
-i\Sigma^{--}_{\rm RPA}=
\label{diag:sigjimm}
\parbox{10cm}{\includegraphics[width=10cm]{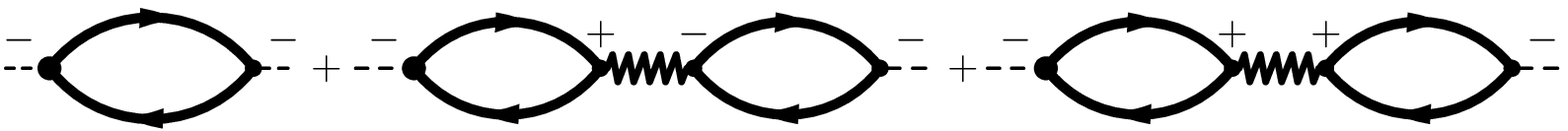} }\,.
\nonumber
\ee
Thus
\begin{align}
\Sigma^{--}_{\rm RPA}&=
\frac{\Sigma_{0}^{--}}{1-\widetilde{\mathcal{G}}^{--}_{0}{\Sigma_0}^{--}}
+\frac{\Sigma_0^{--}}{1-\widetilde{\mathcal{G}}^{--}_{0} {\Sigma_0}^{--}}
\widetilde{\mathcal{G}}^{-+}\Sigma_0^{+-}
+\frac{\Sigma_0^{-+}}{1-\widetilde{\mathcal{G}}^{--}_{0}{\Sigma_0}^{--}}
\widetilde{\mathcal{G}}^{++}\Sigma_0^{+-}
\nonumber\\
&=\frac{\Sigma_{0}^{--}+\widetilde{\mathcal{G}}_0(\Sigma_0^{--}\Sigma_0^{++}-\Sigma_0^{-+}\Sigma_0^{+-})}
{[1-\widetilde{\mathcal{G}}_0\Sigma_0^{--}] [1+\widetilde{\mathcal{G}}_0\Sigma_0^{++}]
+ \widetilde{\mathcal{G}}_0^2\Sigma_0^{+-}\Sigma_0^{-+}}.
\label{smm}
\end{align}
\end{widetext}
Obviously  $\Sigma^{++}_{\rm RPA}$ is found from (\ref{smm}) after the simultaneous replacements
$(--)\leftrightarrow (++)$ and $(-+)\leftrightarrow (+-)$. Also $\Sigma^{--}_{\rm RPA}$
(\ref{smm}) and $\Sigma^{++}_{\rm RPA}$ can be recovered from (\ref{smp}) and  (\ref{F.l1})  with
the help of the relations (\ref{Fretarded}).

\subsubsection{Equilibrium systems}

Let us now consider an equilibrium system. Using the equilibrium relations between $\Sigma_0^{ij}$
and the retarded loop $\Sigma_0^R$ self-energies:
\begin{eqnarray}
&&
\Sigma_0^{-+}=2i\Im \Sigma_0^{R}n_b,\,\,
\Sigma_0^{+-}=2i\Im \Sigma_0^{R}(1+ n_b),
\nonumber\\
&&\Sigma_0^{--}=\Re \Sigma_0^R +i\Im \Sigma_0^{R}(2n_b +1),
\nonumber\\
&&\Sigma_0^{++}=-\Re \Sigma_0^R +i\Im \Sigma_0^{R}(2n_b +1)
 \label{loopr}
\end{eqnarray}
and comparing  (\ref{F.l}) and (\ref{renormSmp}) we see that they
satisfy exactly the standard equilibrium self-consistency relation (\ref{standr}).
Adding (\ref{renormSmp}) and (\ref{smm}) we recover (\ref{F.l1}).

Now consider $N=1$ term
\be
\Sigma^{-+}_{N=1}=
\frac{\Sigma_{0}^{-+}}{[1-\widetilde{\mathcal{G}}_0 \Sigma_0^{--}][1+\widetilde{\mathcal{G}}_0 \Sigma_0^{++}]}
,
\label{smpone}\ee
where we took into account opposite signs of the bare interaction with the $+$ and $-$ signs.
Using the equilibrium relations (\ref{loopr}) we obtain
\begin{align}
&\Sigma^{-+}_{N=1}  \nonumber\\
&\,\,
=\frac{ -2i\Im{\Sigma_{0}}^{R} n_b}
{
\big(1-\widetilde{\mathcal{G}}_0\Re\Sigma_0^{R}\big)^2
+ \widetilde{\mathcal{G}}_0^2(\Im \Sigma_0^{R})^2-(\widetilde{\mathcal{G}}_0)^2 \Sigma_0^{-+}\Sigma_0^{+-}
}
\nonumber\\
&\,\,
=\frac{ -2i\Im {\Sigma_{0}}^{R} n_b}
{
\big(1-\widetilde{\mathcal{G}}_0\Re \Sigma_0^{R}\big)^2 +\big[ (2n_b+1)\widetilde{\mathcal{G}}_0\Im
\Sigma_0^{R}\big]^2
}. \label{sigm1l}
\end{align}
Replacing (\ref{F.l}) and  (\ref{sigm1l}) in  (\ref{standr}) we see that the latter condition may
hold for $N=1$ term only approximately, e.g. for $n_b \ll 1$. Also $\Sigma_{N=1}^{-+}\simeq
\Sigma^{-+}_{\rm RPA}$ provided $| (2n_b+1)\widetilde{\mathcal{G}}_0\Im \Sigma_0^{R}|\ll
|1-\widetilde{\mathcal{G}}_0\Re \Sigma_0^{R}|$ that holds for $T\ll \epsilon_{\rm F}$ in most
interesting for us cases. However as we have stressed, it might be of principal importance to keep
exact conservation laws, like the condition of the conservation of the vector current. As we show
below, the latter condition is exactly satisfied provided one uses $\Sigma_{\rm RPA}$ rather than
$\Sigma_{N=1}$.


\section{Renormalization of interaction}\label{sec:Fermi-liquid}

The results of the previous section show that the interaction $\mathcal{G}_0$ enters the resulting
expression for the resummed interaction $\mathcal{G}^{ij}$ [Eq.~(\ref{Gcalsol})] and the
self-energy $\Sigma^{-+}$ [Eq.~(\ref{Sigmamp-res})] together with the $\widetilde{L}^{--}$ or
$\widetilde{L}^{++}$ functions in specific combinations given by the residual interaction
$\mathcal{G}^{\pm\pm}_{\rm res}$ [Eq.~(\ref{Gres})] and the residual vertex $\tau_{\rm res}^{\pm}$
[Eq.~(\ref{tau-res})]. The structure of Eqs.~(\ref{Gres},\ref{tau-res}) suggests that if one
is able to split an {\it a priori} complicated function $\widetilde{L}^{ii}$ into some part, which we
call ``known'', $\widetilde{L}^{ii}_{\rm known}$, and a reminder $\widetilde{L}^{ii}_{\rm rem}=
\widetilde{L}^{ii}-\widetilde{L}^{ii}_{\rm known} $, then the quantities $\mathcal{G}^{ii}_{\rm
res}$ and $\tau_{\rm res}^{i}$  can be expressed through the ``known'' part
$\widetilde{L}^{ii}_{\rm known}$,
\be
\mathcal{G}^{\pm\pm}_{\rm res} &=&\big[1-\mathcal{G}_{\rm ren}^{\pm\pm}\cdot \widetilde{L}^{\pm\pm}_{\rm known}\big]^{-1}
\cdot\mathcal{G}_{\rm ren}^{\pm\pm}\,,
\nonumber\\
\tau_{\rm res}^{\pm} &=& \big[1-\mathcal{G}_{\rm ren}^{\pm\pm}\cdot \widetilde{L}^{\pm\pm}_{\rm known}\big]^{-1}
\cdot\tau_{\rm ren}^\pm\,,
\label{renorm-res}
\ee
and the renormalized interaction and vertex
\be
\mathcal{G}_{\rm ren}^{\pm\pm}&=&
\big[1- \mathcal{G}_{0}^{\pm\pm}\cdot\widetilde{L}^{\pm\pm}_{\rm rem}\big]^{-1}\cdot \mathcal{G}_{0}^{\pm\pm},
\nonumber\\
\tau_{\rm ren}^\pm &=&
\big[1-\mathcal{G}_{0}^{\pm\pm}\cdot\widetilde{L}^{\pm\pm}_{\rm rem}\big]^{-1}\cdot\tau_0^\pm\,.
\label{renorm-Gtau}
\ee
By a cunning choice of the ``known'' part one can account in $\widetilde{L}^{ii}_{\rm known}$
for the most rapid variations with the energy and momentum in the interval of interest.
Then the renormalized quantities (\ref{renorm-Gtau}) will possess a weak energy-momentum dependence
and can be cast in terms of phenomenological parameters adjusted to some empirical data.
For nuclear physics such a renormalization program was conducted by A.B.~Migdal in his seminal
paper~\cite{M67a}.

\subsection{Fermi-liquid renormalization}

Simplifying consideration we focus below on the description of equilibrium
systems. Then we may deal with only one, e.g. retarded Green's
function. The other Green's functions $G^{--}_n$, $G^{-+}_n$, $G^{+-}_n$, and $G^{++}_n$ are expressed
in equilibrium through the retarded Green's function, see Appendix~\ref{eq.rel}.

At low temperatures of our interest ($T \lsim \Delta \ll \varepsilon_{{\rm F},n},
\varepsilon_{\rmF,p}$) neutrons and protons are only slightly excited above their Fermi seas and
all the processes occur in a narrow vicinity of $\varepsilon_{\rmF, n}$ and
$\varepsilon_{\rmF,p}$. In such a situation the Fermi-liquid approach seems to be the most
efficient one. The basic assumption of the Fermi-liquid renormalization is that in a fermion
system there is some mechanism of single-particle excitations. For normal systems at $T=0$ this
manifests itself in a jump in the particle momentum distribution\footnote{There is a special class
of Fermi systems in which the jump is absent even in the normal state. They are called singular
Fermi liquids or Non-Fermi liqiuds~\cite{SFL}.}. According to the A.B.~Migdal's paper~\cite{Mjump}
this jump indicates the presence of a pole in the fermion Green's function. Thus, the full
retarded Green's function in the momentum representation is given by a sum of a pole term and a
regular part $G_{\rm reg}$,
\be
G^{R}_n(\epsilon,\vec{p}\,)=
\frac{a}{\epsilon-\epsilon_p+i\,\gamma\, \epsilon^2\, }
+ G^{R}_{\rm reg}(\epsilon,\vec{p}\,)\,.
\label{Gn-FL}
\ee
Here and below the energy $\epsilon=p_0-\epsilon_\rmF$ is counted from the Fermi energy and the
kinetic energy is
$
\epsilon_p =\frac{p^2 - p_{\rmF,N}^2 }{2\,m_N^*}
$\,.
The residue  of the pole term, $a$, quantifies the A.B. Migdal's jump in the momentum distribution
\be
&&
a^{-1}=1-\left(\frac{\partial \Sigma_n^R (\epsilon,\vec{p})}{\partial \epsilon}\right)_{0,p_{\rmF,N}}\,,
\label{resid-a}
\ee
where one can put $T=0$ which is correct up to higher order contribution $O(T^2/\epsilon^2_{\rmF,N})$.

The in-medium mass of the fermion is given by
\be
\frac{1}{m^{*}_{N}}=a\, \left(\frac{1}{m_N} + 2\, \frac{\partial \Sigma_n^R (\epsilon,\vec{p})}{\partial p^2}
\right)_{0, p_{\rmF,N}}\,,
\label{eff-mass}
\ee
and the pole width is because of the coupling to the two-particle--hole
mode,
\be
\gamma =- \Im\Sigma_n^R (\epsilon, p_{\rmF,N})/\epsilon^2 \simeq const,
\label{gamm-def}
\ee
for $\epsilon\ll \epsilon_{\rmF, N}$.
At finite temperature we have to replace $\epsilon_{\rmF, N}\to
\mu_N\simeq \epsilon_{{\rm F},N}+O(T^2/\epsilon_{{\rm F},N}^2)$ and $p_{\rmF,N}\to
\sqrt{2\, m_N\, \mu_N}$ but these corrections are small for $T\lsim \Delta$ and can be neglected.

The typical averaged outgoing neutrino energy ($\om \sim T$) is larger then the nucleon particle width $\Gamma_N
(\epsilon\sim T)\sim \gamma\,T^2 \sim T^2/\varepsilon_{{\rm F},N}$. Therefore, one can neglect the
width in the pole term in Eq.~(\ref{Gn-FL}) and work within the quasi-particle Green's function
\be
G_{n,{\rm q.p.}}^{R}(p)=G_{n,{\rm q.p.}}^{R}(\epsilon,\vec{p}\,)=\frac{a}{\epsilon-\epsilon_p+i\, 0\,}.
\label{Gn-QPR}
\ee
Following \cite{M67} only the pole part of $G^{R}_n$ is relevant for descriptions of processes
happening in a weakly excited Fermi system. The regular part can be absorbed by the
renormalization of the particle-particle and particle-hole interactions on the Fermi surface.

In the momentum representation the particle-particle interaction (\ref{Vdef}) depends on spins and
momenta of incoming ($a,\,q/2-p$; $b,\, q/2+p$) and outgoing ($c,\,q/2-p'$; $d,\, q/2+p'$)
particles $\big[\widehat{{V}}^{--}({\textstyle \frac{q}{2}}+p,{\textstyle \frac{q}{2} }-p;
{\textstyle \frac{q}{2}}+p',{\textstyle \frac{q}{2}}-p')\big]_{cd,ab} =
\big[\widehat{V}^{--}(p,p',q)\big]_{cd,ab}\,,
$
where we introduce the total momentum of two particles $q$ and relative momenta in incoming and
outgoing channels, $p$ and $p'$.
The particle-hole interaction (\ref{Udef}) depends on spins and
momenta of incoming ($b,\,p+q/2$; $a,\, p-q/2$) and outgoing ($d,\,p'+q/2$; $c,\, p'-q/2$)
particles and holes, respectively,
$
\big[\widehat{{U}}^{--}({\textstyle p+\frac{q}{2}},p-{\textstyle
\frac{q}{2} }; p'+{\textstyle \frac{q}{2}},p'-{\textstyle
\frac{q}{2}})\big]_{dc,ab} =
\big[\widehat{{U}}^{--}(p,p',q)\big]_{dc,ab}\,.
$
Graphical Eqs. (\ref{Tpp-diag}), (\ref{Tph-diag}) can be written  for causal
functions:
\begin{align}
&\widehat{T}^{--}_{\rm pp}(p,p';q)=\widehat{V}^{--}(p,p';q)+ \intop\frac{\rmd^4 p''}{(2\pi)^4\, i}
\nonumber\\
&\times\widehat{V}^{--}(p,p'';q)\,\widehat{G}^{--}_n(p''_+)\,\widehat{G}^{--}_n(-p''_-)\, \widehat{T}^{--}_{\rm pp}(p'',p';q),
\label{Tpp-p}\\
&\widehat{T}^{--}_{\rm ph}(p,p';q)=\widehat{U}^{--}(p,p';q)+ \intop\frac{\rmd^4 p''}{(2\pi)^4\, i}
\nonumber\\
&\times\widehat{U}^{--}(p,p'';q) \,\widehat{G}^{--}_n(p''_+)\,\widehat{G}^{--}_n(p''_-)\,
\widehat{T}^{--}_{\rm ph}(p'',p';q)\,,
\label{Tph-p}
\end{align}
where $p''_\pm=(\epsilon\pm\om/2,\vec{p}\,''\pm \vec{q}/2)$. The integration over $p''$ involves energies far off the Fermi
surface. One may renormalize interactions $\widehat{V}$ in (\ref{Tpp-p}) and $\widehat{U}$ in
(\ref{Tph-p}) such that integration will go over the region near the Fermi surface and one may use
simple quasi-particle expressions for the Green's functions. Integral over internal momenta can be
written now as
\be
\int \frac{\rmd^4 p}{(2\pi)^4\,i}\big(\dots\big) &\simeq&
\intop \rmd\Phi_0 \langle \dots\rangle_{\vec{n}}\,,
\nonumber
\ee
where for $T=0$ the energy momentum integral is as follows
\be
\intop \rmd\Phi_0 &=& \rho
\intop_{-\infty}^{+\infty} \frac{\rmd \epsilon}{2\,\pi\,i}
\intop_{-\infty}^{+\infty}\rmd \epsilon_p\,,\quad
\ee
and the brackets stand for averaging over the momentum direction $\vec{n}=\vec{p}/|\vec{p}\,|$
\be
\langle\dots\rangle_{\vec{n}} &=& \intop\frac{\rmd \Omega_{\vec n}}{4\,\pi} \big(\dots\big)
\label{angl-aver}
\ee
with $\rho=\frac{m_N^*\,p_{\rmF,N}}{\pi^2}$ being the density of states at the Fermi surface. If the
interaction $\widehat{V}$ is not singular at $q\to 0$, then for small $q$ and $|\vec{p\,}|\simeq
p_{\rm F,N}\simeq |\vec{p\,}{'}|$ the amplitude $\widehat{T}_{pp}$ is  a function of
$\cos\theta_{pp'}= (\vec{n}\vec{n}\,')$,
 $\vec{n}{'}=\vec{p\,}'/|\vec{p\,}'|$,
and Eq.~(\ref{Tpp-p}) can be written as
\begin{align}
&\widehat{T}^{\small--}_{\rm pp}(\vec{n}\,',\vec{n}) =
\widehat{\Gamma}^{\xi}(\vec{n}\,',\vec{n})
\nonumber\\
&\qquad + \mathcal{A}_{\rm pp}\,
\big\langle \widehat{\Gamma}^{\xi}(\vec{n}\,',\vec{n}\,'')\,\widehat{T}^{--}_{\rm pp}(\vec{n}\,'',\vec{n})\big\rangle_{\vec{n}\,''}
.
\label{Tpp-FL}
\end{align}
The information about the bare interaction and the structure of Green's functions in regions far
from the Fermi surface is encoded in the effective interactions
\begin{multline}
\widehat{\Gamma}^{\xi}(\vec{n}\,',\vec{n}) =
\widehat{V}^{--}(\vec{p}\,'_\rmF,\vec{p}_\rmF)+
\intop\frac{\rmd^4 p''}{(2\pi)^4\, i}\theta(\epsilon_{p''}-\xi)
\nonumber\\
\times\widehat{V}^{--}(\vec{p}\,'_\rmF,\vec{p}\,''_\rmF)\widehat{G}^{--}_n(p'')\,\widehat{G}^{--}_n(-p'')
\widehat{\Gamma}^{\xi}(\vec{n}\,'',\vec{n}).
\nonumber
\end{multline}
The quantity $\xi$ ($\Delta\ll \xi\ll \epsilon_{\rm F}$) does not appear
in final expressions, except for the equation expressing the gap in terms
of the phenomenological Landau-Migdal parameter, see Eq. (\ref{gapzero}) below.
The $\mathcal{A}_{\rm pp}$ function in the quasi-particle approximation for the Green's
function (\ref{Gn-FL}),
\be
G_{n,{\rm q.p.}}^{--}(p)=\frac{a}{\epsilon-\epsilon_p+i\, 0\, {\rm sign}
\epsilon}\,,
\label{Gn-QP}
\ee
renders
\be
\mathcal{A}_{\rm pp}= \intop \rmd \Phi_0
\,{G}^{--}_{n,{\rm q.p.}}(+p)\,{G}^{--}_{n,{\rm q.p.}}(-p)\, \theta(\xi-\epsilon_p)\,.
\nonumber
\ee
The similar renormalization can be performed with Eq.~(\ref{Tph-p}). Here we have to note that
even if $\widehat{U}$ changes weakly for small $q$, the particle-hole scattering amplitude is a
sharp function of $q$ because poles in ${G}^{--}_n(p_+)\,{G}^{--}_n(p_-)$ approach each other
producing $\delta(\epsilon)$:
\begin{multline}
{G}^{--}_n(p_+)\,{G}^{--}_n(p_-)
\nonumber\\
\simeq \delta(\epsilon)\intop
\rmd \epsilon\, {G}^{--}_{n,{\rm q.p.}}(p_+)\,{G}^{--}_{n,{\rm q.p.}}(p_-) + \mathcal{B}(p,q),
\nonumber
\end{multline}
and $\mathcal{B}(p,q)$ is a smooth function of $p$ and  $q$. Then,
Eq.~(\ref{Tph-p}) can be rewritten near the Fermi surface, $|\vec{p\,}|\simeq p_{\rmF,N}\simeq
|\vec{p\,}'|$, as
\be
&&\widehat{T}^{--}_{\rm ph}(\vec{n}\,',\vec{n};q) =
\widehat{\Gamma}^{\om}(\vec{n}\,',\vec{n})
\label{Tph-FL}\\
&&\quad+ \langle \widehat{\Gamma}^{\om}(\vec{n}\,',\vec{n}'')
\mathcal{A}^{--}_{\rm ph}(\vec{n}\,'';q)\, \widehat{T}^{--}_{\rm ph}(\vec{n}\,'',\vec{n};q)
\rangle_{\vec{n}\,''}\,.
\nonumber
\ee
The renormalized interaction is determined by the equation
\begin{multline}
\widehat{\Gamma}^{\om}(\vec{n}\,',\vec{n})=
\widehat{U}^{--}(\vec{p}\,'_\rmF,\vec{p}_\rmF)
+ \intop\frac{\rmd^4 p''}{(2\pi)^4\, i}
\nonumber\\
\times\widehat{U}^{--}(\vec{p}\,'_\rmF,\vec{p}\,''_\rmF) \,\mathcal{B}(p_\rmF,q=0)\,
\widehat{\Gamma}^{\om}(\vec{n}\,'',\vec{n})\,,
\end{multline}
where we assume
$|\vec{q}\, \vec{p}\,|/m_N^*<<\om<<\mu$\, and
\begin{align}
\mathcal{B}(p_\rmF,q=0)=&\lim_{\om\to 0}\, \lim_{\vec{q}\to 0}\left[
{G}^{--}_n(p_{\rmF +})\,{G}^{--}_n(p_{\rmF -})\right]\big|,
\nonumber\\
&p_{\rmF \pm}=(\pm\om/2,p_\rmF\,\vec{n}\pm\vec{q}/2)\,.
\nonumber
\end{align}
The particle-hole loop function in Eq.~(\ref{Tph-FL}) is defined
as
\be
\mathcal{A}^{--}_{\rm ph}(\vec{n};q)=\!\! \intop\!  \rmd \Phi_0
\widehat{G}^{--}_{n,{\rm q.p.}}(p''_{\rmF+})\,\widehat{G}^{--}_{n,{\rm q.p.}}(p''_{\rmF-}).
\ee
We emphasize that the amplitudes $\hat\Gamma^{\xi}$ and $\hat\Gamma^{\om}$ are not calculable
within a Fermi liquid approach and should be considered as phenomenological quantities.

The renormalized amplitudes $\hat{\Gamma}^{\xi,\om}$ can serve as a bare interaction
$$\mathcal{G}_0^{--}=\hat{\Gamma}^{\xi,\om}\,, \quad \mathcal{G}_0^{++}=-\hat{\Gamma}^{\xi,\om} $$
in the analysis performed in the previous section.

The renormalization procedure outlined in this section can be applied to a superfluid system
without any modification, if the system is only slightly excited, $T\lsim \Delta \ll
\epsilon_\rmF$. The regular quantities are subtracted at $T=0$. The assumption of thermal
equilibrium, which we used, allows to deal with the only one Green's function, here with $G^{--}$.
If we deal with slightly non-equilibrium system for excitation energies $\epsilon^{*}\lsim
\Delta\ll \epsilon_\rmF$, the same renormalization procedure should be performed  for $G^{-+}$.
Actually, it is more convenient to chose the spectral function $A=-\Im G^R$  as another
independent quantity, rather than $G^{-+}$, and perform relormalization of $A$ ($G^{-+}=A\,f$,
where $f$ is the distribution function satisfying the generalized kinetic equation, see
Ref.~\cite{IKV00}). After the renormalization one may deal only with the quasiparticle Green's
functions and with the renormalized interaction.

\subsection{Landau-Migdal parameters for the nuclear matter}

Because of the diagonal spin structure of the normal Green's functions, the spin structure
of the amplitudes reflects the structure of bare interactions $\hat{V}$ and $\hat{U}$:
\begin{align}
\big[\hat\Gamma^{\xi}(\vec{n}\,',\vec{n})\big]_{cd,ab}&=
  \Gamma^{\xi}_0(\vec{n}\,',\vec{n})\, (i\sigma_2)_{dc}\,
  (i\sigma_2)_{ab}
\label{Gammaxi-spin}\\
&+ \Gamma^{\xi,\alpha\beta}_1(\vec{n}\,',\vec{n})\, (\vec{\sigma}\,^\alpha\, i\sigma_2)_{dc}\,
(i\sigma_2\,\vec{\sigma}\,^\beta)_{ab}\,,
\nonumber\\
\big[\hat\Gamma^{\om}(\vec{n}\,',\vec{n})\big]_{dc,ab} &=
  \Gamma^{\om}_0(\vec{n}\,',\vec{n})\, (\sigma_0)_{dc}\, (\sigma_0)_{ab}
\label{Gammaom-spin}\\
&+ \Gamma^{\om,\alpha\beta}_1(\vec{n}\,',\vec{n})\, (\vec{\sigma}\,^\beta)_{dc}\,(\vec{\sigma}\,^\alpha)_{ab}
\,.
\nonumber
\end{align}
In the graphical form  the quantity $\Gamma^{\om}$  corresponds to the empty block in
Eq.~(\ref{irred}). In the absence of the spin-orbit coupling the interaction is invariant under
independent rotation in spin and orbital spaces. In this case $\Gamma^{\xi
(\om),\alpha\beta}_1=\Gamma^{\xi(\om)}_1\, \delta^{\alpha\beta}$. Oppositely, in a isotropic
system  the interaction containing the spin-orbit coupling is invariant under the combined
rotations in both spaces and the spin structure of the interaction is more involved. According to
Ref.~\cite{Fujita87} the general structure of the interaction in the system with one type of
fermions can be written as
\begin{align}
\hat\Gamma^{\om}(\vec{n}\,',\vec{n}) &=
\Gamma^{\om}_0\, \sigma'_0 \sigma_0 + \Gamma^{\om}_1\, (\vec{\sigma}\,'\vec{\sigma})
\nonumber\\&+\Gamma^{\om}_{\rm T}\,
\big[ 3\,(\vec{\sigma}\,'(\vec{n}-\vec{n}\,'))(\vec{\sigma}(\vec{n}-\vec{n}\,'))-
(\vec{\sigma}\,'\vec{\sigma})\big]
\nonumber\\
&\phantom{\Gamma^{\om}_{\rm T}\,}\times (\vec{n}-\vec{n}\,')^2
\nonumber\\
&+\Gamma^{\om}_{+} \big[(\vec{\sigma}\,'\vec{n})(\vec{\sigma}\vec{n}\,') +
(\vec{\sigma}\,'\vec{n}\,')(\vec{\sigma}\vec{n})\big]
\nonumber\\
&+\Gamma^{\om}_{-} \big[(\vec{\sigma}\,'\vec{n})(\vec{\sigma}\vec{n}\,') -
(\vec{\sigma}\,'\vec{n}\,')(\vec{\sigma}\vec{n})\big]
\nonumber\\
&+\Gamma^{\om}_{\rm K}\,\big(\vec{\sigma}\,'[\vec{n}\times\vec{n}\,']\big)
\big(\vec{\sigma}[\vec{n}\times\vec{n}\,']\big)\,.
\label{Gom-fullspin}
\end{align}
Each pre-factor here is a function of $(\vec{n}\,'\vec{n})$. The spin index assignment is the same
as in Eq.~(\ref{Gammaom-spin}). Matrices $\sigma_j=(\sigma_j)_{ab}$ with $j=0,\dots,3$, act on the
incoming fermions, while matrices $\sigma'_j=(\sigma_j)_{dc}$ act on the outgoing fermions. The
similar decomposition can be written also for $\hat\Gamma^{\xi}(\vec{n}\,',\vec{n})$ with the only
replacements $\sigma_j\rightarrow i\sigma_2\,\sigma_j$ and $\sigma'_j\rightarrow \sigma'_j\,
i\sigma'_2$.

The renormalized amplitudes, $\Gamma^{\om,\xi}_{a}$ in
Eq.~(\ref{Gom-fullspin}),  where $a=0,1,\pm,T,K$, can be expanded in Legendre polynomials
\be
\Gamma^{\om(\xi)}_{a}(\vec{n},'\vec{n})= \sum_l\, \Gamma^{\om(\xi)}_{a;l}\,
P_l(\vec{n}\,'\vec{n}).
\label{exp-gen}
\ee
The structure with $\Gamma^{\om}_{\rm T}$ is the  tensor interaction considered  in
Ref.~\cite{Dabrowski76,Backman79,Friman81}. The effect of the tensor terms on some of the static
properties of nuclear and neutron matter was found to be very small~\cite{Dabrowski76}. However
they could play an important role in the condition for stability of the ground state of nuclear
matter~\cite{Backman79}. The spin-orbit and tensor terms in $\Gamma^{\xi}$ are important for the
description of $3P_2$ pairing \cite{Takatsuka93}. A non-local contribution of the pion exchange to
$\Gamma^\xi$ and its effect on  nucleon superfluidity were studied in Ref.~\cite{Sedr}. For the
description of the majority of nuclear phenomena including the $1S_0$ pairing it is sufficient to
consider only the terms with $\Gamma^{\om(\xi)}_0$ and $\Gamma^{\om}_1$. In further we consider
only these terms.

Following Ref.~\cite{M67} we introduce the dimensionless parameters
\be
&&\bar{f}_l^\om=\frac{\Gamma^{\om}_{0;l}}{a^2 (n_0)\, \rho (n_0)} \, ,
\,\,
\bar{g}_l^\om=\frac{\Gamma^{\om}_{1;l}}{a^2 (n_0)\, \rho (n_0)}\, ,
\nonumber\\
&&\bar{f}_l^\xi=\frac{\Gamma^{\xi}_{0;l}}{a^2 (n_0)\, \rho (n_0)}\, ,
\,\,
\bar{g}_l^\xi=\frac{\Gamma^{\xi}_{1;l}}{a^2 (n_0)\, \rho (n_0)} ,
\label{LM-param}
\ee
where $\bar f$'s and $\bar g$'s are constants.
We considered the system with the one type of particles, e.g. neutron matter.
In general case of the nuclear matter of an arbitrary isotopic composition,
we need to know the $nn$, $pp$ and $np$ interaction amplitudes. These quantities can be
parameterized as
\begin{align}
a^2 (n_0)\rho (n_0) \hat{\Gamma}_{N_1 N_2}^\om &= \bar{f}_{N_1 N_2}^\om\,\sigma'_0\,\sigma_0
+ \bar{g}_{N_1 N_2}^\om (\vec{\sigma}\,' \vec{\sigma}) ,
\nonumber\\
a^2 (n_0)\rho (n_0) \hat{\Gamma}_{N_1 N_2}^\xi &=
\bar{f}_{N_1 N_2}^\xi (i\sigma'_2)\,(i\sigma_2)
\nonumber\\
&+ \bar{g}_{N_1 N_2}^\xi
(\vec{\sigma}\,'\, i\sigma'_2)\, ( i\sigma_2\,\vec{\sigma})\,,
\label{LMP-bar}
\end{align}
where $N_1 ,N_2=n, p$. The calculation of the Landau-Migdal parameters is a formidable task and
the results vary essentially, depending on a calculation scheme and a model for the bare
nucleon-nucleon interaction, see, e.g., Refs.~\cite{APW,Backman68,Backman73,Babu73,Celenza82}.
Another possible path is to try to extract the Landau-Migdal parameters from the analysis
of phenomena in atomic nuclei. Starting from Ref.~\cite{M67} one traditionally presents the nucleon-nucleon
interaction amplitudes in the form, cf. \cite{Khodel82},
\begin{align}
\hat{\Gamma}^\om &= \frac{C}{a^2(n_0) }\big[f^\om\, \sigma'_0\,\sigma_0 + g^\om\, (\vec{\sigma}\,' \vec{\sigma})
\nonumber\\
&\quad+f'^{\om}\vec{\tau}\,' \vec{\tau}\, \sigma'_0\,\sigma_0 +g'^{\om}\,(\vec{\tau}\,' \vec{\tau}\,)\, (\vec{\sigma}\,'
\vec{\sigma})\big],
\label{TKFS-om}\\
\hat{\Gamma}^\xi &= \frac{C}{a^2(n_0)}\big[f^\om\, \sigma'_0\,\sigma_0 + g^\xi (\vec{\sigma}\,' \vec{\sigma})
\nonumber\\
&\quad+f'^{\om}(\vec{\tau}' \vec{\tau})\,\sigma'_0\,\sigma_0 +g'^{\om}(\vec{\tau}\,' \vec{\tau})\,  (\vec{\sigma}\,'
\vec{\sigma})\big].
\label{TKFS-xi}
\end{align}
The constant $C=1/\rho (n_0)$ is introduced as a dimensional
normalization factor. One usually fixes its value as $a^2 (n_0)\, C = 300$~MeV$\cdot$fm$^3$, see \cite{MSTV90}.
Making use of the parameterizations (\ref{TKFS-om}) and (\ref{TKFS-xi}) one implicitly assumes that
the Fermi-liquid renormalization preserves  isospin symmetry of the strong interaction.
Then, instead of six independent amplitudes  $f^\om_{N_1N_2}$ and  $g^\om_{N_1N_2}$
for $nn$, $np$ and $pp$ channels  one deals with  four amplitudes $f^\om , g^\om$ and
$f'^\om,
g'^\om$. The others follow from the relations
\be
\bar{f}^\om_{nn} = \bar{f}^\om_{pp}=f^\om+f'^\om,
\,\,
\bar{f}^\om_{np} = \bar{f}^\om_{pn}=f^\om-f'^\om,
\nonumber\\
\bar{g}^\om_{nn} = \bar{g}^\om_{pp}=g^\om+g'^\om,
\,\,
\bar{g}^\om_{np} = \bar{g}^\om_{pn}=g^\om-g'^\om.
\label{iso-relat}
\ee
This assumption can be justified only for the nucleon matter with {\it a small isospin
asymmetry}. For strongly asymmetrical nuclear matter, like the neutron star matter, the
application of the relations (\ref{iso-relat}) is questionable and should not hold {\it a priori}.

We note that for the particle-particle channel we use the spin parameterization (\ref{LMP-bar})
different from that in Eq.~(\ref{TKFS-xi}). The two sets of parameters are related as
\begin{align}
\bar{f}^\xi_{nn} &= (f^\xi +f'^\xi) - 3\, (g^\xi + g'^\xi)\,,
\nonumber\\
\bar{f}^\xi_{np} &= (f^\xi -f'^\xi) - 3\, (g^\xi - g'^\xi)\,,
\nonumber\\
-\bar{g}^\xi_{nn} &= (f^\xi +f'^\xi) + (g^\xi +g'^\xi)\,,
\nonumber\\
-\bar{g}^\xi_{np} &= (f^\xi -f'^\xi) + (g^\xi -g'^\xi)\,.
\end{align}
The values of the zero-th and first Legendre harmonics of $f$, $f'$, $g$, $g'$ are extracted
from analysis of many data on atomic nuclei. Unfortunately, there are essential uncertainties in
numerical values of some of these parameters. These uncertainties are, mainly, due to attempts to
get the best fit to experimental data in each specific case, modifying parameterization of the
residual part of the $NN$ interaction.  Numerical values of the parameters extracted in
Ref.~\cite{M67} are $f_0^\om\simeq 0.25$, $f'^{\om}_0 \simeq 1$, $g_0^\om\simeq 0.5$, $g'^{\om}_0
\simeq 1$.  Calculations in Ref.~\cite{Saperstein} give the values $f_0^\om\simeq 0$, $f'^{\om}_0
\simeq 0.5$--0.6, $g_0^\om\simeq 0.05\pm 0.1$, $g'^{\om}_0 \simeq 1.1 \pm 0.1$. In
Ref.~\cite{RFF} the value $g_{pp;0}$ was fixed by the data on the two-neutrino double $\beta$ decays
and the single $\beta$ decays, as $g_{pp;0}^{\om}\simeq 1$. This is in favor of the choice of
Ref.~\cite{Saperstein}. First harmonics $f_1^\om$, $f'^{\om }_1$ are related to the value of the
effective nucleon mass. The values $g^\om_{pp;1}=-g^\om _{pn;1} \simeq -0.11$ are estimated from
analysis of the decay energies and the Gamov-Teller strength distributions in neutron-rich
short-lived nuclides~\cite{BFKZ}. In Ref.~\cite{Borzov03} the values $f_0^\xi \simeq
-(0.25$--$0.33)$ and $g'^{\xi}_0\simeq -(0.3$--$0.5)$ are extracted, whereas Ref.~\cite{SMS}
calculated $f_0^\xi \simeq -0.47$ and $g_0^{\xi}\simeq +0.46$ using Cogny DSI force.

Pairing gaps  depend on the density and are very sensitive to values of parameters
in particle-particle channel because of exponential dependence on  the interaction amplitude,
see Refs.~\cite{FTTZ} and~\cite{SF,KCTZ}. In application to the pairing in neutron stars
it seems to be preferable to use the values of the Landau-Migdal parameters
adjusted to reproduce the pairing gaps obtained in micorscopic calculations, like in
Refs.~\cite{APW,Tamagaki70,Amundsen85,Takatsuka93, KKC96,Schulze96,Elgaroy98,Khodel01,SF,KCTZ,Hebeler07,Chen08,SMS}.

\section{Equilibrium systems with pairing at $T\neq 0$}\label{sec:eqpairing}

In Section~\ref{sec:White} we have demonstrated
that set of the diagrams for the one-nucleon process rate built up with the non-equlibrium Green's
functions can be rewritten as the RPA series of the retarded self-energies. For the latter
we may exploit the standard Fermi liquid approach. Thus further we follow
the lines of \cite{KV2}.


\subsection{Green functions and response of a system with pairing}

As argued in the previous section, for $T\ll \epsilon_{\rm F}$ we can use the quasiparticle
approximation for the normal Green's function, once studied process occurs near the Fermi surface.

We assume that the renormalization procedure is properly done. Thus, we may deal only with the
pole parts of the Green's functions characterized by the effective mass $m^*$ and the residue $a$.

Then, neglecting $(T/\epsilon_{\rm F})^2$  corrections, for the retarded Green's function we
write
\be
\hat{G}_n^{R}(p)={G}_n^{R}(p)\,\sigma_0\,,
\,\,
G_n^{R}(p)=\frac{a}{\epsilon -\epsilon_p+i0 }\,.
\label{GnR}
\ee
The Green's function of the hole is then given by
\be
\hat{G}_n^{h,R}(p)={G}_n^{h,R}(p)\,\sigma_0\,,
\,\, G^{h,R}(p)=G^A (-p).
\label{GhR}
\ee
Recall, here $p=(\epsilon,\vec{p}\,)$.
We will use the approximation $\epsilon_{\vec{p}+\vec{q}/2}\approx \epsilon_p+\vec{v}\,\vec{q}/2$, where $\vec{v}$
is the nucleon velocity at the Fermi surface, $\vec{v}=v_\rmF\, \vec{n}(1+O(T^2/\epsilon_{\rm
F}^2))$. Actually, the denominators of the Green's
functions are $|\om \pm (\epsilon_{\vec{p}+\vec{q}/2}-\epsilon_{\vec{p}-\vec{q}/2})|\ll
\epsilon_{{\rm F}}$. Moreover, the terms $\propto\vec{v}\,\vec{q}$ may vanish under the angular
integrations. Taking this into account we estimate that the neglected terms are at most of the
order of $(\Delta/\epsilon_{\rm F})^2\ll 1$ compared to the remained terms. Such corrections are
usually omitted in most calculations within the Fermi-liquid theory for superfluids.

We need the gap function nearby the Fermi surface, hence, it is a function of the direction, $\vec{n}$, of the relative
momentum of pared fermions
\be
\hat{\Delta}^{(1)}(\vec{n}\,) &=&
\big(\Delta^{(1)}_0(\vec{n}\,)\,\sigma_0+\vec{\Delta}^{(1)}_1(\vec{n}\,)\,\vec{\sigma}\big)i\sigma_2\,,\quad
\nonumber\\
\hat{\Delta}^{(2)}(\vec{n}\,) &=&
i\sigma_2\,\big(\Delta^{(2)}_0(\vec{n}\,)\, \sigma_0 + \vec{\Delta}^{(2)}_1(\vec{n}\,)\,\vec{\sigma}\big)\,.
\nonumber
\ee
As we argued in Section~\ref{sec:LM} the spin structure of the anomalous Green's functions
repeats
the spin structure of the gap functions
\be
\hat{F}^{(1)}(p) &=&
\big(F^{(1)}_0(p)\,\sigma_0+\vec{F}^{(1)}_1(p)\,\vec{\sigma}\big)i\sigma_2\,,\quad
\nonumber\\
\label{Fspin-p}
\hat{F}^{(2)}(p) &=&
i\sigma_2 \big(F^{(2)}_0(p)\, \sigma_0 + \vec{F}^{(2)}_1(p)\,\vec{\sigma}\big)\, .
\ee
Since, as we will see
the gap is a sharp function of the temperature, we should retain
this temperature dependence omitting it in other quantities.

In momentum representation Gor'kov's equations (\ref{Gorgap}) for
the retarded quantities render
\be
\hat{G}^R (p) &=& \hat{G}^R_n (p)+\hat{G}^R_n (p)\,\hat{\Delta}^{(1)R}(p,T)\, \hat{F}^{(2)R}(p)\,,
\nonumber\\
\hat{F}^{(2)R}(p) &=& \hat{G}^{hR}_n (p)\,\hat{\Delta}^{(2)R}(p,T)\,\hat{G}^R (p)\,.
\ee
Since $\hat G_n\propto \sigma_0$ one easily finds the solution
\be
\hat{G}^R (p)&=&\frac{\sigma_0\,[G^{hR}_n (p)]^{-1}}{[G_n^R
(p)\,G^{hR}_n (p)]^{-1}+ \Delta^2(p,T)/a^2},
\nonumber\\
\hat{F}^{(1,2)R}(p)&=&\hat\Delta^{(1,2)R}(p,T)\,G^{hR}_n (p)G^R (p)
\nonumber\\
&=&\frac{\hat\Delta^{(1,2)R}(p,T)}{[G_n^R(p)\,G^{hR}_n (p)]^{-1}+\Delta^2(p,T)/a^2}.
\nonumber
\ee
We denote here
\be
\Delta^2(p,T) &=& -a^2\,{\textstyle\frac12}{\rm Tr}\big\{\hat{\Delta}^{(2)R}(p,T)\hat{\Delta}^{(1)R}(p,T)\big\}
\nonumber\\
&=&\phantom{-} a^2\, [\Delta_0^2(p,T)+\vec{\Delta\,}_1^2(p,T)]\,.
\ee
Then using that
\be
[G_n (p)\,G^h_n (p)]^{-1}=[-(\epsilon+i0)^2+\epsilon_p^2 ]/a^2\,,
\ee
we arrive at explicit expressions for the quasi-particle
retarded Green's functions in the presence of pairing
\begin{align}
G^R (p)&=a\frac{\epsilon +\epsilon_p}{(\epsilon +i0)^2 -E_p^2 }
\nonumber
\\
&=\frac{a\,u_p^2}{(\epsilon -E_p +i0 )} +\frac{a\,v_p^2}{(\epsilon +E_p +i0 )},
\nonumber\\
\widehat{F}^{(1)R} (p) &= F^R (p)\, \frac{\widehat{
\Delta}^{(1)}}{\Delta}\,, \,\,
\widehat{F}^{(2)} (p) =F^R (p) \, \frac{\widehat{\Delta}^{(2)}}{\Delta},
\nonumber
\\
F^R (p)&=\frac{-a\Delta(p,T)}{(\epsilon +i0)^2 -E_p^2 }
\nonumber
\\
&=\frac{-a\, u_p v_p}{(\epsilon -E_p +i0 )} +\frac{ a\,u_p v_p}{(\epsilon +E_p +i0 )},
\nonumber
\end{align}
where the Bogolyubov's factors are
\be
u_p^2 =\frac{E_p +\epsilon_p}{2E_p},\quad v_p^2 =\frac{E_p
  -\epsilon_p}{2E_p},
\ee
and the quasi-particle excitations  possess the gapped spectrum
\be
E_p^2 = \epsilon_p^2 +\Delta^2(p,T)\,.
\label{gapspec}
\ee
After the retarded Green's functions are known other Green's functions
can be expressed through them, see Appendix~\ref{eq.rel}.
In the rest of the paper we consider the singlet 1S$_0$ pairing, hence $\vec{\Delta}^{(1,2)}_1=0$ and
we will denote $\Delta=\Delta_0^{(1)}=\Delta_0^{(2)}$.
So the causal Green's function at $T=0$ reads
\be
G^{--}(p)=\frac{a\,(\epsilon+\epsilon_p)}{\epsilon^2-E_p^2+i\,0{\rm
sgn}\epsilon}\,,
\nonumber\\
F^{--}(p)=\frac{-a\,\Delta}{\epsilon^2-E_p^2+i\,0{\rm sgn}
\epsilon}\,.
\ee

In further we will calculate the production rate using the equilibrium relation (\ref{standr}).
Hence we have to calculate the retarded self-energy at finite temperature. We can use the
Matsubara technique with
\be
G(p)&=& \frac{a\,(\epsilon_n +\epsilon_p)}{\epsilon_n^2-E_p^2},
\,\,
F(p)=\frac{-a\, \Delta}{\epsilon_n^2-E_p^2}\,,
\label{GFpole}
\ee
where $\epsilon_n =i\,\pi (2n+1)$, $n$ is the integer number running form $-\infty$ to $\infty$.

The energy-momentum integration at arbitrary temperature is defined as
\be
&&\int \rmd\Phi_T \,f(\epsilon,\epsilon_p)
\\&&=\left\{
\begin{array}{cc}
\dsp \rho\,
\int_{-\infty}^{+\infty} \frac{\rmd \epsilon}{2\pi i}
\int_{-\infty}^{+\infty}\rmd\epsilon_p f(\epsilon,\epsilon_p) &
\mbox{for}~T=0\\ \dsp \rho \,T \sum_{n=-\infty}^{\infty}
\int_{-\infty}^{+\infty}\rmd\epsilon_p f(i\epsilon_n,\epsilon_p) &
\mbox{for}~T\neq 0
\end{array}\right. .
\nonumber
\ee

The singlet-pairing gap is determined by the $\Gamma_{0}^\xi$ term in the
particle-particle interaction, and the gap equation reads
\be
&&\Delta(\vec{n}) = - \langle \Gamma_0^\xi(\vec{n},\vec{n}')\,A_0(\Delta(\vec{n\,}')) \, \Delta(\vec{n\,}') \rangle_{\vec{n}'}\,,
\label{gapeq}\\
&&A_0(\Delta)= \int \rmd\Phi_T G_0(p)\, G^h (p)\, \theta(\xi-\epsilon_p),
\label{A0}
\ee
where $G_0(p)=1/(\epsilon_n -\epsilon_p)$ is the Matsubara Green's function for the Fermi system
without pairing ($\Delta=0$).

Note that the same value $A_0$ can be introduced as $A_0= \int
\rmd \Phi_T\,(G(p)\, G^h(p)+F(p)\, F(p))\,
\theta(\xi-\epsilon_p)$, cf.
Ref.~\cite{Leg65}.\footnote{Definition of the value $A_0$ is here
the same as in Ref. \cite{LM63}  and differs in sign from that
used in Ref.~\cite{Leg65a,Leg65}.}
For vanishing temperature the direct calculation gives
\be
\frac{A_0(\Delta)}{a^2\, \rho} &=&
\intop_{-\xi}^{+\xi}\rmd \epsilon_p \intop_{-\infty}^{+\infty}
\frac{\rmd \epsilon}{2\,\pi\,i}\,
\frac{1}{\epsilon-\epsilon_p+i\, 0}\,
\frac{-\epsilon+\epsilon_p}{\epsilon^2-E_p^2+i\,0}
\nonumber\\ &=&
\intop_{-\xi}^{+\xi} \frac{\rmd \epsilon_p}{2\, \sqrt{\epsilon_p^2+\Delta^2}}
= \ln\Big[\frac{\xi}{\Delta}+\sqrt{\frac{\xi^2}{\Delta^2}+1}\Big]
\nonumber\\
&\approx& \ln\Big[\frac{2\,\xi}{\Delta}\Big]\,,\quad
{\xi}\gg{\Delta}.
\ee
In case of the attractive interaction $f^\xi_0<0$ the fermions are able to pair in $1S_0$
state
and  the relation between $\Delta$, $\xi$ and $f^\xi_0$ is as
follows,
\be
\Delta=\xi
\exp\Big[-\frac{\rho(n_0)}{\rho(n)\,\bar{f}^\xi_0}\Big]\,.\label{gapzero}
\ee
In the renormalization procedure we used that $\xi \ll \epsilon_{\rm F}$.
However, fixing the gap, one usually puts $\xi =\epsilon_\rmF$
bearing in mind the weak logarithmic dependence.
For $T\neq 0$,
\be
A_0(\Delta,T) =a^2\, \rho\,\intop_{-\xi}^{+\xi} \frac{\rmd \epsilon_p
(1-2\,n_f) }{2\, \sqrt{\epsilon_p^2+\Delta^2}}\,,
\ee
where $n_f$ is the Fermi distribution.

The diagrams for the self-energy are shown in Eq.~(\ref{chi-graph}).
In momentum representation they produce the following equation:
\be
&&\Sigma =\Sigma_0+\Sigma_1\,,
\nonumber\\
&&\Sigma_0 = \Big\langle\intop \rmd \Phi_T \, t_0^\om(\vec{n},q)\, \Big( G_+\, G_-\, t_0(\vec{n},q)
\nonumber\\
&&- F_+\, F_- \,t_0^h(\vec{n},q) +(G_+\, F_- - F_+\, G_-)\,\widetilde{t}_0(\vec{n},q) \Big) \Big\rangle_{\vec{n}} \,,
\nonumber\\
&&\Sigma_1 = \Big\langle\intop \rmd \Phi_T \, \vec{t}_1^{\,\om}(\vec{n},q)\, \Big( G_+\, G_-\, \vec{t}_1(\vec{n},q)
\nonumber\\
&&+ F_+\, F_- \,\vec{t}_1^{\,h}(\vec{n},q) +(G_+\, F_- - F_+\, G_-)\,\widetilde{\vec{t}}_1(\vec{n},q)\Big)
\Big\rangle_{\vec{n}}
\,.
\nonumber
\ee
The left vertices here are the bare vertices (\ref{bare-tau}) after the Fermi-liquid renormalization, like in
Eq.~(\ref{renorm-Gtau}). Their spin decomposition is
$\hat{\tau}^{\om}(\vec{n},q)=t_0(\vec{n},q)\sigma_0+ \vec{t}_1(\vec{n},q)\,\vec{\sigma}$\,.
Doing calculations in Matsubara technique, we use  notations: $G_{+}=G(p_0+\om,\vec{p}+\vec{q}/2)$
and $G_{-}=G(p_0,\vec{p}-\vec{q}/2)$, and similarly for $F_\pm$ functions. For continues
frequencies we use the symmetrical 4-vector notations, i.e. $G_\pm=G(p\pm q/2)$ and, analogously,
$F_\pm =F(p\pm q/2)$.

\begin{figure*}
\begin{align}
&\includegraphics[height=2.5cm]{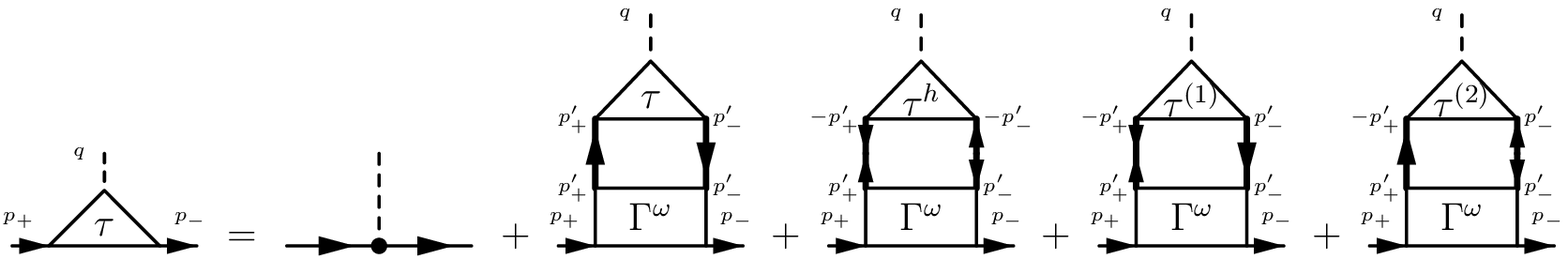}
\nonumber\\
&\includegraphics[height=2.5cm]{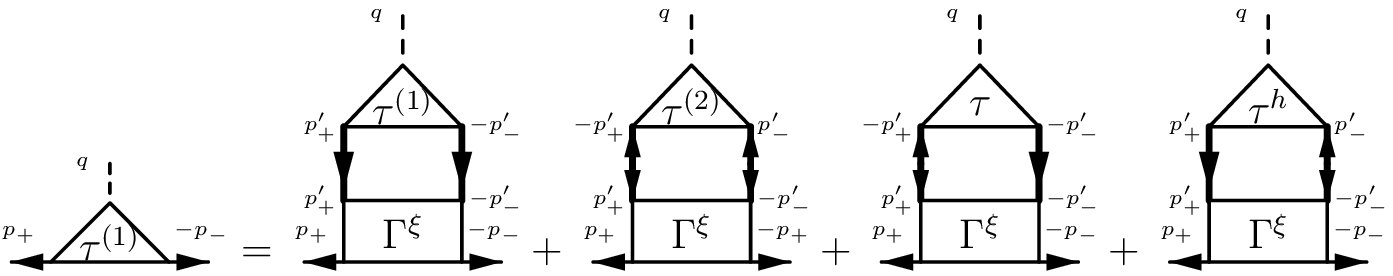}
\nonumber
\end{align}
\caption{Equations for the dressed vertices $\hat{\tau}$ and $\hat{\tau}^{(1)}$
from Eq.~(\ref{dr-vert}) in the momentum representation.
}
\label{fig:verteq}
\end{figure*}

The full vertices, $t_0(\vec{n},q)$ and $\vec{t}_1(\vec{n},q)$, are determined by the diagramatic
Eqs. (\ref{dr-vert}). In Fig.~\ref{fig:verteq} two of these equations are written explcitly
in the momentum representation. These graphical equations were first introduced by A.I.~Larkin and
A.B.~Migdal in Ref.~\cite{LM63}.

\subsection{ Larkin-Migdal equations}\label{subsec:LME}

The right vertices in Eq.~(\ref{chi-graph}) are the full in-medium-dressed vertices, which are
functions of the out-going frequency $\om$ and momentum $\vec{q}$, and the nucleon velocity
$\vec{v}\simeq v_\rmF\, \vec{n}$, $\vec{n}=\vec{p}/p$\,. Their spin structure is
\be
\hat{\tau}(\vec{n},q) &=& t_0(\vec{n},q)\, \sigma_0 +
\vec{\sigma}\,\vec{t}_1(\vec{n},q)\,,
\\
\hat{\tau}^h(\vec{n},q)&=& t_0(-\vec{n},q)\, \sigma_0 + \vec{\sigma}^{\,\rm T}\vec{t}(-\vec{n},q),
\nonumber\\
\hat{\tau}^{(1)}(\vec{n},q)&=& \big(t_0^{(1)}(\vec{n},q)\, \sigma_0
   + \vec{\sigma}\,\vec{t}_1^{\,\,(1)}(\vec{n},q)\big)\, i\sigma_2,
\nonumber\\
\hat{\tau}^{(2)}(\vec{n},q)&=& i\sigma_2\,\big(t_0^{(2)}(\vec{n},q)\,\sigma_0
   + \vec{\sigma}\,\vec{t}_1^{\,\,(2)}(\vec{n},q)\big)\,.
\nonumber
\label{taut}
\ee
After opening the spin structure of the diagrams in
Fig.~\ref{fig:verteq} we arrive at the following set of equations
for $t_0$, $t_0^h$, $t_0^{(1)}$ and $t_0^{(2)}$ (for brevity we
omit the dependence of the vertices on $\vec{n}$, $\om$ and
$\vec{q\,}$), cf. Eq.~(\ref{tsol:0}),
\begin{subequations}
\label{LManalytic}
\be
t_0 -t_0^\om &=&\phantom{-} \Big\langle \int \rmd\Phi_T\,
\Gamma_0^\om\, \Big[G_{+}\, G_{-}\, t_0 - F_{+}\, F_{-}\,t^h_0
\nonumber\\
&-& G_{+}\, F_{-}\, t^{(1)}_0 - F_{+}\, G_{-}\,
t^{(2)}_0\Big] \Big\rangle_{\vec{n\,}'}\,,
\label{LMan:t0}\\
t^h_0-t^{h\om}_0 &=& \phantom{-} \Big\langle
\int\rmd\Phi_T\,\Gamma_0^\om\, \Big[G^h_+\, G^h_-\,t^h_0 - F_+\,
F_-\, t_0 \nonumber\\&-&F_+\,G^h_-\, t^{(1)}_0 - G_+^h\,
F_-\,t^{(2)}_0 \Big] \Big\rangle_{\vec{n\,}'}\,,
\label{LMan:th0}\\
t^{(1)}_0 &=& -\Big\langle\int\rmd\Phi_T\,\Gamma_0^\xi\, \Big[G_{+}\, G^h_{-}\,t^{(1)}_0
-F_{+}\,F_{-}\, t^{(2)}_0 \nonumber\\&+& G_{+}\, F_{-}\, t_0 +F_{+}\, G^h_{-}\, t^h_0 \Big]
\Big\rangle_{\vec{n\,}'}\,,
\label{LMan:t10}\\
t^{(2)}_0 &=&-\Big\langle \int\rmd\Phi_T\,\Gamma_0^\xi\, \Big[G^h_+\,G_-\, t^{(2)}_0 -
F_+\,F_-\,t^{(1)}_0
\nonumber\\
&+& F_+\, G_- \, t_0 + G^h_+\,F_-\, t^h_0\Big] \Big\rangle_{\vec{n\,}'}\,.
\label{LMan:t20}
\ee
\end{subequations}
The similar set of equations for 3-vector vertices $\vec{t}_1$,
$\vec{t}_1^{\,h}$, $\vec{t}_1^{\,(1)}$ and $\vec{t}_1^{\,(2)}$ is
written with the only differences that $\Gamma_0^{\om,\xi}$ is
replaced by $\Gamma_1^{\om,\xi}$ and in front of all terms with
$t^h$ the sign must be changed, cf. Eq.~(\ref{tsol:1}).
For the sake of convenience we introduce brief notations, e.g.,
\be
\int \rmd\Phi_T\,  G^{h}_{+}\,F_{-}=G^{h}_{+}\cdot F_{-}.
\label{dotoper}
\ee
The details of calculations of these products within the Matsubara technique
are given in Ref.~\cite{KV2}. For instance the useful relations introduced in Ref.~\cite{Leg65}
\be
G_{+}\!\cdot\! G^h_{-}&=&   G^h_{+}\!\cdot\! G_{-}\,,\quad
G_{+}\!\cdot\! F_{-}  = - F_+\!\cdot\!
G_-\,,\nonumber\\F_{+}\!\cdot\! G^h_{-} &=&-  G^h_+ \!\cdot\!
F_-\,, \label{relat1}
\ee
are recovered for arbitrary temperature. From (\ref{LManalytic}) we can immediately find relations
between the vertices $t^{(1)}_0$ and $t^{(2)}_0$. Taking the sum of Eqs.~(\ref{LMan:t10}) and
(\ref{LMan:t20}) and making use of Eq.~(\ref{relat1}) we obtain the homogeneous equation for the
sum $t^{(1)}_0+t^{(2)}_0$,
\be
t^{(1)}_0+t^{(2)}_0 &=& -\Big\langle \int\rmd\Phi_T\,\Gamma^\xi\,
\Big[G_{+}\, G^h_{-}\,-F_{+}\,F_{-}\Big]\,
\nonumber\\
&\times&(t^{(1)}_0+t^{(2)}_0) \Big\rangle_{\vec{n\,}'}\,,
\ee
which implies
\be
t^{(1)}_0+t^{(2)}_0=0\, \label{t12relat}
\ee
for frequencies relevant for PBF processes, which we will consider below.
The latter relation justifies the parameterization of the full in-medium vertices used below in
Eqs.~(\ref{full:ttv}) -- (\ref{full:tta}). The same relation is valid for $\vec{t\,\,}^{(1)}_1$
and $\vec{t\,\,}^{(2)}_1$ vertices.

In their original paper~\cite{LM63} A.I.~Larkin and A.B.~Migdal presented Eqs.~(\ref{LManalytic})
in somewhat different form.  They noted that the vertices for the holes, $t^{h}$, can be obtained
from the particle vertices with the replacement $\vec{n}\to -\vec{n}$,
 \be
t_0^h(\vec{n},q)=t_0(-\vec{n},q)\,, \quad
\vec{t\,}_1^h(\vec{n},q)=\vec{t}_1(-\vec{n},q)\,.
 \ee
Therefore, one can introduce the operator $\piproj$, which performs this change of $\vec{n}$ in
the vertex
 \be
\piproj\, t_0(\vec{n},q)=t_0^h(\vec{n},q)\,,\quad
\piproj\,\vec{t}_{1}(\vec{n},q)=\vec{t\,}_{1}^h(\vec{n},q)\,. \ee
Because of relation (\ref{t12relat}),
Eqs.~(\ref{LMan:t10},\ref{LMan:t20}) reduce to one equation for
the vertex
 \be
\widetilde{t}_0=-t_0^{(1)}=t_0^{(2)}. \nonumber
 \ee
Analogously we introduce $\widetilde{\vec{t}}_1=-\vec{t}_1^{\,(1)}=\vec{t\,}_1^{(2)}$\,. Then
four Eqs.~(\ref{LManalytic}) for scalar vertices '0' and four equations for 3-vector vertices
$\vec{t}_1$ can be cast in terms of four equations
\begin{subequations}
\label{LMeq}
\begin{align}
&t_0-t_0^\om = \Big\langle \Gamma_0^\om \big( L(\piproj)\,t_0
+M\,\widetilde{t}_0\big) \Big\rangle_{\vec{n\,}'}\, ,
\label{SwP:t0}
\\
&\widetilde{t}_0 = -\Big\langle \Gamma_0^\xi\, \Big( \big(N+A_0
\big)\,\widetilde{t}_0+O(\piproj )\,t_0 \Big) \Big\rangle_{\vec{
n\,}'} \label{SwP:tt0}\, ,
\\
&\vec{t}_1 -\vec{t}^\om_1 = \Big\langle \Gamma_{1}^{\om} \big(
L(-\piproj )\,\vec{t}_1 +M\,\widetilde{\vec{t\,}}_1\big)
\Big\rangle_{\vec{n\,}'} \label{SwP:t1}\, ,
\\
&\widetilde{\vec{t}}_1 = -\Big\langle \Gamma^\xi_1\,
\Big(\big(N+A_0\big)\,\widetilde{\vec{t\,}}_1+O(-\piproj
)\,\vec{t}_1 \Big) \Big\rangle_{\vec{n\,}'}\,. \label{SwP:tt1}
\end{align}
\end{subequations}
We shall call this set of equations {\it the Larkin-Migdal equations}. In Ref.~\cite{LM63} these
four equations are further reduced to only two equations with the help of the operator $\hat{P}$
(defined by Eq.~(31) in Ref.~\cite{LM63}), which includes additionally the change of the sign
between Eqs.~(\ref{SwP:t0}) and (\ref{SwP:t1}) and between Eqs.~(\ref{SwP:tt0}) and
(\ref{SwP:tt1}). Functions $L$, $M$, $N$, and $O$ are defined as in Ref.~\cite{LM63}:
 \be
L(\vec{n},q;\piproj ) &=&  G_+\!\cdot\! G_- - F_+\!\cdot\!
F_-\,\piproj \,, \nonumber\\ M(\vec{n},q)&=&  G_+\!\cdot\! F_- -
F_+\!\cdot\! G_- \,, \nonumber\\ N(\vec{n},q) &=& G_+\!\cdot\!
G^h_- +F_+\!\cdot\! F_- - A_0\,, \nonumber\\ O(\vec{n},q;\piproj )
&=& - G_+\!\cdot\! F_- - F_+\!\cdot\! G^h_- \piproj\, .
\label{depend}
 \ee
We emphasize that Eqs.~(\ref{LMeq}) are valid at arbitrary temperature. The temperature dependence
is hidden in the convolutions of the Green's functions (\ref{depend}). In Ref.~\cite{LM63} the
latter ones were calculated explicitly only for $T=0$ using the method of Ref.~\cite{VGL62}. The
extension to $T\neq 0$ is done in Ref.~\cite{KV2}, see Appendix~\ref{app:loop}.


\section{Current-current correlators for the PBF process $n\to n+\nu\bar{\nu}$}
\label{sec:neutrino}

Below we demonstrate how to apply the above results to  calculate the neutrino emissivity in the
PBF process $n\to n+\nu\bar{\nu}$ from the superfluid neutron star interior. To perform this
calculation \cite{KV1,KV2} we start with the diagrammatic presentation of the current-current
correlator (\ref{chi-graph}) for weak neutral currents, where enter the vertices shown in Fig.
~\ref{fig:verteq},  calculated  the previous section. In Eqs.  (\ref{chi-graph}), (\ref{chi-ss}) we used the
contour Green's functions, whereas now we will treat the same   Eq.  (\ref{chi-graph}) for  the retarded
self-energy. We use the Matsubara technique at $T\neq 0$ and then perform appropriate analytic
continuation to obtain the retarded self-energy. Thus we recover the RPA current-current correlator.

The bare vertices generated by the weak currents (\ref{currents}) are equal to
\begin{subequations}
\label{tauom}
\be
\hat\tau^\om_V(\vec{n},q) &=& g_V\,\big(\tau^\om_{V,0}\,l_0-\vec{\tau}^\om_{V,1}\, \vec{l}\,\,\big)\hat{1}\, ,
\\
\tau^\om_{V,0}&=&\frac{e_V}{a}\,\,,
\quad
\vec{\tau\,}^\om_{V,1}=\frac{e_V}{a}\, \vec{v},
\nonumber
\label{VC:tauom}\\
\hat\tau^\om_A(\vec{n},q) &=& -g_A\,\big(\vec{\tau\,}^\om_{A,1}\,\vec{\sigma}\,l_0
                           -\tau^\om_{A,0}\, \vec{\sigma}\vec{l}\,\,\big)\, ,
\\
\tau^\om_{A,0}&=&\frac{e_A}{a}\,, \quad
\vec{\tau\,}^\om_{A,1}=\frac{e_A}{a}\, \vec{v}\,.
\nonumber\label{AC:tauom}
\ee
\end{subequations}
Here $e_V$ and $e_A$ are effective charges of the vector and axial-vector currents.
For the vector current $e_V =1$ and for the axial-vector current $e_A \simeq 0.8 - 0.95$,
as it follows from studies of the Gamov-Teller transitions in nuclei, see
Refs. \cite{M67,Pyatov,Borzov03} and references  therein. The
corresponding vertices for holes are defined as
\begin{subequations}
\label{tauomh}
\begin{align}
\hat\tau^{\om,h}_V(\vec{n},q) &= [\hat\tau^{\om}_V(-\vec{n},q)]^{\rm T}
\nonumber\\
&=g_V\, \big(\tau^\om_{V,0}\,l_0+\vec{\tau\,}^\om_{V,1}\, \vec{l}\,\,\big)\sigma_0,
\label{VC:tauomh}\\
\hat\tau^{\om,h}_A(\vec{n},q) &=[\hat\tau^{\om}_A(-\vec{n},q)]^{\rm T}
\nonumber\\
&=-g_A\, \big(-\vec{\tau\,}^\om_{A,1}\,\vec{\sigma\,}^{\rm T} \,l_0-
\tau^\om_{A,0}\, \vec{\sigma\,}^{\rm T}\vec{l}\,\,\big)\,.
\label{AC:tauomh}
\end{align}
\end{subequations}
We used here explicitly that
$\tau^\om_{a,0}(-p,q)=\tau^\om_{a,0}(p,q)$ and
$\vec{\tau\,}^\om_{a,1}(-p,q)=-\vec{\tau\,}^\om_{a,1}(p,q)$.

The structure of the full vertices reads
\begin{subequations}
\label{full}
\be
\hat{\tau}_V&=&g_V\, \big(\tau_{V,0}\, l_0-\vec{\tau}_{V,1}\,\vec{l\,}\big)\,\sigma_0\,,
\nonumber\\
\hat{\tau}^h_V&=&g_V\, \big(\tau_{V,0}^h\, l_0-\vec{\tau\,}_{V,1}^h\,\vec{l}\,\,\big)\,\sigma_0\,,
\label{full:tv}\\
\hat{\tau}_V^{(1)}&=&(\hat{\tau}_V^{(2)})^\dag
\nonumber\\
&=& -g_V\,\big(\widetilde{\tau}_{V,0}\, l_0-\vec{\widetilde{\tau}}_{V,1}\,\vec{l}\,\,\big)\, i\, \sigma_2\,,
\label{full:ttv}\\
\hat{\tau}_A&=&-g_A\, \big(\vec{\tau}_{A,1}\vec{\sigma}\, l_0-\tau_{A,0}\,\vec{\sigma}\vec{l}\,\,\big)\,,
\nonumber\\
\hat{\tau}^h_A&=&-g_A\, \big(\vec{\tau\,}_{A,1}^h\vec{\sigma}^{\rm T}\, l_0 - \tau_{A,0}^h\,\vec{\sigma\,}^{\rm T}\vec{l}\,\,\big)\,,
\label{full:ta}\\
\hat{\tau}^{(1)}_A &=&(\hat{\tau}^{(2)}_A)^\dag
\nonumber\\
&=&g_A\, \big(\vec{\widetilde\tau}_{A,1}\, \vec{\sigma}\, l_0-\widetilde{\tau}_{A,0}\, \vec{\sigma}\,
\vec{l}\,\,\big)\, i\, \sigma_2\,.
\label{full:tta}
\ee
\end{subequations}

We rewrite the retarded current-current correlator corresponding to (\ref{optpi})
as
\be
\Sigma^R_{{\rm[nucl]},\mu\nu}\,l^\mu l^\nu=\chi(q)=\chi_V(q)+\chi_A(q)\,,
\ee
where the contributions of vector and axial currents are
\begin{subequations}
\be
\chi_V(q)&=& g_V^2\, \big\langle\big(\tau_{V,0}^\om\,l_0-\vec{\tau\,}^\om_{V,1}\, \vec{l}\,\big)\,
\\
&\times&\big(l_0\, \chi_{V,0}(\vec{n},q) - \vec{\chi}_{V,1}(\vec{n},q)\,
\vec{l}\,\,\big)\big\rangle_{\vec{n}},
\label{chiV}\nonumber\\
\chi_A(q) &=& g_A^2\, \big\langle\big(\vec{\tau\,}^\om_{A,1}\,l_0-\tau_{A,0}^\om\,\vec{l}\,\big)\,
\\
&\times&\big(l_0\, \vec{\chi}_{A,1}(\vec{n},q) - \chi_{A,0}(\vec{n},q) \vec{l}\,\,\big)
\big\rangle_{\vec{n}}\,
\nonumber\label{chiA} \ee
\end{subequations}
with the scalar and vector response functions
\begin{subequations}
\be
&&\chi_{a,0}(\vec{n},q) = \intop \rmd \Phi_T \big(G_+\, G_-\,
\tau_{a,0} - F_+\, F_- \,\tau^h_{a,0}
\nonumber\\
&&+(G_+\, F_- - F_+\, G_-)\,\widetilde{\tau}_{a,0}\big)\,,
\label{chi0}\\
&&\vec{\chi}_{a,1}(\vec{n},q)=\intop \rmd \Phi_T \big(G_+\, G_-\,
\vec{\tau}_{a,1} + F_+\, F_- \,\vec{\tau\,}^h_{a,1}
\nonumber\\
&&+(G_+\, F_- - F_+\, G_-)\,\widetilde{\vec{\tau}}_{a,0}\big)\,.
\label{chi1} \ee
\end{subequations}

Then we apply the Larkin-Migdal Eqs. (\ref{LMeq}) for the case of the weak-current vertices
(\ref{tauomh},\ref{full}). For the  vector-current vertices we use
Eqs.~(\ref{SwP:t0},\ref{SwP:tt0}) and for the axial-vector--current vertices,
Eqs.~(\ref{SwP:t1},\ref{SwP:tt1}). Then we separate the parts proportional to the scalar $l_0$ and
the vector $\vec{l}$ and obtain altogether eight equations for the vector and  axial-vector current
vertices. These sets of equations are cast in the following form~\cite{KV1},
\begin{widetext}
\begin{subequations}
\label{LMeqWeak}
\be
\tau_{a,0}(\vec{n},q)=\tau^{\om}_{a,0}(\vec{n},q)
&+&\big\langle\Gamma_a^{\om}(\vec{n},\vec{n'})\, \big[
L(\vec{n'},q;\hat{\mathcal{P}}_{a,0})\, \tau_{a,0}(\vec{n'},q) + M(\vec{n'},q)\,\widetilde{\tau}_{a,0}(\vec{n'},q)
\big]\big\rangle_{\vec{n}'}\,,
\label{LMEw1}\\
\widetilde{\tau}_{a,0}(\vec{n},q)=&-&
\big\langle\Gamma_a^{\xi}(\vec{n},\vec{n'})\, \big[
(N(\vec{n'},q)+A_0)\, \widetilde{\tau}_{a,0}(\vec{n'},q) + O(\vec{n'},q;\hat{\mathcal{P}}_{a,0})\,{\tau}_{a,0}(\vec{n'},q)
\big]\big\rangle_{\vec{n}'}\,,
\label{LMEw2}\\
\vec{\tau}_{a,1}(\vec{n},q)=\vec{\tau}^\om_{a,1}(\vec{n},q)&+&
\big\langle\Gamma_a^{\om}(\vec{n},\vec{n'})\, \big[
L(\vec{n'},q;\hat{\mathcal{P}}_{a,1})\,\vec{\tau}_{a,1}(\vec{n'},q)+ M(\vec{n'},q)\,\vec{\widetilde{\tau}}_{a,1}(\vec{n'},q)
\big]\big\rangle_{\vec{n}'}\,,
\label{LMEw3}\\
\vec{\widetilde{\tau}}_{a,1}(\vec{n},q)=&-&
\big\langle\Gamma_a^{\xi}(\vec{n},\vec{n'})\, \big[
(N(\vec{n'},q)+A_0)\, \vec{\widetilde{\tau}}_{a,1}(\vec{n'},q) + O(\vec{n'},q;\hat{\mathcal{P}}_{a,1})\,\vec{\tau}_{a,1}(\vec{n'},q)
\big]\big\rangle_{\vec{n}'}\,,
\label{LMEw4}
\ee
 \end{subequations}
\end{widetext}
where  $a=V$ for the vector and  $a=A$ for the axial-vector
weak currents. The notation for the effective
interaction is $\Gamma_V^{\om,\xi}=\Gamma_0^{\om,\xi}$ and
$\Gamma_A^{\om,\xi}=\Gamma_1^{\om,\xi}$. Operators
$\hat{\mathcal{P}}_{a,i}$ are defined as follows
\be
\hat{\mathcal{P}}_{a,i}=(-1)^i\, P_{a,i}\, \piproj\,,\quad i=0,1\,
\ee
with  parameters
\be
P_{V,0}=1\,,\, P_{V,1}=-1\,,\, P_{A,0}=-1\,,\, P_{A,1}=1,
\ee
which are the eigenvalues of the operators $\hat{\mathcal{P}}_{a,i}$,
acting on the bare vertices
\be
\hat{\mathcal{P}}_{a,0}\,\tau_{a,0}^\om=P_{a,0}\,\tau_{a,0}^\om\,,
\quad
\hat{\mathcal{P}}_{a,1}\,\vec{\tau}_{a,1}^\om=P_{a,1}\,\vec{\tau}_{a,1}^\om\,.
\ee
To proceed,  let us for simplicity assume that $\Gamma_a^{\om}$ and $\Gamma_a^{\xi}$ contain only
zero-th Legendre harmonics. From (\ref{LMEw2}) we find
\be
\widetilde{\tau}_{a,0}(q) =-\eta^\xi_a\, \frac{\big\langle
O(\vec{n},q;P_{a,0})\big\rangle_{\vec{n}} } {\big\langle
N(\vec{n},q)\big\rangle_{\vec{n}}}\, {\tau}_{a,0}(q),
\label{widetilde}
\ee
 where
\be
\eta^\xi_a=\frac{\Gamma_a^\xi\,\langle
N(\vec{n},q)\rangle_{\vec{n}} } {1+\Gamma_a^\xi [A_0+ \langle
N(\vec{n},q)\rangle_{\vec{n}}]}.
\ee
For the channel $a=V$, for which the gap equation $1=-A_0\,\Gamma_{V}^{\xi} $ is valid, we obtain
$\eta_V^\xi=1$. For the other channel, $\eta_A^\xi\neq 1$. Substituting Eq.~(\ref{widetilde}) in
Eq.~(\ref{LMEw1}) we obtain
\be
&&\tau_{a,0}(q) = \gamma_a(q;P_{a,0})\,\tau^\om_{a,0}\,,
\nonumber\\ &&\gamma_a^{-1}(q;P)=1 -
\Gamma^\om_a\,\langle\mathcal{L}_a(\vec{n},q;P)\rangle_{\vec{n}}\,,
\label{solLMeq}
\ee
where we introduced the notation
\be
\mathcal{L}_a(\vec{n},q;P)&=& L(\vec{n},q;P)
\nonumber\\
&-& \eta_a^\xi\, \frac{\langle  O(\vec{n},q;P)  \rangle_{\vec{n}}}
{\langle N(\vec{n},q)\rangle_{\vec{n}}}\, M(\vec{n},q)\, .
\label{curleL}
\ee
Solving the second pair of the Larkin-Migdal equations (\ref{LMEw3},\ref{LMEw4}) we first note
that for the constants $\Gamma_a^{\om}$ and $\Gamma_a^{\xi}$ the angular averages on the right-hand
sides of  equations do not depend on $\vec{n}$. Therefore, the component of the bare vertex
proportional to $\vec{v}$ is not renormalized in medium. However, in view of the identity
\be
\langle f(\vec{n},\vec{q\,})\, \vec{n}\rangle_{\vec{n}}=
\langle f(\vec{n},\vec{q\,})\, (\vec{n\,}\vec{q})\rangle_{\vec{n}}\, \vec{q}/\vec{q\,}^2
\label{ident}
\ee
valid for an arbitrary scalar function $f$ of $\vec{n}$ and $\vec{q}$, the full vertices gain a
component proportional to $\vec{q}$. Thus, we decompose 3-vectors $\vec{\tau}_{a,1}(\vec{n},q)$ and
$\vec{\widetilde{\tau}}_{a,1}(\vec{n},q)$\, into the parts proportional to the
$\vec{n}$ and
$\vec{n}_q =\vec{q}/|\vec{q}|$ vectors and introduce new scalar form-factors
\be
\vec{\tau}_{a,1}(\vec{n},q)&=&{\tau}^{\om}_{a,1}\, \vec{n}+ {\tau}^{(q)}_{a,1}(q)\, \vec{n}_q ,
\nonumber\\
\vec{\widetilde{\tau}}_{a,1}(\vec{n},q)&=& {\widetilde{\tau}}^{(q)}_{a,1}(q)\, \vec{n}_q\,
\label{vecvert}
\ee
with the bare vertex ${\tau}^{\om}_{a,1}=(\vec{n}\,\vec{\tau\,}^{\om}_{a,1})$\,. Action of the
operator $\hat{\mathcal{P}}_{a,1}$ on the vertices (\ref{vecvert}) is given by
\be
&&\hat{\mathcal{P}}_{a,1}\, \vec{\tau}_{a,1}(\vec{n},q)=
P_{a,1}\,\tau_{a,1}^\om\,\vec{n}+ (-P_{a,1})\, \vec{n}_q\, \tau_{a,1}^{(q)}(q),
\nonumber\\
&&\hat{\mathcal{P}}_{a,1}\, \vec{\widetilde{\tau}}_{a,1}(\vec{n},q) = (-P_{a,1})\,\vec{n}_q\,
\widetilde{\tau}_{a,1}^{(q)}(q)\, .
\nonumber \ee
Then, from Eq.~(\ref{LMEw4}) we recover
\be
\widetilde{\tau}_{a,1}^{(q)}&=& -\eta_a^\xi\, \frac{\langle
O(\vec{n}, q;-P_{a,1})\rangle_{\vec{n}}}{\langle N(\vec{n},q)\rangle_{\vec{n}}}\, {\tau}_{a,1}^{(q)}
\nonumber\\
&-&\eta_a^\xi\frac{\langle O(\vec{n}, q; P_{a,1})
(\vec{n\,}\vec{n}_q)\rangle_{\vec{n}}} {\langle N(\vec{n}, q)\rangle_{\vec{n}}}\, \tau_{a,1}^\om\,.
\label{widetildevec}
\ee
From Eq.~(\ref{LMEw3}), substituting there Eq.~(\ref{widetildevec}), we find
\be
\tau^{(q)}_{a,1} &=& \gamma_a(q;-P_{a,1})\,\Gamma_a^\om\,
\nonumber\\
&\times&\langle \widetilde{\mathcal{L}}_a(\vec{n}, q;
P_{a,1}) (\vec{n\,}\cdot\vec{n}_q) \rangle_{\vec{n}}\, \tau_{a,1}^\om\,.
\label{LME:sol:vec}
\ee
Here we introduced the quantity
\be
\widetilde{\mathcal{L}}_a(\vec{n}, q; P)&=& L(\vec{n}, q; P)
\nonumber\\&-&\eta^\xi_a\frac{\langle M(\vec{n},q)\rangle_{\vec{n}}} {\langle N(\vec{n}, q)\rangle_{\vec{n}}}\,
O(\vec{n}, q; P)
\label{curleLtilde}
\ee
and used the identity
\be\label{idL} \langle
\widetilde{\mathcal{L}}_a(\vec{n}, q; P_{a,1}) \rangle_{\vec{n}}=
\langle {\mathcal{L}}_a(\vec{n}, q; P_{a,1}) \rangle_{\vec{n}}\, ,
\ee
which allows to use for the vector vertices the same function
$\gamma_a$\, as for the scalar vertex\,.

In terms of the loop-functions (\ref{depend}) the response
functions (\ref{chi0},\ref{chi1}) can be expressed as
\be
\chi_{a,0}(\vec{n},q)=L(\vec{n},q;\hat{\mathcal{P}}_{a,0})\,\tau_{a,0}(\vec{n},q)
+M(\vec{n},q)\, \widetilde{\tau}_{a,0}(\vec{n},q),
\nonumber\\
\vec{\chi}_{a,1}(\vec{n},q)= L(\vec{n},q;\hat{\mathcal{P}}_{a,1})\, \vec{\tau}_{a,1}(\vec{n},q)
+M(\vec{n},q)\, \vec{\widetilde{\tau}}_{a,1}(\vec{n},q).
\nonumber 
\label{chis}
\ee
Using solutions (\ref{widetilde}) and (\ref{solLMeq}), for the scalar vertices we find
\be
\chi_{a,0}(\vec{n},q) = \gamma_a(q;P_{a,0})\, \tau_{a,0}^\om \,\mathcal{L} (\vec{n},q;P_{a,0})\,.
\label{chi0:sol}
\ee
With the help of Eq. (\ref{vecvert}) we construct
\be
\vec{\chi}_{a,1}(\vec{n},q) &=& L(\vec{n},q;P_{a,1})\vec{n}\,\tau_{a,1}^\om
\nonumber\\
&+& L(\vec{n},q;-P_{a,1})\,\vec{n}_q\, \tau^{(q)}_{a,1}(\vec{n},q)
\nonumber\\
&+&
M(\vec{n},q)\,\vec{n}_q\, \widetilde{\tau}^{(q)}_{a,1}(\vec{n},q) \,.
\ee
Using solutions (\ref{widetildevec}) and (\ref{LME:sol:vec}) for the three-vector vertices we
obtain
\begin{align}
&\vec{\chi}_{a,1}(\vec{n},q) =\Big( L(\vec{n},q;P_{a,1})\,\vec{n}
\nonumber\\
&\qquad- M(\vec{n},q)\,\eta^\xi_a\, \frac{\langle O(\vec{n}, q; P_{a,1})
(\vec{n\,}\vec{n}_q)\rangle_{\vec{n}}} {\langle N(\vec{n},
q)\rangle_{\vec{n}}}\,\vec{n}_q\Big)\,\tau_{a,1}^\om \,
\nonumber\\
&\qquad+ \mathcal{L}_a(\vec{n},q;-P_{a,1})\,\gamma_a(q;-P_{a,1})\,
\nonumber\\
&\qquad\times \Gamma_a^\om\, \langle \widetilde{\mathcal{L}}_a(\vec{n}, q;
P_{a,1})(\vec{n\,}\vec{n}_q) \rangle_{\vec{n}}\, \vec{n}_q\,\tau_{a,1}^\om\, ,
\label{chivec1:1}
\end{align}
and, then, rewrite it as follows
\begin{align}
&\vec{\chi}_{a,1}(\vec{n},q)= \vec{\tau}^\om_{a,1}\,\gamma_a(q;-P_{a,1})\,{\mathcal{L}}_a(\vec{n},q;P_{a,1})
\nonumber\\
&\qquad\qquad+ \delta\vec{\chi}_{a,1}(\vec{n},q)\,,
\label{chivec:sol}\\
&\delta\vec{\chi}_{a,1}(\vec{n},q)
\nonumber\\
&\quad = \eta^\xi_a\, \frac{M(\vec{n},q)}{\langle N(\vec{n}, q)\rangle_{\vec{n}}}\,
\langle O(\vec{n'}, q; P_{a,1}) (\vec{n}-\vec{n\,}') \rangle_{\vec{n'}}\, \tau_{a,1}^\om
\nonumber\\
&\quad + \gamma_a(q;-P_{a,1})\,\tau_{a,1}^\om\,\Gamma_a^\om\,\Big(
\mathcal{L}_a(\vec{n},q;-P_{a,1})\,
\nonumber\\
&\quad\times\langle\widetilde{\mathcal{L}}_a(\vec{n}, q; P_{a,1}) \vec{n\,} \rangle_{\vec{n}}
\nonumber\\
&\qquad-\mathcal{L}_a(\vec{n},q;P_{a,1})\,\vec{n}\, \langle
\widetilde{\mathcal{L}}_a(\vec{n},q;-P_{a,1})\rangle_{\vec{n}} \Big)\,.
\nonumber
\end{align}

Relations
\be
\langle \om\, \chi_{V,0}-\vec{q}\,\vec{\chi}_{V,1}\rangle_{\vec{n}}=0 ,
\nonumber\\
\Im\langle (\vec{q}\,\vec{v}\,)\, (\om\, \chi_{V,0}-\vec{q}\,\vec{\chi}_{V,1})\rangle_{\vec{n}} =0\,
\label{conserv}
\ee
proved in Ref.~\cite{KV2} ensure the transversality of the polarization tensor for the weak vector
current. Thus, we have demonstrated that the retarded self-energy given
by the RPA set of diagrams (\ref{chi-ss}) with the vertices shown by Fig.~\ref{fig:verteq}
complies with the vector current conservation.

Below we exploit expressions for the following averages
\begin{subequations}
\label{aver}
\begin{align}
&\langle \chi_{a,0}(\vec{n},q)\rangle_{\vec{n}}
= \gamma_a(q;P_{a,0})\, \tau_{a,0}^\om \,
\nonumber\\
&\qquad
\times \langle \mathcal{L}(\vec{n},q;P_{a,0})\rangle_{\vec{n}}\,,
\label{aver:chi0}\\
&\langle \chi_{a,0}(\vec{n},q)\, (\vec{q\,}\vec{v}) \rangle_{\vec{n}}
= \gamma_a(q;P_{a,0})\, \tau_{a,0}^\om \,
\nonumber\\
&\qquad \times\langle
\mathcal{L}(\vec{n},q;P_{a,0})(\vec{q\,}\vec{v})\rangle_{\vec{n}}\,,
\label{aver:chi0qv}\\
&\langle \vec{q\,} \vec{\chi}_{a,1}(\vec{n},q)\rangle_{\vec{n}}
= \gamma_a(q;-P_{a,1})\,
\nonumber\\
&\qquad\times \langle
\widetilde{\mathcal{L}}_a(\vec{n},q;P_{a,1})\,(\vec{\tau\,}_{a,1}^\om\,\vec{q}\,)\rangle_{\vec{n}}\,.
\label{aver:chi1q}\\
&\langle \vec{v\,} \vec{\chi}_{a,1}(\vec{n},q)\rangle_{\vec{n}} = \langle
L(\vec{n},q;P_{a,1})\, (\vec{v\,}\,\vec{\tau\,}_{a,1}^\om)\rangle_{\vec{n}}
\nonumber\\
&- \langle M(\vec{n},q)\,(\vec{v\,}\vec{n}_q)\rangle_{\vec{n}}\,\eta^\xi_a\,
\frac{\langle O(\vec{n}, q; P_{a,1})
(\vec{\tau\,}_{a,1}^\om\,\vec{n}_q) \rangle_{\vec{n}}} {\langle
N(\vec{n}, q)\rangle_{\vec{n}}}
\nonumber\\
&\quad+ \gamma_a(q;-P_{a,1})\,\Gamma_a^\om\, \langle
\mathcal{L}_a(\vec{n},q;-P_{a,1})\,(\vec{v\,}\vec{n}_q)\rangle_{\vec{n}}\,
\nonumber\\
&\quad\times\langle \widetilde{\mathcal{L}}_a(\vec{n}, q;
P_{a,1})(\vec{\tau\,}_{a,1}^\om\vec{n}_q) \rangle_{\vec{n}} \,.
\label{aver:chi1v}
\end{align}
\end{subequations}

In the region of the $1S_0$ neutron pairing in neutron stars occurs at low densities  ($<n_0$) and the
nucleons are non-relativistic, $v_\rmF\ll 1$. Expanding $\mathcal{L}$ and $L$ at small $v_\rmF$,
see Refs.~\cite{KV1,KV2}, we find that the correlation functions $\gamma_a$ differ from unity only
in the second order in $v_\rmF\,|\vec{q}\,|$, i.e.
\be
\gamma_a(q;P)\approx 1 +O(\Gamma_a^\om\,a^2\,\rho\, \vec{q\,}^2\,
v_\rmF^2/\om^2)\,. \label{gammaexp}
\ee
We obtain that in the expression induced by the vector currents both scalar and vector components,
(\ref{aver:chi0}) and (\ref{aver:chi1v}), contribute at the order $v_\rmF^4$,
\be
\Im\langle\chi_{V,0}(\vec{n},q) \rangle &\approx&
-\frac{4\,\vec{q\,}^4 v_\rmF^4}{45\, \om^4}\,e_V a\, \rho\, \Im g_T(0,\om,0)\,,
\nonumber\\
\Im\langle
\vec{v}\,\vec{\chi}_{V,1}(\vec{n},q) \rangle\, &\approx&
-\frac{2\,\vec{q\,}^2 v_\rmF^4}{9\,\om^2}\,e_V a\, \rho\, \Im g_T(0,\om,0)\,.
\nonumber
\ee
Working in the leading order in $v_\rmF$ we have to put $\gamma_V\to 1$ in view of
Eq.~(\ref{gammaexp}). The function $g_T$ is introduced in Appendix \ref{app:loop}. From (\ref{gT1})
it follows that
\be
\Im g_T(0,\om,0)=-\frac{2\,\pi\, \Delta^2\,
\theta(\om-2\,\Delta)}{\om\, \sqrt{\om^2-4\, \Delta^2}}
\,\left(1-2\,n\big(\frac{\om}{2T}\big)\right)\,.\nonumber
\label{gt0}
\ee


\section{Neutrino emissivity in the neutron PBF process}\label{sec:PBF}

Once the averages (\ref{aver}) are known, expressions (\ref{emis}) and (\ref{Wneut}) solve the
problem of the neutrino emission from the superfluid neutron matter via the
PBF reactions.
The emissivity is given by
\be
\varepsilon_{\nu\nu} &=&
\varepsilon_{\nu\nu,V}+\varepsilon_{\nu\nu,A}\,,
\\
\varepsilon_{\nu\nu,a} &=& \frac{G^2}{48\, \pi^4}\,
\int^\infty_0\!\!\rmd \om\int_0^\om\!\!\rmd |\vec{q\,}|
\frac{\om\, \vec{q\,}^2}{e^{\om/T}-1}\, K_a(q)\,,
\label{emiss}
\nonumber\ee
where $K_a$ is the correlator function summed over the lepton spins and integrated over the
lepton phase space
\be
K_a(q) &=& \frac{3}{4\pi} \intop\rmd \Phi_l\Im \sum_{\rm spin}\chi_a^{\mu\nu}(q)\, .
\label{kappa}
\ee
The integration over the lepton phase space can be performed analytically
with the help of Eqs.~(\ref{Leptens-neut}) and (\ref{conserv}),  and we find, see Ref.~\cite{KV1} for
details,
\begin{align}
K_V(q) &= \frac{e_V\,g_V^2}{a} (\vec{q\,}^2-\om^2)\,
\nonumber\\
&\times\Im \langle\chi_{V,0}(\vec{n},q) - \vec{v}\, \vec{\chi}_{V,1}(\vec{n},q)\rangle_{\vec{n}}\,,
\label{kappaV:Ward}\\
K_A(q) &= \frac{e_A\,g_A^2}{a} \Im\big[
\vec{q\,}^2 \langle \vec{v}\, \vec{\chi}_{A,1}(\vec{n},q)\rangle_{\vec{n}}
\nonumber\\
&+ (3\,\om^2-2\, \vec{q\,}^2)\, \langle \chi_{A,0}(\vec{n},q)\rangle_{\vec{n}}
\nonumber\\
&-\om \langle \vec{q}\, \vec{\chi}_{A,1}(\vec{n},q)\rangle_{\vec{n}}
-\om\,\langle(\vec{q}\,\vec{v}\,)\, \chi_{A,0}(\vec{n},q)\rangle_{\vec{n}} \big]\,.
\label{kappaA}
\end{align}

For the neutron PBF emissivity on the vector current we obtain~\cite{KV1} (for one neutrino
flavor)
\be\label{eV}
\epsilon_{\nu\nu,V}^{\rm nPBF}&\simeq&\epsilon_{\nu\nu}^{(0n)}\,
g_V^2\,e_V^2\frac{4}{81}v_{\rmF,n}^4 \,,
\,\, g_V^2 e_V^2 =1.
\ee
Here
\be
&&\epsilon_{\nu\nu}^{(0n)}= \frac{4 \rho\,G^2 \Delta^7_n}{15\, \pi^3} I(\frac{\Delta}{T}),\,
\\
&&I(z)=\!\intop^\infty_1\! \frac{\rmd y\, y^5}{\sqrt{y^2-1}} \frac{1}{(e^{zy}+1)^2}\,
\nonumber
\label{emiss0}
\ee
is the result derived previously in Ref.~\cite{FRS76,VS87,SV87}.

Now let us turn to neutrino emissivity induced by the axial-vector current. In the expansion
$v_\rmF\ll 1$ the leading term contributing to the emissivity is of the order $v_\rmF^2$. Keeping
only the leading terms we cast Eq.~(\ref{kappaA}) as
\be
&&K_A \approx -g_A^2 e_A^2\,\rho\,v_\rmF^2\, \vec{q\,}^2\,
\nonumber\\
&&\times\big[1+(1-{\textstyle\frac23\, \frac{\vec{q}^{\,2}}{\om^2}})- {\textstyle \frac23}\big]\,
\Im g_T (0,\om,0).
\label{kappaA:2}
\ee
The correlation factors $\gamma_a$ contribute at the sub-leading order $\sim v_\rmF^4$, therefore
we neglected these terms in the approximate expression (\ref{kappaA:2}). As the result,
the neutron PBF emissivity induced by the axial-vector current is  (for one neutrino flavor)
given by
\be\label{eA}
\epsilon_{\nu\nu,A}^{\rm nPBF}&\simeq&
\left(1+\frac{11}{21}-\frac{2}{3}\right)\, g_A^{\,2}\,e_A^2
v_{\rmF,n}^2\,\epsilon_{\nu\nu}^{(0n)}\,.
\ee
The resulting emissivity is the sum of contributions (\ref{eV}) and (\ref{eA}). We stress that
Eqs.~(\ref{eV}) and (\ref{eA}) are approximate expressions obtained in the leading order in
$v_{\rmF}$. General result looks more cumbersome but it is easily recovered with the help of
Eqs.~(\ref{aver}). The latter equations are derived  in \cite{KV2} and here  at arbitrary temperature.


\section{Conclusion}

A.I.~Larkin and  A.B.~Migdal extended the Landau's Fermi-liquid  theory onto superfluid systems.
In this paper we re-formulated their approach for systems out of equilibrium. For that we used
Schwinger-Kadanoff-Baym-Keldysh formalism. Important improvements of the Larkin-Migdal approach
compared to the Nambu-Gorkov one are  that the former approach allows to deal with strong
interactions different  in the particle-hole and particle-particle channels. These achievements
have been used by A.J.~Leggett who generalized the Larkin-Migdal approach to describe strongly
interacting fermion superfluids at finite temperatures and applied it to description of superfluid
$^3$He. He used Matsubara diagram technique. The use of the  Schwinger-Keldysh diagram technique
allows to consider variety of non-equilibrium problems. In application to nucleon systems, in
general, the considered in this paper formalism can be applied to the paring in the states with an
arbitrary angular momentum; it operates with various forms of a nucleon-nucleon interaction: scalar,
spin-spin, spin-orbit and tensor interactions. As argued by A.B.~Migdal the tensor forces mediated
by the pion exchange should  enhance (pion softening) with increase of the nucleon density.
Inclusion of this effect might be important in the case of the $P$-pairing in neutron star
interiors.

We considered  the neutrino radiation from a finite piece of the nuclear matter bearing in mind
the problem of the neutron-star cooling. We used optical theorem formalism formulated in terms of
closed diagrams with the full fermion and boson Green's functions and the full nucleon-nucleon
interaction. The series of the diagrams is constructed with respect to the number, $N$, of the
full $G^{-+}$ fermion Green's functions. For simplification we considered a system which evolves
slowly in time and has small spatial gradients. This allowed us to perform the gradient expansion
after the Wigner transformation and keep only gradient-independent terms in calculations of
reaction rates. We demonstrated that in order to exactly satisfy the vector current conservation in the
nucleon pair breaking and formation processes it is not sufficient to include only one $N=1$ term
of the series, rather one needs to re-sum the RPA series including multi-piece diagrams. (The
multi-piece diagrams decay in more than two pieces, being cut through $(-+)$, $(+-)$ lines). This
demonstration shows, how one should separate one-nucleon, two-nucleon, etc. processes, in
accordance with exact conservation law of the vector current. Comparison of the RPA $\Sigma^{-+}$
self-energy with the $N=1$ contribution shows the accuracy with which one may deal, using only one
$N=1$ diagram.

Then we demonstrated how the developed formalism allows to calculate neutrino emissivity from the
piece of a warm nucleon matter in presence of the nucleon pairing. As simplest example we
calculated neutrino emissivity in the neutron pair breaking and formation  processes. These
processes are of one-nucleon origin. To simplify consideration, we focused on the case of the
ordinary $1S_0$ pairing of neutrons. More difficult is to calculate the emissivity of the
two-nucleon ($N=2$) processes, and $N\geq 3$ processes in the presence of pairing. The existing
nowadays results for the reaction rates in nucleon systems with pairing are based on the so-called
$R$ phase-space suppression factors used to reduce the production rates calculated without
pairing. Such an approach can be used only for rough estimations. The formalism formulated
in the present paper is fully suited to properly perform the calculations. It is also interesting
to search for new processes which might be open in the non-equilibrium and equilibrium medium
because of the interaction between different reaction channels. These questions require a separate
consideration. In the present paper we  focused on the neutrino radiation problem. However, the white
body radiation of other quanta can be considered in  similar way.

The calculated rate $\Sigma^{-+}$ can be considered as the gain term in the generalized kinetic
equation for the virtual $W/Z$ boson or for the anti-neutrino. For consistency then one needs to
include first-order gradient memory terms into the collision term.

Another important question is how to go beyond the quasi-particle approximation for fermions in
strongly interacting fermion systems with pairing.

Within the quasi-particle approximation for fermions, the formalism based on the Fermi-liqiud
renormalization is developed. To quantify the results it remains to know the Landau-Migdal
parameters. For the problems under consideration one needs to know them as functions of the
density, isospin composition, frequency and momentum. The information extracted from analysis of
atomic nucleus experiments is definitely insufficient for these purposes. Existing calculations of
the Landau-Migdal parameters are still incomplete. We hope that the present study will motivate
further attempts to extract these parameters. In spite of all difficulties, it seems to be the
cheapest way to achieve  understanding of many new interesting phenomena occurring in the strongly
interacting fermion systems in the presence of the pairing.

The direction of the research was shown in the works done in 50th--70th years of the
XXth century. The pioneering contribution to the development of the methods of the
quantum many-body theory including the problem of fermion pairing
belongs to Arkady~Migdal.


\begin{acknowledgments}
The work was partially supported by the DFG grant WA431/8-1, by the VEGA grant of the Slovak Ministry of
Education and by COMPSTAR, an ESF Research
Networking Programme.
\end{acknowledgments}
\appendix
\section{Matrix notation}\label{Contour}

The Schwinger-Keldysh contour in Fig.~\ref{fig:contour} consists of two branches: time-ordered and
anti-time ordered. For a given space-time coordinate $x$ the contour coordinates take two values,
$x^+$ and $x^-$, depending on the branch of the contour. The closed real-time contour integration can
be written as
\begin{eqnarray}
\label{C-int}
\int_{\mathcal{C}} d x^{\mathcal{C}} \dots  =\int_{t_0}^{\infty}d
x^-\dots -\int_{t_0}^{\infty}d x^+\dots,
\end{eqnarray}
where only the time limits are explicitly given and the spatial integration $\rmd^3 x$ is assumed.
The folding of two two-point functions defined on the contour reads as
\begin{eqnarray}\label{H=FG1}
&&H(x^i,y^k)=\int d z^{\cal C} \zeta(x^i,z^{\cal C})G(z^{\cal C},y^k)\,.
\end{eqnarray}

Any two-point function $\zeta=\left< \mathcal{T}_C
\widehat{A}(x)\widehat{B}(y)\right>$ being a function of two
contour variables $x$ and $y$ can be viewed as a matrix, which
elements are defined in dependence on their belonging to the
branches of the contour
\begin{eqnarray}
\zeta^{ij}(x,y)=\zeta(x^i,y^j), \quad i,j\in\{-,+\}\,.
\label{Fij}
\end{eqnarray}
For Green's functions this convention produces the following
matrices
\begin{align}
&i G^{ik}(x,y) =
\left(
\begin{array}{cc}
iG^{--}(x,y)& iG^{-+}(x,y)\\
iG^{+-}(x,y)& iG^{++}(x,y)
\end{array}\right)
\nonumber\\
&\qquad=
\left(\!\!\begin{array}{cc}
\langle\mathcal{T}\widehat{\Psi}(x)\widehat{\Psi}^{\dagger}(y)\rangle
&\mp \langle\widehat{\Psi}^{\dagger}(y)\widehat{\Psi}(x)\rangle\\
\langle\widehat{\Psi}(x)\widehat{\Psi}^{\dagger}(y)\rangle
&\langle\widetilde{\mathcal{T}}\widehat{\Psi}(x)\widehat{\Psi}^{\dagger}(y)\rangle
\end{array}\!\!\right),
\label{Fxy}
\end{align}
where $\mathcal{T}$ and $\widetilde{\mathcal{T}}$ are the usual time and anti-time ordering
operators and the upper sign is for fermions, the lower one is for bosons. Eq.~(\ref{Fxy}) implies
the following relations between non-equilibrium and usual retarded and advanced Green's functions
\begin{align}
\label{Fretarded}
G^R(x,y)&=G^{--}(x,y)-G^{-+}(x,y)
\nonumber\\
        &=G^{+-}(x,y)-G^{++}(x,y)
\nonumber\\
        &=\Theta(x_0-y_0)\big[G^{+-}(x,y)-G^{-+}(x,y)\big],
\nonumber\\
G^A(x,y)&=G^{--}(x,y)-G^{+-}(x,y)
\\
        &=G^{-+}(x,y)-G^{++}(x,y)
\nonumber\\
&=\Theta(y_0-x_0)\big[G^{-+}(x,y)-G^{+-}(x,y)\big],
\nonumber
\end{align}
where $\Theta(x_0-y_0)$ is the step function of the time
difference. The similar relations hold for the self-energies
\begin{align}
\label{Fretarded1}
\Sigma^R(x,y)&=\Sigma^{--}(x,y)+\Sigma^{-+}(x,y)
\nonumber\\
&=-\Sigma^{+-}(x,y)-\Sigma^{++}(x,y)
\nonumber\\
&=\Theta(x_0-y_0)\big[-\Sigma^{+-}(x,y)+\Sigma^{-+}(x,y)\big],
\nonumber\\
\Sigma^A(x,y)&=\Sigma^{--}(x,y)+\Sigma^{+-}(x,y)
\\
&=-\Sigma^{-+}(x,y)-\Sigma^{++}(x,y) \nonumber\\
&=\Theta(y_0-x_0)\big[\Sigma^{+-}(x,y)-\Sigma^{-+}(x,y)\big]\,.
\nonumber
\end{align}
The difference in signs  compared to (\ref{Fretarded}) is due to the fact that $\Sigma$ includes
vertices, and $-$ and $+$ vertices differ by the sign.

In terms of matrices defined as in Eq.~(\ref{Fij}) the contour integration in the folding of two
functions (\ref{H=FG1}) turns into the usual space-time coordinate integration applied to the
product of  matrix-valued functions
\begin{eqnarray}
H(x^i,y^k)=\sum_{j=+,-}\int d z \zeta^{ij}(x,z)\,\eta^{ij}\,G^{jk}(z,y)\,,
\label{H=FG2}
\end{eqnarray}
where $\eta^{ij}$ is the diagonal matrix with the elements $\eta^{--}=-\eta^{++}=1$.
It takes into account the extra minus sign of the anti-time-ordered branch of the contour.

Applying hermitian conjugation operation to the Green's function definitions (\ref{Fxy}) we obtain
the following relations
\begin{eqnarray}
&&\big[G^{\pm\mp}(x,y)\big]^\dag = -G^{\pm\mp}(y,x)\,,
\nonumber\\
&&\big[G^{--}(x,y)\big]^\dag =  -G^{++}(y,x)\,,
\nonumber\\
&&\big[G^R(x,y)\big]^\dag = G^A(y,x)\,,
\label{Gconjugat}
\end{eqnarray}
and the analogous ones for the self-energies.

Instead of purely coordinate representation for two-point functions it can be more convenient to
perform the Wigner transformation and operate in the mixed coordinate-momentum representation. For
any two-point function $\zeta(x,y)$ one introduces the relative coordinate $\xi=x-y$ and the middle
coordinate $X=\frac12(x+y)$ and makes the Fourier transformation from four-space coordinate $\xi$ to
four-momentum $p$:
\be
\zeta^{ij}(X;p)=\intop \rmd \xi e^{i\, p\xi}\,
\zeta^{ij}(X+\xi/2,X-\xi/2) \label{Wigner-def} \ee
with $i,j\in\{+,-\}$\,. The quantities $G^{+-}$ and $G^{-+}$ are called the Wigner
densities in the eight dimensional $(X,p)$ phase-space.
In the mixed Wigner representation Eqs.~(\ref{Gconjugat}) read
\begin{eqnarray}
&&\big[G^{\pm\mp}(X;p)\big]^\dag = -G^{\pm\mp}(X;p)\,,
\nonumber\\
&&\big[G^{--}(X;p)\big]^\dag =  -G^{++}(X;p)\,,
\nonumber\\
&&\big[G^R(X;p)\big]^\dag = G^A(X;p)\,.
\label{Gconjugat-p}
\end{eqnarray}
In particular, these relations imply that the functions $i\,G^{\pm\mp}$ are always real.

The merit of the Wigner transformation is that it allows to set up an approximation scheme for
treating the system being not to far out of equilibrium. For slightly inhomogeneous and slowly
evolving systems, the degrees for freedom can be subdivided into rapid and slow ones. Then, the
variable $\xi$ relates to rapid and short-ranged microscopic processes and the variable $X$ refers
to slow and long-ranged collective motions. A gradient expansion with respect to slow degrees of
freedom can be applied since the Wigner transformation converts any convolution of two-point
functions into a product of the corresponding Wigner functions plus first-order and higher-order gradient terms
\begin{multline}
\intop \rmd \xi e^{ip\xi}\intop\rmd z \zeta(x,z)\, G(z,y)
\\
=
\Big[e^{\frac{i}{2}\hbar (\partial_p\partial_{X'}-\partial_X\partial_{p'})} \zeta(X;p)G(X';p')
\Big]\Big|_{p'=p,X'=X}
\\
\approx \zeta(X;p)G(X;p)+\frac{i\hbar}{2}\Big[
\frac{\partial\zeta}{\partial p^\mu}
\frac{\partial G}{\partial X_\mu}
-
\frac{\partial \zeta}{\partial X^\mu}
\frac{\partial G}{\partial p_\mu}
\Big].
\label{grad-exp}
\end{multline}
Including local and only first-order gradient terms, one derives from the set of the  Dyson
equations the Kaddanoff-Baym generalized kinetic equation describing slow   evolution of slightly
spatially inhomogeneous system of particles having non-zero mass widths, see \cite{KB62,IKV00}.
Although the gradient terms are assumed to be smaller than the local terms they should be kept not
only in the Vlasov part of the generalized kinetic equation but also in the collision term giving
rise to the memory effects \cite{IKV00}.

With the help of the relations (\ref{Gconjugat-p})
we can prove the convenient relation between the functions
$\widetilde{L}^{ij}$ introduced in Eq.~(\ref{Ldef}):
\be
&&\big[\widetilde{L}^{\pm\mp}(p;q)\big]^\dag=-\widetilde{L}^{\pm\mp}(p;q),
\nonumber\\
&&\big[\widetilde{L}^{--}(p;q)\big]^\dag=-\widetilde{L}^{++}(p;q)\,.
\label{Lconjugat}
\ee
From Eq.~(\ref{Fretarded}) follows, {\it inter alia}, that among four quantities, e.g.,
$G^{\pm\mp}$ and $G^{\pm\pm}$, not all are independent, because of the
relation\footnote{Actually in generalized kinetics only two real
quantities are independent, e.g. $iG^{-+}$ and $A=-2\Im G^R$.
Other quantities, $G^{+-}$, $G^{--}$, $G^{++}$ are expressed
through them \cite{IKV00}. In equilibrium only one quantity is
independent.}
$G^{++}+G^{--}-G^{+-}-G^{-+}=0$. The similar holds for self-energies
$\Sigma^{++}+\Sigma^{--}+\Sigma^{+-}+\Sigma^{-+}=0$, see. Eq.~(\ref{Fretarded1}).
Let us check whether the similar completeness relation holds also for the $\widetilde{L}$
functions. Applying Eqs.~(\ref{Fretarded}) and~(\ref{Fretarded1}) recursively we find
\begin{align}
&i(\widetilde{L}^{++}+\widetilde{L}^{--}-\widetilde{L}^{+-}-\widetilde{L}^{-+})=
\nonumber\\
&\,\,\,=G_{p+q/2}^{R}\, \big[G_{p-q/2}^A+G_{p-q/2}^{+-}-G_{p-q/2}^{-+}\big]
\nonumber\\
&
\,\,\,+\big[G_{p+q/2}^R+G_{p+q/2}^{-+}-G_{p+q/2}^{+-}\big]\, G^A_{p-q/2}
\nonumber\\
&\,\,\,=G^R_{p+q/2}\, G^R_{p-q/2} + G^A_{p+q/2}\, G^A_{p-q/2}.
\label{completeL}
\end{align}
Note that the last two terms  vanish after the integration over $p_0$,
\be
\intop\rmd p_0 G^R_{p+q/2}\, G^R_{p-q/2}=\intop\rmd p_0 G^A_{p+q/2}\, G^A_{p-q/2}=0,
\nonumber\\
\label{RR-LL}
\ee
since the poles of both integrated functions lie in one and the same semi-plane  of the complex plane --- $\Im
p_0<0$ for the retarded functions and $\Im p_0>0$ for the advanced ones --- and the integration
contour can be close in the opposite semi-plane. Thus the relation
\be
\intop\rmd p_0
(\widetilde{L}^{++}+\widetilde{L}^{--}-\widetilde{L}^{+-}-\widetilde{L}^{-+})=0\,
\label{Lcompl-int}
\ee
holds for $\widetilde{L}$'s but only after the $p_0$-integration.


\section{Lepton tensor} \label{app:lept-tensor}

Here we calculate the leptonic tensors entering the reaction probabilities
(\ref{Wneut},\ref{Wcharg}).
We will assume that lepton 1 is massive
$\om_1=\sqrt{m_1^2+|\vec{q}_1|^2}$,
whereas  lepton 2 is massless $\om_2=|\vec{q}_2\,|$.
The trace of lepton currents is given by
\begin{multline}
\sum_{\rm spin}(l_\mu\, l_\nu^\dagger) =
2\,{\rm Tr}\{q_1\!\!\!\!\!/\,\gamma_\mu\, q_2\!\!\!\!\!/\,\gamma_\nu\,(1-\gamma_5)\}
\nonumber\\
=
8\,\big( q_{1\mu}\,q_{2\nu}  + q_{2\mu}\,q_{1\nu}- (q_1\cdot
q_2)\, g_{\mu\nu}
\nonumber\\
 + i\, \epsilon_{\alpha\mu\beta\nu} q_1^\alpha\, q_2^\beta\big)\,.
\nonumber
\end{multline}

The integral over the lepton phase space
\be
I^{\mu\nu}=\intop\frac{\rmd^3 q_1}{2\,\om_1}\, \frac{\rmd^3 q_2}{2\,\om_2}
q_1^\mu\, q_2^\nu\, \delta^{(4)}(q_1+q_2-q)
\nonumber
\ee
can be calculated in the standard
way.  The result is as follows:
\begin{align}
I^{\mu\nu} =\frac{\pi}{24}&
\Big[\Big(1+2\frac{m_1^2}{q^2}\Big)\,\big(2\,q^\mu\, q^\nu+
q^2\,g^{\mu\nu}\big) -3\,m_1^2\, g^{\mu\nu} \Big] \nonumber\\
&\times\Big(1-\frac{m_1^4}{q^4}\Big)\theta(\om)\,\theta(q^2-m^2_1)\,.
\\ \nonumber
\end{align}
The lepton tensor in Eq.~(\ref{Wneut}) is given by
\begin{align}
T_{\rm lept}^{\mu\nu}(q)=&8\,
\big(I^{\mu\nu}+I^{\nu\mu}-g^{\mu\nu}\, I^\mu_\mu\big) \nonumber\\
=&\frac{4\pi}{3} \Big[\Big(1+2\frac{m_1^2}{q^2}\Big)\,\big(q^\mu\,
q^\nu-q^2\,g^{\mu\nu}\big) +\frac32\,m_1^2\, g^{\mu\nu} \Big]
\nonumber\\
&\times\Big(1-\frac{m_1^4}{q^4}\Big)\theta(\om)\,\theta(q^2-m^2_1)\,.
\label{Leptens-neut}
\end{align}
Setting $m_1 =0$ we obtain expression for the $T_{\rm
lept}^{\mu\nu}(q)$ valid for neutral currents (for $\nu\bar\nu$).

For the process occurring on the charged current the reaction
rates enters the integral:
\begin{multline}
\intop\frac{\rmd^3 q_1}{2\,\om_1}\, \frac{\rmd^3 q_2}{2\,\om_2}
q_1^\mu\, q_2^\nu\, [1-n_f(q_1\cdot u)]\delta^{(4)}(q_1+q_2-q)
\\
=I^{\mu\nu}-J^{\mu\nu}\,, \nonumber
\end{multline}
where second term appeared  due to the Pauli blocking for the
charged leptons (electrons or muons). Here $u$ is a four-vector of
a collective motion of the medium, in the co-moving frame
$u=(1,0,0,0)$\,, $q_1\cdot u =q_1^\mu u_\mu$. Using two four-vectors $q$
and $u$ we write the general structure of $J^{\mu\nu}$
as
\begin{align}
J^{\mu\nu}&=
A\, q^\mu\, q^\nu+ B\, q^\mu\, u^\nu+ C\, u^\mu\, q^\nu
\nonumber\\
&+ D\, u^\mu\, u^\nu+ E\, g^{\mu\nu}\,,
\label{J-AB}
\end{align}
or alternatively as
\begin{align}
J^{\mu\nu}&= C_{qq}\, t_q^{\mu\nu}+C_{uu}\, t_u^{\mu\nu}+ C_{uq}\,
j_{u}^\mu\,j_q^\nu \nonumber\\ &+ C_{qu}\, j_q^\mu\, j_u^\nu+
C_g\, g^{\mu\nu}\,, \label{J-CC}
\end{align}
where we use tensors and four-vectors
\begin{align}
&t_q^{\mu\nu}=q^2\,g^{\mu\nu}-q^\mu\, q^\nu\,, \quad
t_u^{\mu\nu}=g^{\mu\nu}-u^\mu\, u^\nu\,,
\nonumber\\
&j_q^\mu=q^\mu-(u\cdot q)\, u^\mu\,,\quad
j_u^\mu= u^\mu\, q^2-(u\cdot q)\, q^\mu\,,\quad
\nonumber
\end{align}
which satisfy relations
\begin{align}
& t_q^{\mu\nu}\, q_\nu=q_\nu\, t_q^{\nu\mu}=0
\,,\,\, t_u^{\mu\nu}\, u_\nu=u_\nu\, t_u^{\nu\mu}=0
\nonumber\\
&  t^{\mu\nu}_q u_\mu u_\nu=t^{\mu\nu}_u q_\mu q_\nu = q^2-(u\cdot q)^2\,,
\nonumber\\
& t_{q,\mu}^\mu=3\, q^2\,,\,\, t_{u,\mu}^\mu=3\,,\,\, j^\mu_q u_\mu=j_u^\mu q_\mu=0\,,
\nonumber\\
& j^\mu_q q_\mu=j_u^\mu u_\mu= q^2-(u\cdot q)^2\,,\,\,
\nonumber\\
& j_q^\mu j_{u,\mu}= (u\cdot q)\, [(u\cdot q)^2-q^2]\,.
\label{tj-relat}
\end{align}
Coefficients in Eqs.~(\ref{J-AB}) and (\ref{J-CC}) are related as
\be
A&=& -C_{qq} - (C_{qu} - C_{uq}) (u\cdot q),
\nonumber\\
B&=& C_{qu}\, q^2 + C_{uq} (u\cdot q)^2,
\nonumber\\
C&=& C_{uq}\, q^2 + C_{qu} (u\cdot q)^2,
\nonumber\\
D&=& -C_{uu} - (C_{qu}+C_{uq})\, q^2\, (u\cdot q),
\nonumber\\
E&=& C_g + C_{uu} + C_{qq}\, q^2\,.
\ee
Using Eq.~(\ref{tj-relat})  from Eq.~(\ref{J-CC}) we derive:
\begin{align}
J^{\mu}_{\mu}=&3\,( q^2\, C_{qq} + C_{uu}) +
\Omega\, Q^2\, ( C_{uq}+ C_{qu}) + 4\,C_g\,,
\nonumber\\
& J^{\mu\nu} q_{\mu}q_{\nu}=J_{qq} = -C_{uu}\, Q^2 + C_g\, q^2\,,
\nonumber\\
& J^{\mu\nu} u_{\mu}q_{\nu}=J_{uq} = \phantom{-}C_{uq}\, Q^4 + C_g\,
\Omega\,,
\nonumber\\
& J^{\mu\nu} q_{\mu}u_{\nu}=J_{qu} = \phantom{-}C_{qu}\, Q^4 + C_g\,
\Omega\,,
\nonumber\\
& J^{\mu\nu} u_\mu u_\nu =J_{uu} = -C_{qq}\, Q^2 + C_g\,,
\nonumber\\
&Q^2=(u\cdot q)^2-q^2\,,\quad \Omega =(u\cdot q)\,.
\label{syst-CC}
\end{align}
Explicit expressions for all  $C_{ij}$-coefficients look clumsy
and we do not present them. However, they are easily written through
the coefficient $C_g$, which equals to
\be
C_g=\frac{3 (J_{qq}+q^2\, J_{uu})+Q^2 J^\mu_\mu -\Omega
(J_{qu}+J_{uq})} {2\,(2\,\Omega^2+ Q^2)}\,. \label{Cg} \ee
From the definition of the tensor $J^{\mu\nu}$ we can easily find
that the various convolutions with the vectors $q$ and $u$ and the
trace $J^\mu_\mu$ can be expressed as
\be
&&J_\mu^\mu={\textstyle\frac12}(q^2-m_1^2)\, I_0\,,
\,\, J_{uu}=\Omega\, I_1-I_2\,,
\nonumber\\
&&J_{qq}= {\textstyle\frac12}(q^4-m_1^4)\,I_0\,,\,\,
J_{uq}= {\textstyle\frac12}(q^2-m_1^2)\,I_1\,,
\nonumber\\
&&J_{qu}={\textstyle\frac12}(q^2+m_1^2)\big[\Omega\,I_0 -I_1\big]\,,
\label{Js-scalInt}
\ee
where there appeared  tree scalar integrals
\begin{align}
I_n&=
\intop\frac{\rmd^3 q_1}{2\,\om_1}\, \frac{\rmd^3 q_2}{2\,\om_2}
(u\cdot q_1)^n\, n(u\cdot q) \delta^{(4)}(q_1+q_2-q)\,
\nonumber\\
&=\pi\,\theta(q^2-m_1^2)\theta(\om)
\nonumber\\
&\quad\times\intop_{\max\{m_1,\om_-(q)\}}^{\min\{\om,\om_+(q)\}}
\!\!\!\!\rmd \om_1 n(\om_1)\, \om_1^n\sqrt{\om_1^2-m_1^2}\,,
\nonumber\\
&\quad
\om_\pm(q)=\frac{\om\, (q^2-m_1^2)}{2\,q^2} \pm
|\vec{q}\,|\,\sqrt{\frac{q^4+m_1^4}{4\,q^4}}\,,
\label{In}
\end{align}
$n=0,1,2$.
Finally, we can construct the lepton tensor for charged lepton currents
\be
T_{\rm lept}^{\mu\nu}&=&8\, (I^{\mu\nu}+I^{\nu\mu}-g^{\mu\nu}\, I^\mu_\mu)
\nonumber\\
&-&8\, (J^{\mu\nu}+J^{\nu\mu}-g^{\mu\nu}\, J^\mu_\mu),
\nonumber\ee
where
\begin{align}
& 8\, (J^{\mu\nu}+J^{\nu\mu}-g^{\mu\nu}\, J^\mu_\mu)
= 2\,C_{qq}\, t_q^{\mu\nu}+2\, C_{uu}\, t_u^{\mu\nu}
\nonumber\\
&\quad+ (C_{uq}+C_{qu})\, (j_{u}^\mu\,j_q^\nu+j_q^\mu\, j_u^\nu) + 2\,C_g\, g^{\mu\nu}
\nonumber\\
&\quad -\big(C_{qq}\, 3\, q^2 +C_{uu}\, 3 + (C_{uq}+C_{qu})\,\Omega\, Q^2 + C_g\, 4\big)\, g^{\mu\nu}\,.
\nonumber
\end{align}
For the tensor $\widetilde{T}_{\rm lept}^{\mu\nu}$ we can use the same expression and the
relations for the $C_{ij}$-coefficients with the only replacements
$J^\mu_\mu\to \widetilde{J}^\mu_\mu$ and $J_{ab}\to \widetilde{J}_{ab}$, $a,b=q,u$, where
\be
&&\widetilde{J}_\mu^\mu={\textstyle\frac12}(q^2-m_1^2)\, I_1\,,
\,\, \widetilde{J}_{uu}=\Omega\, I_2-I_3\,,
\nonumber\\
&&\widetilde{J}_{qq}= {\textstyle\frac12}(q^4-m_1^4)\,I_1\,,\,\,
\widetilde{J}_{uq}= {\textstyle\frac12}(q^2-m_1^2)\,I_2\,,
\nonumber\\
&&\widetilde{J}_{qu}={\textstyle\frac12}(q^2+m_1^2)\big[\Omega\,I_1
-I_2\big]\,.
\label{Js-scalInt-tilde}
\ee


\section{Equilibrium Relations}\label{eq.rel}

The equilibrium Kubo-Schwinger-Martin relations between $(-+)$ and $(+-)$
fermion Green's functions and the boson self-energies  are
\begin{align}
G^{-+}(p) &= \mp G^{+-}(p) e^{-\varepsilon/T},
\nonumber\\
\Sigma^{-+}(q) &= \pm \Sigma^{+-}(q) e^{-\epsilon/T},
\label{KMS-G}
\end{align}
see Ref.~\cite{KMS}, where $\varepsilon = p_0 - \mu$ with the chemical potential
related to the conserved charge. In the case considered in the given
paper $\mu \neq 0$ for fermions except  for
neutrinos/antineutrinos which freely escape from the piece of
matter and $\mu =0$ for bosons.
All the  Green's functions can be expressed through the retarded and advanced Green's functions
\begin{subequations}
\begin{align}
G^{--}(p)&= G^R(p) \pm i\,n(\varepsilon) A(p),
\label{Geq:mm}\\
G^{++}(p)&=-G^A(p) \pm i\,n_f(\varepsilon) A(p),
\label{Geq:pp}\\
G^{-+}(p)&=\pm i\, n(\varepsilon)\, A(p),
\label{Geq:mp}\\
G^{+-}(p)&=-i\, [1\mp n(\varepsilon)]\, A(p),
\label{Geq:pm}
\end{align}
\label{Geq}
\end{subequations}
where  $A=2\,\Im G^A=-2\,\Im G^R $ is the spectral density and
\be
n(\epsilon) =1/(\exp(\epsilon/T)\pm 1)\,.
\label{F-dist}
\ee
Analogously for the  self-energies we have
\begin{subequations}
\begin{align}
\Sigma^{--}(p)&= \Sigma^R(p) \pm i\, n(\epsilon)\, \Gamma(p),
\label{Seq:mm}\\
\Sigma^{++}(p)&=- \Sigma^A(p)\pm i\, n(\epsilon)\, \Gamma(p),
\label{Seq:pp}\\
\Sigma^{-+}(p)&= \mp i\, n(\epsilon)\, \Gamma(p),
\label{Seq:mp}\\
\Sigma^{+-}(p)&= i\, [1\mp n(\epsilon)]\,\Gamma(p),
\label{Seq:pm}
\end{align}
\label{Seq}
\end{subequations}
where $\Gamma=2\, \Im\Sigma^A=-2\, \Im\Sigma^R$ is the width.

Now, using (\ref{Lconjugat})  we derive the equilibrium relations among the products of the
fermion Green's functions $\widetilde{L}^{ij}$  and the functions introduced as
$\widetilde{L}^R=\widetilde{L}^{--}-\widetilde{L}^{-+}$ and
$\widetilde{L}^A=\widetilde{L}^{-+}-\widetilde{L}^{++}=[\widetilde{L}^R]^\dag$, see
Eqs~(\ref{LR},\ref{LA}). Making use of Eq.~(\ref{KMS-G}) we immediately find
\begin{align}
i\widetilde{L}^{-+}(p;q)&=G^{-+}_{p+q/2}\,G^{+-}_{p-q/2}=
G^{+-}_{p+q/2}\,G^{-+}_{p-q/2}\, e^{-\frac{\om}{T}}
\nonumber\\
&=i\widetilde{L}^{+-}(p;q)\, e^{-\frac{\om}{T}}\,.
\label{Lpm-KMS}
\end{align}
From Eq.~(\ref{LcomplG0}) follows
\begin{align}
\mathcal{G}_0\cdot (\widetilde{L}^{-+}-\widetilde{L}^{+-})\cdot \mathcal{G}_0&=
\mathcal{G}_0\cdot (\widetilde{L}^{A}-\widetilde{L}^{R})\cdot \mathcal{G}_0
\nonumber\\
&=
-2i\Im(\mathcal{G}_0\cdot\widetilde{L}^R\cdot\mathcal{G}_0)\,.
\label{LA-LR}
\end{align}
Combining Eqs.~(\ref{Lpm-KMS}) and (\ref{LA-LR}) we obtain
\be
\mathcal{G}_0\cdot\widetilde{L}^{-+}\cdot \mathcal{G}_0=
\frac{2\, i\, \Im(\mathcal{G}_0\cdot\widetilde{L}^R\cdot\mathcal{G}_0)}{
e^{\frac{\om}{T}}-1}
\label{Lmp-KMS}
\ee
and recover Eq.~(\ref{loop-equil:mp}). Equation~(\ref{loop-equil:pm}) follows then from
Eqs.~(\ref{Lmp-KMS}) and (\ref{Lpm-KMS}). From the definition of the retarded function and
Eq.~(\ref{LcomplG0}) we have
\begin{align}
\mathcal{G}_0\cdot\widetilde{L}^{--}\cdot \mathcal{G}_0 &=
\mathcal{G}_0\cdot\widetilde{L}^{R}\cdot \mathcal{G}_0 +
\mathcal{G}_0\cdot\widetilde{L}^{-+}\cdot \mathcal{G}_0\,,
\nonumber\\
\mathcal{G}_0\cdot\widetilde{L}^{++}\cdot \mathcal{G}_0 &=
\mathcal{G}_0\cdot\widetilde{L}^{+-}\cdot \mathcal{G}_0 -
\mathcal{G}_0\cdot\widetilde{L}^{R}\cdot \mathcal{G}_0\,,
\label{Lrelat}
\end{align}
which together with Eqs.~(\ref{Lmp-KMS}) and (\ref{Lpm-KMS}) translate into
Eqs.~(\ref{loop-equil:mm}) and (\ref{loop-equil:pp}).

For completeness,
we list also the  fermionic quasi-particle Green's functions for a system with paring in equilibrium
\begin{align}
\widehat{G}^{-+}(\epsilon,\vec{p}\,) &= 2\pi i\, a\, n_f(\epsilon)
\nonumber\\
&\times
[u_p^2 \delta (\epsilon -E_p )+v_p^2 \delta (\epsilon +E_p )]\sigma_0 ,
\nonumber\\
\widehat{F}^{(1,2)-+}(\epsilon,\vec{p}\,) &= 2\pi i \,a\,n_f(\epsilon) \frac{\widehat{\Delta}^{(1,2)}(p)}{4\, E_p^2}
\nonumber\\
&\times [\delta (\epsilon +E_p )- \delta (\epsilon -E_p )],
\nonumber\\
\widehat{G}^{+-}(\epsilon,\vec{p}\,) &=-2\pi i\, a\, [1-n_f(\epsilon)]
\nonumber\\
&\times[u_p^2 \delta (\epsilon -E_p )+v_p^2 \delta (\epsilon +E_p )]\sigma_0 ,
\nonumber\\
\widehat{F}^{(1,2)+-}(\epsilon,\vec{p}\,) &=2\pi i a [1-n_f(\epsilon)] \frac{\widehat{\Delta}^{(1,2)}(p)}{4\, E_p^2}
\nonumber\\
&\times [\delta (\epsilon -E_p )- \delta(\epsilon +E_p )] .
\nonumber
\end{align}
Here the quasi-particle spectrum is given by $E_p^2 = \epsilon_p^2 +\Delta^2(p,T)$ with
$$\Delta^2(p)=-a^2\frac12{\rm Tr}\{\hat{\Delta}^{(1)}(p)\hat{\Delta}^{(2)}(p)\}$$ and
$\epsilon_p =(p^2-p^2_{\rm F})/(2\,m^*_N)$\,.
The Bogolyubov's factors are
\be
u_p^2 =\frac{E_p +\epsilon_p}{2E_p},\quad v_p^2 =\frac{E_p
  -\epsilon_p}{2E_p}.
\ee
For the Green's functions for holes
\be
\widehat{G}^{h\,\pm\mp}(p)=\sigma_2
\widehat{G}^{\pm\mp}(-p)\sigma_2\,.
\nonumber
\ee
Expressions for normal
systems are obtained from here for $v_p\to 0$.


\section{The loop functions}
\label{app:loop}

At zero temperature the loop functions (\ref{depend}) were
calculated in Ref.~\cite{VGL62} using the Feynman method for the
integral of the Green's function products
 \be
L(\vec{n},q; P) &=& a^2\, \rho\, \Big[\frac{\vec{q}\, \vec{v}}{\om- \vec{q}\,\vec{v}}\,(1- g(z))-\frac{g(z)}{2}\,
(1+P)\Big]\,,
\nonumber\\
M(\vec{n},q) &=& -a^2\,\rho\,\frac{\om+\vec{q}\,\vec{v}}{2\, \Delta}\, g(z)\,,
\nonumber\\
N(\vec{n},q) &=& a^2\, \rho\, \frac{\om^2-(\vec{q}\,\vec{v})^2}{4\, \Delta^2}\, g(z)\,,
\nonumber\\
O(\vec{n},q;P) &=& a^2\, \rho\, \Big[ \frac{\om+\vec{q}\,\vec{v}}{4\, \Delta}+
\frac{\om-\vec{q}\,\vec{v}}{4\, \Delta}\, P \Big]\, g(z)\,.
\label{loopsT0}
 \ee
Here the variable $P$ can be an operator as in Eq.~(\ref{depend}) or $(\pm 1)$ value as in
Eq.~(\ref{widetilde}).
The universal function $g$ is  given by
\be
g&=&2\intop\rmd \Phi_0\, \frac{F_{+}\,F_-}{a^2\, \rho}
=\intop \frac{2\,\Delta^2\,\rmd\Phi_0}{[\epsilon_+^2-E_+^2]\,[\epsilon_-^2-E_-^2]}\,,
\nonumber
\ee
where $\intop\rmd \Phi_0$ is $\intop\rmd \Phi_T$ at $T=0$, see Eq. (\ref{dotoper});
$\epsilon_\pm=\epsilon\pm{\textstyle \frac12}\om$ and $E_\pm=E_{p\pm q/2}$\,.
Evaluation of the integral yields
\be
g(z)&=& -\frac{{\rm arcsinh}\sqrt{z^2-1}}{z\, \sqrt{z^2-1}}
-\frac{i\, \pi\, \theta(z^2-1)}{2\, z\,\sqrt{z^2-1}}\,,
\label{gfunc} \\
z^2&=&\frac{\om^2-(\vec{q}\, \vec{v})^2}{4\, \Delta^2}>1\,,
\quad \vec{v}=v_\rmF \, \vec{n}.
\nonumber
\ee
In various limiting cases one obtains
\begin{align}
&g(z)\simeq 1-z^2 /2 \quad {\rm{for}}\quad |z| \ll 1 ,
\nonumber\\
&g(z)\rightarrow  -i\frac{\pi}{2|z|\sqrt{-1-z^2}} \quad{\rm for}\quad z^2 \rightarrow -1,
\nonumber\\
&g(z)\rightarrow \frac{\ln (2z)}{z^2} \quad {\rm{for}}\quad z^2
\rightarrow \infty  ,
\nonumber\\
&g(z)\rightarrow -\frac{\ln (2|z|)}{|z^2|} -i\frac{\pi}{2|z^2|} \quad {\rm{for}} \quad -z^2
\rightarrow \infty,
\label{glim}
\end{align}
and for
$$ \frac{(\vec{q}\,\vec{v}_{\rm F} )^2
}{8\Delta} <|\om -2\Delta |\ll 2\Delta$$
we  have
\begin{align}
g(z)\simeq& -i\frac{\pi\sqrt{\Delta}}{2|z|\sqrt{\om -2\Delta
  }}\Big(1+\frac{(\vec{q}\,\vec{v}_{\rm F} )^2}{8\Delta (\om -2\Delta )}\Big)
-1
\nonumber\\
&+ O \left[ \Big(\frac{(\vec{q}\,\vec{v}_{\rm F} )^2}{\Delta (\om -2\Delta)}\Big)^2 \right].
\end{align}

At finite temperatures
the Feynman method does not work~\cite{W93} and the Matsubara
technique can be used instead. In Ref.~\cite{KV2} it was shown that the functions
$L$, $M$, $N$, and $O$ can be expressed through one universal temperature-dependent function
\be
&&g_T(\vec{n},i\om_m,\vec{q}\,)=2\intop\rmd \Phi_T\,\frac{F_{+}\,F_-}{a^2\, \rho}=2\,\Delta^2
\\&&\times
\intop_{-\infty}^{+\infty}\rmd\epsilon_p\sum_{n=-\infty}^{+\infty}
\frac{T}{[(i\epsilon_n+i\om_m)^2-E_+^2]\,[(i\epsilon_n)^2-E_-^2]}.
\nonumber\label{gT}
\ee
Here $\epsilon_n=(2\, n+1)\, \pi\, T$ and $\om_m=2\, m\, \pi\, T$\,.
After the summation over $n$ we obtain
\be
&&g_T(\vec{n},i\om_n,\vec{q}\,)= \Delta^2 \intop_{-\infty}^{+\infty}\rmd\epsilon_p
\nonumber\\
&&\times\Bigg[
\frac{(E_+ - E_-)}{E_+\, E_-}\,
\frac{(n_f(E_-)-n_f(E_+))}{(i\om_m)^2-(E_+ - E_-)^2}
\nonumber\\
&&- \frac{(E_+ + E_-)}{E_+\, E_-}\, \frac{(1- n_F(E_-)-
n_f(E_+))}{(i\om_m)^2 - (E_+ + E_-)^2}
\Bigg]\,.
\label{gT1}
\ee
After the replacement $i\om_m\to
\om_+=\om+i\, 0$ we obtain the analytical continuation to the
retarded $g_T$ function in the $\om$-complex plain.
The expressions for the loop functions read
\begin{align}
M(\vec{n\,},\om,\vec{q\,}) &=
-a^2\,\rho\,\frac{\om+\vec{q}\,\vec{v}}{2\, \Delta}\,g_T(\vec{n\,},\om,\vec{q\,})\,,
\nonumber\\
N(\vec{n\,},\om,\vec{q\,}) &=
a^2\, \rho\, \frac{\om^2-(\vec{q}\,\vec{v})^2}{4\, \Delta^2}\,g_T(\vec{n\,},\om,\vec{q\,})\,,
\nonumber\\
O(\vec{n\,},\om,\vec{q\,};P) &=
a^2\, \rho\, \Big[ \frac{\om+\vec{q}\,\vec{v}}{4\, \Delta}+ \frac{\om-\vec{q}\,\vec{v}}{4\, \Delta}\, P \Big]\,
\nonumber\\
&\times
g_T(\vec{n\,},\om,\vec{q\,})\,,
\nonumber\\
L(\vec{n\,},\om,\vec{q\,};P)&=
a^2\, \rho\, \Big[\frac{\vec{v}\, \vec{q}}{\om- \vec{v}\,\vec{q}}\,
\big( g_T(\vec{n\,},(\vec{v\,}\vec{q\,}),\vec{q\,})
\nonumber\\
& - g_T(\vec{n\,},\om,\vec{q\,})\big)
  -\frac{1+P}{2}\,g_T(\vec{n\,},\om,\vec{q\,})\Big]\, .
\nonumber\\
\label{loopsT}
\end{align}
For $T=0$, $g_T(\vec{n\,},(\vec{v\,}\vec{q\,}),\vec{q\,})=1$ and the old result (\ref{loopsT0}) is
recovered.



\begin{thebibliography}{99}
\bibitem{Landau58}
L.D. Landau, Zh.\ Eksp.\ Teor.\ Fiz.\ {\bf 30}, 1058 (1956) [Sov. Phys. JETP {\bf 3}, 920 (1956)];
Zh.\ Eksp.\ Teor.\ Fiz.\ {\bf 32}, 59 (1957) [Sov. Phys. JETP {\bf 5}, 1011 (1957)];
Zh.\ Eksp.\ Teor.\ Fiz.\ {\bf 35}, 97 (1958) [Sov. Phys. JETP {\bf 8}, 70 (1959)];
also in {\it Collected Papers of Landau}, ed. Ter Haar (Gordon \& Breach, 1965) papers 75--77.

\bibitem{LP1981}
E.M. Lifshitz and L.P. Pitaevskii, {\it Statistical Physics, Part 2}, Pergamon, 1980.

\bibitem{Mjump}
A.B. Migdal,
Zh.\ Eksp.\ Teor.\ Fiz.\ {\bf 32}, 399 (1957) [Sov. Phys. JETP 5, 333 (1957)].

\bibitem{Mqp}
V.M. Galitsky and A.B. Migdal, Sov. Phys. JETP {\bf 7}, 96 (1958);
A.B. Migdal, {\it Nuclear Theory: the Quasiparticle Method}, W.A. Benjamin, N.Y., 1968.

\bibitem{M67a}
A.B.~Migdal, Zh.\ Eksp.\ Teor.\ Fiz.\ {\bf 43}, 1940 (1962) [Sov. JETP {\bf 16}, 1366 (1963)];

\bibitem{M67}
A.B. Migdal, {\it Theory of Finite Fermi Systems and properties of Atomic Nuclei},
Wiley and Sons, N.Y., 1967 (Russian edition 1965),
2nd edition, Nauka, Moscow, 1983 (in Russian).


\bibitem{BCS57} J. Bardeen,  L.N. Cooper, and J.R. Schriffer,
Phys. Rev. {\bf 106}, 162 (1957); {\it ibid}. {\bf 108}, 1175 (1957).

\bibitem{Schriffer} J.R.~Schriffer, {\it Theory of Superconductivity}, Benjamin, N.Y., 1964.

\bibitem{Bog58}
N.N. Bogoliubov, Zh.\ Eksp.\ Teor.\ Fiz.\ {\bf 34}, 58 (1958) [Sov. Phys. JETP {\bf 34}, 41 (1958)],
Nuovo Cim. {\bf 7}, 794 (1958).

\bibitem{Gor58}
L.P. Gorkov, Zh.\ Eksp.\ Teor.\ Fiz.\ {\bf 34}, 735 (1958) [Sov. Phys. JETP {\bf 7}, 505 (1958)].

\bibitem{Nam60} Y. Nambu, Phys. Rev. {\bf 117}, 648 (1960).

\bibitem{El60} G.M. Eliashberg, Zh.\ Eksp.\ Teor.\ Fiz.\ {\bf 38}, 966 (1960)  [Sov. Phys. JETP {\bf 11}, 696 (1960)].

\bibitem{MigEl} A.B. Migdal,
Zh.\ Eksp.\ Teor.\ Fiz.\ {\bf 34}, 1438 (1958) [Sov. Phys. JETP {\bf 7}, 996 (1958)].

\bibitem{M59} A.B. Migdal,
Zh.\ Eksp.\ Teor.\ Fiz.\ {\bf 37}, 249 (1959) [Sov. Phys. JETP {\bf 10}, 176 (1960)].


\bibitem{LM63}
A.I. Larkin and A.B. Migdal,
Zh.\ Eksp.\ Teor.\ Fiz.\ {\bf 44}, 1703 (1963) [Sov. Phys. JETP {\bf 17}, 1146 (1963)].

\bibitem{Leg65a}
A.J. Leggett, Phys. Rev. {\bf 140}, A1869 (1965).

\bibitem{Leg65}
A.J. Leggett, Phys. Rev. {\bf 147}, 119 (1966).

\bibitem{Schw61}
J. Schwinger, J. Math. Phys. {\bf 2}, 407 (1961).

\bibitem{KB62}
L.P. Kadanoff and G. Baym, {\it Quantum Statistical Mechanics}, Benjamin, 1962.

\bibitem{Kel64}
L.P. Keldysh, Zh.\ Eksp.\ Teor.\ Fiz.\ {\bf 47}, 1515 (1964) [Sov. Phys. JETP {\bf 20}, 1018 (1965)].

\bibitem{LandauP}  L. D. Landau and I. Pomeranchuk,
Dokl. Akad. Nauk SSSR {\bf 92} 553 (1953); {\it ibid.} 735.

\bibitem{LPM} A.B. Migdal,  Phys. Rev. {\bf 103}, 1811 (1956);
Zh.\ Eksp.\ Teor.\ Fiz.\ {\bf 32}, 633 (1957) [Sov. Phys. JETP {\bf 5}, 527 (1957)].

\bibitem{eSLAC}  P.L. Anthony, et al., Phys. Rev. Lett. {\bf 75}, 1949 (1995).

\bibitem{SKlein}S. Klein, Rev. Mod. Phys. {\bf 71}, 1501 (1999).

\bibitem{M71} A.B. Migdal, Zh.\ Eksp.\ Teor.\ Fiz.\ {\bf 61}, 2210 (1971) [Sov. Phys. JETP {\bf 34}, 1184 (1972)].

\bibitem{SS72} R. Sawyer, Phys. Rev. Lett. {\bf 29}, 382 (1972);
               R.F. Scalapino, Phys. Rev. Lett. {\bf 29}, 386 (1972).

\bibitem{M78}  A.B. Migdal, Rev. Mod. Phys. {\bf{50}}, 107 (1978).

\bibitem{MSTV90} A.B. Migdal, E.E. Saperstein, M.A. Troitsky, and
 D.N. Voskresensky, Phys. Rep. {\bf 192}, 179 (1990).

\bibitem{BLRT94} V.P.~Berezovoy, I.V.~Krive, and  E.M.~Chudnovsky,
Sov.\ J. Nucl.\ Phys.\ {\bf 30}, 581 (1979);
D.~Kaplan and A.~Nelson, Phys.\ Lett.\ {\bf B 175}, 57 (1986);
T.~Muto and T.~Tatsumi, Phys.\ Lett.\ B {\bf 283}, 165  (1992);
G.E.~Brown, V.~Thorsson, K.~Kubodera, and M.~Rho, Phys.\ Lett.\ {\bf B291}, 355 (1992);
G.E. Brown, C.H. Lee, M. Rho, and V. Thorsson, Nucl. Phys. {\bf A567}, 937 (1994).
E.E. Kolomeitsev, D.N. Voskresensky, and B. K\"ampfer, Nucl. Phys. A {\bf 588}, 889 (1995);
E.E. Kolomeitsev and D.N. Voskresensky, Phys. Rev. {\bf C68}, 015803 (2003).

\bibitem{V97} D.N. Voskresensky, Phys. Lett. {\bf B392}, 262 (1997).

\bibitem{KV05} E.E. Kolomeitsev and D.N. Voskresensky, Nucl. Phys. A {\bf 759}, 373 (2005).

\bibitem{VM78}
D.N. Voskresensky and I.N. Mishustin, Pis'ma Zh.\ Eksp.\ Teor.\ Fiz.\ {\bf 28}, 486 (1978)  [JETP Lett. {\bf 28}, 449 (1978)].

\bibitem{VM82}
D.N. Voskresensky and I.N. Mishustin, Pis'ma Zh.\ Eksp.\ Teor.\ Fiz.\ {\bf 34}, 317 (1981)
[JETP Lett. {\bf 34}, 303 (1981)];
Yad. Fiz. {\bf 35}, 1139 ( 1982) [Sov. J. Nucl. Phys. {\bf 35}, 667 (1982)].

\bibitem{D82}
A.M. Dyugaev, Pisma v Zh.\ Eksp.\ Teor.\ Fiz.\. {\bf 35}, 341 (1982);\\
              Zh.\ Eksp.\ Teor.\ Fiz.\ {\bf 83}, 1005 (1982) [Sov. JETP {\bf 56}, 567 (1982)];\\
              Yad. Fiz. {\bf 38}, 1131 (1983)[ Sov. J. Nucl. Phys. {\bf 38}, 680 (1983)].

\bibitem{VS87} D.N. Voskresensky and A.V. Senatorov,
Yad. Fiz. {\bf 45}, 657 (1987) [Sov. J. Nucl. Phys. {\bf 45}, 411 (1987)].

\bibitem{V93}
D. N. Voskresensky, Yad. Fiz. 50, 1583 (1989) [Sov. J. Nucl. Phys. {\bf 50}, 983 (1989)];
{\it ibid} {\bf 55}, 368 (1992) [{\it ibid.} {\bf 55}, 202 (1992)];
Nucl. Phys. A {\bf 555}, 293 (1993).

\bibitem{VBRS95}
D.N. Voskresensky, D. Blaschke, G. R\"opke, and H. Schulz, Int. J. Mod. Phys. E {\bf 4}, 1 (1995).

\bibitem{V01} D.N. Voskresensky, Lect. Notes Phys. {\bf 578}, 467 (2001) [eprint: astro-ph/0101514].

\bibitem{RW}
R. Rapp, G. Chanfray, and J. Wambach, Nucl. Phys. A {\bf 617}, 472 (1997).


\bibitem{MMMS77} A.B. Migdal, Phys. Lett. {\bf 52}, 172 (1974);
A.B. Migdal, O.A. Markin, I.N. Mishustin, and G.A. Sorokin, Phys. Lett. B {\bf 65}, 423 (1977).

\bibitem{VSC77} A.B. Migdal, V.S. Popov, and D.N. Voskresensky,
Zh.\ Eksp.\ Teor.\ Fiz.\ {\bf 72}, 834 (1977) [Sov. Phys. JETP {\bf 45}, 436 (1977)];
D.N. Voskresensky, G.A. Sorokin, and A.I. Chernoutsan, Pis'ma Zh.\ Eksp.\ Teor.\ Fiz.\ {\bf 25}, 495 (1977)
   [JETP. Lett {\bf 25}, 465 (1977)];
D.N. Voskresensky and A.I. Chernoutsan, Yad. Fiz. {\bf 27}, 1411 (1978)
   [Sov. J. Nucl. Phys. {\bf 27}, 742 (1978)].



\bibitem{W84} E. Witten, Phys. Rev. D {\bf 30}, 272 (1984).

\bibitem{AFO} Ch. Alcock, E. Farhi, and  A. Olinto, Astrophys. J. {\bf 310}, 261 (1986) .

\bibitem{Bombaci} Madappa Prakash, I. Bombaci, Manju Prakash, P.J. Ellis, J.M.
Lattimer, and R. Knorren, Phys. Rept. {\bf 280}, 1 (1997).

\bibitem{BKV00} D. Blaschke, T. Kl\"ahn, and D.N. Voskresensky, Astrophys. J. {\bf 533}, 406 (2000);
H. Grigorian, D. Blaschke, and D. Voskresensky, Phys. Rev. C {\bf 71}, 045801 (2005).

\bibitem{VS86} D.N. Voskresensky and A.V. Senatorov,
Zh.\ Eksp.\ Teor.\ Fiz.\. {\bf 90}, 1505 (1986) [Sov. Phys. JETP {\bf 63}, 885 (1986)].

\bibitem{KV95} J. Knoll and  D. N. Voskresensky, Phys. Lett. B {\bf 351}, 43 (1995);
               Ann. Phys. {\bf 249}, 532 (1996).

\bibitem{KV99}  E.E. Kolomeitsev and D.N. Voskresensky, Phys. Rev. C {\bf 60}, 034610 (1999).


\bibitem{BW65}  J.N. Bahcall and R.A. Wolf, Phys. Rev. B {\bf 140}, 1445 (1965).

\bibitem{TC65}  S. Tsuruta and A.G.W. Cameron, Canad. J. Phys. {\bf 43}, 2056 (1965).

\bibitem{FM79}  B. Friman and O.V. Maxwell, Astrophys. J. {\bf 232}, 541 (1979).

\bibitem{M79}   O.V. Maxwell, Astrophys. J. {\bf 231}, 201 (1979).

\bibitem{T79}   S. Tsuruta, Phys. Rept. {\bf 56}, 237 (1979).

\bibitem{NT81}  K. Nomoto and S. Tsuruta, Astrophys. J. Lett. {\bf 250}, 19 (1981).

\bibitem{SWWG96} Ch. Schaab, F.Weber, M.K. Weigel, and N.K. Glendenning, Nucl. Phys. A {\bf 605}, 531 (1996).

\bibitem{FSB75} G. Flowers, P.G. Sutherland, and J.R. Bond, Phys. Rev. D {\bf 12}, 315 (1975).

\bibitem{LPPH91}  J.M. Lattimer, C.J. Pethick, M. Prakash, and P. Haensel, Phys. Rev. Lett. {\bf 66}, (1991).

\bibitem{MBCDM77} O. Maxwell, G.E. Brown, D.K. Campbell, R.F. Dashen, and J.T. Manassah, Astrophys. J. {\bf 216}, 77 (1977).

\bibitem{VS84} D.N. Voskresensky and A.V. Senatorov, Pis'ma Zh.\ Eksp.\ Teor.\ Fiz.\ {\bf 40}, 395 (1984)
                [JETP Lett. {\bf 40}, 1212 (1984)].

\bibitem{BKPP88}  G.E. Brown,  K. Kubodera, D. Page, and P. Pizzochero, Phys. Rev. D {\bf 37}, 2042 (1988).

\bibitem{T88} T. Tatsumi, Prog. Theor. Phys. {\bf 88}, 22 (1988).

\bibitem{Iwamoto} N. Iwamoto, Ann. Phys. {\bf 141}, 1 (1982).

\bibitem{VKZC00} D.N.~Voskresensky, V.A.~Khodel, M.V.~Zverev, and J.W.~Clark,
                 Apstrophys. J. Lett. {\bf 533}, 127 (2000).
\bibitem{APR98}
A. Akmal, V.R. Pandharipande, and D.G. Ravenhall, Phys. Rev. C {\bf 58}, 1804 (1988).

\bibitem{BGV} D. Blaschke, H. Grigorian, and D.N. Voskresensky, Astron. Astrophys. {\bf 424}, 979 (2004).

\bibitem{GV} H. Grigorian and D. N. Voskresensky, Astron. Astrophys. {\bf 444}, 913 (2005).

\bibitem{Army}T. Kl\"ahn {\it et al.}, Phys. Rev. C {\bf 74}, 035802 (2006).

\bibitem{SV87} A.V. Senatorov and D.N. Voskresensky, Phys. Lett. B {\bf 184}, 119 (1987).

\bibitem{BRSSV}
D. Blaschke, G. R\"opke, H. Schulz, A.D. Sedrakian, and D.N. Voskresensky,
      Mon. Not. R. Astron. Soc. {\bf 273}, 596 (1995).

\bibitem{HPR} C. Hanhart, D.R. Phillips, and S. Reddy,  Phys. Lett. B {\bf 499}, 9 (2001).

\bibitem{Schwenk04} A.~Schwenk, P.~Jaikumar, and Ch.~Gale, Phys. Lett B {\bf 584}, 241 (2004).

\bibitem{RPLP} S.~Reddy, Madappa~Prakash, J.M.~Lattimer, and J.A.~Pons, Phys. Rev. C {\bf 59}, 2888 (1999).

\bibitem{ST83} S.~Shapiro and S.A.~Teukolsky,
{\it Black Holes, White  Dwarfs and Neutron Stars: The Physics of Compact Objects}, Wiley, New York 1983,
 chapter 11.

\bibitem{KCZ} V.A.~Khodel, J.W.~Clark, and M.V.~Zverev. Phys. Rev. Lett. {\bf 87}, 031103 (2001).

\bibitem{YKL99}  D.G.~Yakovlev, A.D.~Kaminker, and K.P.~Levenfish, Astron. Astrophys. {\bf 343}, 650 (1999).


\bibitem{APW} J.~Wambach, T.L.~Ainsworth, and D.~Pines, Nucl. Phys. A {\bf 555}, 128 (1993).

\bibitem{Tamagaki70} R.~Tamagaki, Prog. Theor. Phys. {\bf 44}, 905 (1970);

\bibitem{Amundsen85} L.~Amundsen and E. \O{}stgaard Nucl. Phys. A {\bf 437}, 487 (1985); {it ibid.}
{\bf 442}, 163 (1985)

\bibitem{Takatsuka93} T.~Takatsuka and  R.~Tamagaki, Prog. Theor. Phys. Suppl. {\bf 112}, 27 (1993).

\bibitem{KKC96} V.A~Khodel, V.V~Khodel, and J.W.~Clark, Nucl. Phys. A {\bf 598}, 390 (1996).

\bibitem{Schulze96} H.-J.~Schulze, J. Cugnon, A.~Lejeune, M.~Baldo, and U.~Lombardo, Phys. Lett. B
{\bf 375}, 1 (1996).

\bibitem{Elgaroy98} \O{}.~Elgar\o{}y and M.~Hjorth-Jensen, Phys. Rev. C {\bf 57}, 1174 (1998).

\bibitem{Khodel01} V.A.~Khodel, Yad. Fiz. {\bf 64}, 447 (2001) [Phys. At. Nucl. {\bf 64}, 393 (2001)].

\bibitem{SF} A. Schwenk and B. Friman, Phys. Rev. Lett. {\bf 92}, 082501 (2004).

\bibitem{KCTZ} V.A.~Khodel, J.W.~Clark, M.~Takano, and M.V.~Zverev, Phys. Rev. Lett. {\bf 93}, 151101 (2004).

\bibitem{Hebeler07} K.~Hebeler, A.~Schwenk, and B.~Friman, Phys. Lett. B {\bf 648}, 176 (2007).

\bibitem{Chen08} W.~Chen, B.~Li, D.~Wen, and L.~Liu, Phys. Rev. C {\bf 77}, 065804 (2008).

\bibitem{SC06} A.~Sedrakian and J.W.~Clark,
in {\it Pairing in Fermionic Systems: Basic Concepts and Modern Applications},
Series on Advances in Quantum Many-Body Theory Vol. 8, eds. A. Sedrakian, J.W. Clark, and M.
Alford, World Scientific, Singapore, 2006, 135 [eprint:  nucl-th/0607028].

\bibitem{PPLS10}
  D.~Page, M.~Prakash, J.M.~Lattimer, and  A.W.~Steiner, arXiv:1011.6142.

\bibitem{YLS99} D.G.~Yakovlev, K.P.~Levenfish, and Yu.A.~Shibanov,
Usp. Fiz. Nauk {\bf 169}, 825 (1999) [Sov. Phys. Usp. {\bf 42}, 737 (1999)].

\bibitem{FRS76} E.~Flowers, M.~Ruderman, and P.~Sutherland,  Astrophys. J. {\bf 205 }, 541 (1976).

\bibitem{KHY} A.D.~Kaminker, P.~Haensel, and D.G.~Yakovlev, Astron. Astrophys. {\bf 345}, L14 (1999).

\bibitem{LP} L.B.~Leinson and A.~Perez, Phys. Lett. B {\bf 638}, 114 (2006).

\bibitem{KV1} E.E.~Kolomeitsev and D.N.~Voskresensky, Phys.\ Rev.\ C {\bf 77}, 065808 (2008).

\bibitem{KV2}E.E.~Kolomeitsev and D.N.~Voskresensky, Phys.\ Rev.\ C {\bf 81}, 065801 (2010).

\bibitem{LPpairing} L.B.~Leinson, Phys. Rev. C {\bf 81}, 025501 (2010).

\bibitem{SMS} A.~Sedrakian, H.~Muther, and P.~Schuck, Phys. Rev. C {\bf 76}, 055805 (2007);
              A.~Sedrakian and J.~Keller, Phys. Rev. C {\bf 81}, 045806 (2010).

\bibitem{SR09} A.W.~Steiner and S.~Reddy, Phys. Rev. C {\bf 79}, 015802 (2009).


\bibitem{KR} J.~Kundu and S.~Reddy, Phys. Rev. C {\bf 70}, 055803 (2004).

\bibitem{VKK} D.N.~Voskresensky, E.E.~Kolomeitsev, and B.~K\"ampfer,
              Zh.\ Eksp.\ Teor.\ Fiz.\ {\bf 114}, 385 (1998) [JETP {\bf 87}, 211 (1998)].

\bibitem{L00} L.B.~Leinson, Phys. Lett. B {\bf 473}, 318 (2000).

\bibitem{GBSMK} S.~Gupta, E.F.~Brown, H.~Schatz, P.~M\"oller, and K.-L.~Kratz,
Astrophys. J {\bf 662}, 1188 (2007).

\bibitem{SVSWW97} Ch.~Schaab, D.~Voskresensky, A.D.~Sedrakian, F.~Weber, and  M.K.~Weigel,
Astron. Astrophys. {\bf 321}, 591 (1997).

\bibitem{P98} D.~Page, {\it Many Faces of Neutron Stars},
eds. R. Buccheri, J. van Paradijs, and M.A. Alpar, Kluwer,Dordrecht, 1998, p. 538.

\bibitem{Page} D.~Page, J.M.~Lattimer, M.~Prakash, and A.W.~Steiner, Astrophys. J. Suppl
{\bf 155}, 623 (2004)

\bibitem{IKV99} Yu.B.~Ivanov, J.~Knoll, and D.N.~Voskresensky, Nucl. Phys. A {\bf 657}, 413 (1999).

\bibitem{IKV00} Yu.B.~Ivanov, J.~Knoll, and D.N.~Voskresensky, Nucl. Phys. A {\bf 672}, 314 (2000).

\bibitem{KIV01} J.~Knoll, Yu.B.~Ivanov, and D.N.~Voskresensky, Ann. Phys. {\bf 293}, 126 (2001);
Yu.B.~Ivanov, J.~Knoll, and D.N.~Voskresensky,
Yad. Fiz. {\bf 66}, 1950 (2003) [Phys. Atom. Nucl. {\bf 66}, 1902 (2003)].

\bibitem{Bogolyubov}
N.N. Bogoliubov, {\it Problems of Dynamical Theory in Statistical Physics}, Gostekhisdat, Moscow, 1946;\\
N.N. Bogoliubov, J. Phys. (USSR) {\bf 10}, 256 (1946).

\bibitem{Klimontovich}
Yu.L. Klimontovich, {\it Statisticheskaya Fizika}, Nauka, Moscow, 1982
[transl. {\it Statistical Physics}, Gordon and  Breach, New York, 1986].

\bibitem{Dan84} P.~Danielewicz, Ann. Phys. (N.Y.) {\bf 152}, 239 (1984); {\it ibid.} 305.

\bibitem{Lif81} E.M.~Lifshitz and L.P.~Pitaevskii, {\it Physical Kinetics}, Pergamon, Oxford 1981.

\bibitem{Koshelkin99} A.V.~Koshelkin, Phys. Lett B {\bf 471}, 202 (1999).

\bibitem{Bornath99} Th.~Bornath, D.~Kremp and M.~Schlanges, Phys. Rev. E {\bf 60}, 6382 (1999).

\bibitem{Baym} G.~Baym, Phys. Rev. {\bf 127}, 1391 (1962).

\bibitem{IV09} Yu.B.~Ivanov and D.N.~Voskresensky,  Phys. Atom. Nucl. {\bf 72}, 1168 (2009).

\bibitem{SST99} T.~Suzuki, H.~Sakai, and T.~Tatsumi,
nucl-th/9901097; A.~Arima, W.~Bentz, T.~Suzuki, and T.~Suzuki, Phys. Lett., {\bf 499}, 104 (2001).

\bibitem{LKR}
C.L.~Korpa, M.F.M.~Lutz, and  F.~Riek,  Phys. Rev. C {\bf 80}, 024901 (2009).

\bibitem{Nakano} M.~Nakano {\it et al.}, Int. J. Mod. Phys. E {\bf 10}, 459 (2001).

\bibitem{Ripka} H.~Toki, S.~Sugimoto, and K.~Ikeda, Prog. Theor. Phys. {\bf 108}, 903 (2002);
S.~Sugimoto, K.~Ikeda, and H.~Toki, Phys. Rev. C {\bf 75}, 014317 (2007);
G.~Ripka, Nucl. Phys. A {\bf 814}, 33 (2008).

\bibitem{APR} A.~Akmal, V.R.~Pandharipande, and D.G.~Ravenhall,  Phys. Rev. C {\bf 58}, 1804 (1998).

\bibitem{D75} A.M.~Dyugaev, Pis'ma Zh.\ Eksp.\ Teor.\ Fiz.\ {\bf 22}, 181 (1975) [JETP Lett {\bf 22}, 83 (1975)].

\bibitem{EW} T.E.O.~Ericson and W.~Weise, {\it Pions And Nuclei}, Claredon, Oxford, 1988.

\bibitem{Fauser} R. Fauser, Nucl. Phys. A {\bf 606}, 479 (1996).

\bibitem{SD99} A. Sedrakian and A. Dieperink, Phys. Lett. B {\bf 463}, 145 (1999);
               Phys. Rev. D {\bf 62}, 083002 (2000).

\bibitem{RLandau83} R.H. Landau, Phys. Rev. C {\bf 27}, 2191 (1983).

\bibitem{SFL} C.M.~Varma, Z.~Nussinov, and W.~van~Saarloos, Phys. Rep. {\bf 361}, 267 (2002).

\bibitem{Fujita87} T.~Fujita and K.F.~Quader, Phys. Rev. B {\bf 36}, 5152 (1987).

\bibitem{Dabrowski76} J.~Dabrowski and P.~Haensel, Ann. Phys. {\bf 97}, 452 (1976).

\bibitem{Backman79} S.-O.~B\"ackman, O.~Sj\"oberg, and A.D.~Jackson, Nucl. Phys. A {\bf 321}, 10
(1979).

\bibitem{Friman81} B.L.~Friman and P. Haensel, Phys. Lett. B {\bf 98}, 323 (1981).

\bibitem{Sedr} A.~Sedrakian, Phys. Rev. C {\bf 68} (2003) 065805.

\bibitem{Backman68} S.-O.~B\"ackman, Nucl. Phys. A {\bf 120}, 593 (1968).

\bibitem{Backman73} S.-O.~B\"ackman, C.-G.~K\"allman, and O.~Sj\"oberg, Phys. Lett. B {\bf 43},
263 (1973).

\bibitem{Babu73} S.~Babu and G.E.~Brown, Ann. Phys. {\bf 78}, 1 (1973).


\bibitem{Celenza82}  L.S.~Celenza, W.S.~Pong, and C.M.~Shakin, Phys. Rev. C {\bf 25}, 3115 (1982).

\bibitem{Khodel82} V.A.~Khodel  and E.E. Saperstein, Phys. Rept. {\bf 92}, 183 (1982).

\bibitem{Saperstein}
I.N.~Borzov, S.V.~Tolokonnikov, and S.A.~Fayans, Yad. Fiz. {\bf 40}, 1151 (1984)
[Sov. J. Nucl. Phys. {\bf 40} (1984) 732].
E.E.~Saperstein and S.V.~Tolokonnikov, Pis'ma Zh.\ Eksp.\ Teor.\ Fiz.\ {\bf 68}, 529 (1998) [JETP Lett. {\bf 68}, 553 (1998)];
S.A.~Fayans and D.~Zawischa, Phys. Lett. B {\bf  363}, 12 (1995).

\bibitem{RFF}
V.A.~Rodin, A.~Faessler, F.~\v{S}imkovic, and P.~Vogel,
Nucl. Phys. A {\bf 766}, 107 (2006); Erratum: {\it ibid.} {\bf 793}, 213 (2007).


\bibitem{BFKZ} I.N.~Borzov, S.A.~Fayans, E.~Kr\"omer, and  D.~Zawischa, Z. Phys. A {\bf 355}, 117 (1996).

\bibitem{Borzov03} I.N.~Borzov, Phys. Rev. C {\bf 67}, 025802 (2003).

\bibitem{FTTZ} S.A.~Fayans, S.V.~Tolokonnikov, E.L.~Trykov, and D.~Zawischa, Nucl. Phys. A {\bf 676}, 49 (2000).

\bibitem{VGL62} V.G.~Vaks, V.M.~Galitsky, and A.I.~Larkin, Sov. Phys. JETP {\bf 14}, 1177 (1962).

\bibitem{Pyatov} N.I.~Pyatov and S.A.~Fayans, Sov. J. Part. Nuclei, {\bf 14}, 401 (1983).

\bibitem{KMS}
R.~Kubo, J. Phys. Soc. Jap. {\bf 12}, 570 (1957);
P.C.~Martin and J.~Schwinger,  Phys. Rev. {\bf 115}, 1342 (1959).

\bibitem{W93} H.A.~Weldon,  Phys. Rev. D {\bf 47}, 594 (1993).




\end{thebibliography}
\end{document}